\documentclass[
reprint,
prx,
superscriptaddress,
showpacs,
amsmath,
amssymb,
]{revtex4-2}

\usepackage{lineno}
\usepackage{mathtools}
\usepackage{color,soul}
\usepackage{graphicx}
\usepackage{dcolumn}
\usepackage{bm}
\usepackage{feynmp-auto}
\usepackage{amssymb,amsfonts,amsmath}
\usepackage{threeparttable}
\usepackage{textcomp}
\usepackage{float}
\usepackage{wrapfig}
\usepackage{verbatim}
\usepackage{accents}
\usepackage{fancyhdr}
\usepackage{euscript}
\usepackage{blkarray}
\usepackage{multirow}
\usepackage{tikz}
\usepackage{subfigure}
\usepackage{enumitem}
\usepackage{physics}
\usepackage[T1]{fontenc}
\usepackage{mathptmx}
\usepackage{mhchem}
\usetikzlibrary[topaths]
\usepackage[colorlinks,
            linkcolor=red,
            anchorcolor=blue,
            citecolor=blue
            ]{hyperref}

\graphicspath{ {./figs/} } 

\begin{document}
\title{
Interacting topological Dirac magnons}

\author{Hao Sun}
\email{sun.hao@nus.edu.sg}
\affiliation{The Institute for Functional Intelligent Materials (I-FIM), National University of Singapore, 4 Science Drive 2, Singapore 117544}

\author{Dhiman Bhowmick}
\affiliation{School of Physical and Mathematical Sciences, Nanyang Technological University, Singapore}

\author{Bo Yang}
\email{yang.bo@ntu.edu.sg}
\affiliation{School of Physical and Mathematical Sciences, Nanyang Technological University, Singapore}
\affiliation{Institute of High Performance Computing, A*STAR, Singapore}

\author{Pinaki Sengupta}
\email{psengupta@ntu.edu.sg}
\affiliation{School of Physical and Mathematical Sciences, Nanyang Technological University, Singapore}
 
\date{\today}

\begin{abstract}
In this work, we study the magnon-magnon interaction effect in typical honeycomb ferromagnets consisting of van der Waals-bonded stacks of honeycomb layers, e.g., chromium trihalides CrX3 (X = F, Cl, Br and I), that display two spin-wave modes (Dirac magnon). Using Green's function formalism with the presence of the Dzyaloshinskii–Moriya interaction, we obtain a spinor Dyson equation up to the second-order approximation by the cluster expansion method. Numerical calculations show prominent renormalizations of the single-particle spectrum. Furthermore, we propose a tunable renormalization effect using a parametric magnon amplification scheme. By amplifying the magnon population at different k points, the enabled renormalization effect not only reshapes the band structure but also modifies the Berry curvature distribution. Our work demonstrates the interplay between band geometry, interactions, and the external light field in the bosonic system and can potentially lead to new insights into the properties of magnon-based spintronic devices.
\end{abstract}

\maketitle
\section{Introduction}\label{intro}
Following the phenomenal growth in the study of Dirac fermions in graphene and other (quasi-) two dimensional materials~\cite{Novoselov2005,Martino2007,Park2008,Park2008_2,Yan2008,Zhu2010,Castro2009,Borisenko2014,Sun2016,Sun_2016}, there has been growing interest in recent years in investigating bosonic analogs of the same~\cite{Chumak2015,Fransson2016,Sun2021,Xu2016,Pershoguba2018,Cheng2007,Wu2015,Lu2016,Wang2015, Lado_2017},
in such diverse platforms as photonic, phononic, and magnonic systems. Magnons are low-energy quasiparticle excitations in quantum magnets and have long served as a versatile testbed for realizing bosonic analogs of fermionic phases.

Since magnetic properties are easily controlled by external magnetic fields, magnonic bands offer a unique platform to explore the rich and still evolving fundamentals of the band theory. So far, several topological magnon phases have been proposed and material candidates have been identified~\cite{Zhang2013,Owerre_2016,Cai_2019,Chen2018,Cai_2021,Nguyen_2021, Costa_2020, McClarty2022}. Some key results include the characterization of the Chern magnon bands and linear touching points -- for example, the material \ce{Cu[1,3-bdc]} with the kagome lattice exhibits a magnetic field-dependent thermal Hall conductivity, and the material \ce{Cu3TeO6} shows a nodal topology~\cite{Hirschberger2015, Yuan2020}.

Magnonic systems are also ideal platforms for investigating interacting bosons~\cite{Pershoguba2018, Mook2021}. 
However, so far, studies of magnonic band topology have largely neglected magnon-magnon interactions, treating the magnons within the single-particle approximation~\cite{Owerre_2016}. This is typically justified by the argument that interaction effects are frozen out at low temperatures or are negligible as a small perturbative effect. But when the magnon density is enhanced due to increased temperature or parametric amplification (as discussed later), the single-particle picture does not generally apply, and the effects of interactions become important. 
In Dirac fermions, Coulomb interaction leads to a logarithmic renormalization of the Fermi velocity~\cite{Elias2011}, and similar analyses for interacting magnons have shown that both velocity and energy are renormalized due to interaction in the vicinity of the Dirac points~\cite{Pershoguba2018}. In contrast to fermions, magnon number is not conserved and allows for nonconserved many-body interactions such as spontaneous decay of magnons~\cite{Mook2021, Zhitomirsky2013}. 

In this work, we have explored the effects of interactions on {\it topological} magnons in a Heisenberg ferromagnet with Dzyaloshinskii-Moriya interaction (DMI) on a honeycomb lattice. In the absence of DMI, the non-interacting magnon spectrum features two bands that touch at the Dirac points with linear dispersion. DMI opens up a gap in the spectrum and imparts non-trivial topology to the bands. We have investigated the effects of interactions on the magnon band topology using Green's function formalism up to the second order in perturbation theory. Crucially, we are able to systematically probe the interaction-driven renormalization of the magnon bands in different parts of the magnetic Brillouin zone and isolate the effects of interactions from thermal fluctuations, by employing the recently developed magnon amplification scheme to controllably tune magnon density at selective energy-momentum values~\cite{Malz2019}.   Our results show that the magnon-magnon interaction leads to a significant momentum-dependent renormalization of the single-magnon dispersion -- interactions suppress the bandwidth, and lead to dissipative scattering of the magnons~\cite{Pershoguba2018, Klein2018}. More interestingly, interaction effects, in conjunction with DMI and magnon amplification, can introduce non-trivial topological transitions in the magnon spectrum, 
leading to magnon bands with tunable Chern numbers. In the same vein, the Berry curvature distribution of the magnon bands can be tailored by varying the interaction strength. We argue from a microscopic perspective that effective gauge fields emerge in this bipartite system through their underlying interaction, and propose that a magnon-mediated anomalous transport - the thermal Hall effect - allows for the experimental investigation of this many-body effect~\cite{Onose297}.
   
The remainder of this paper is structured as follows. Section~\ref{sec.II} introduces the 
framework to study interacting magnons within the Holstein-Primakoff transformation.
The Green's function formalism is introduced in Sec.~\ref{sec.II_A}. 
A low-temperature approximation is discussed in Sec.~\ref{sec.II_B} as an intuitive introduction to the band renormalization, and finally, the renormalization expression is extended beyond the low-temperature approximation in Sec.~\ref{sec.II_C}. Section~\ref{sec.III} focuses on the main calculation results. The emergence of the gauge field and the modification of band geometry due to many-body interaction is discussed using the quantum geometric tensor in Sec.~\ref{sec.III}, and the evolution of the spinor spectral function and the scattering rate with DMI using second-order self-energy is discussed in Sec.~\ref{sec.IV}. Section~\ref{sec.V} presents the results of magnon amplification via parametric instability and experimental feasibility of demonstrating the theoretical results in a real quantum magnet, with Sec.~\ref{sec.V_A} showing how the magnon population can be amplified by the external light field and Sec.~\ref{sec.V_B} examining the relationship between magnon populations and the thermal Hall conductivities. We end with a summary and conclusions in Sec.~\ref{con}.
\begin{figure}[t!]
\includegraphics[width=8.5 cm]{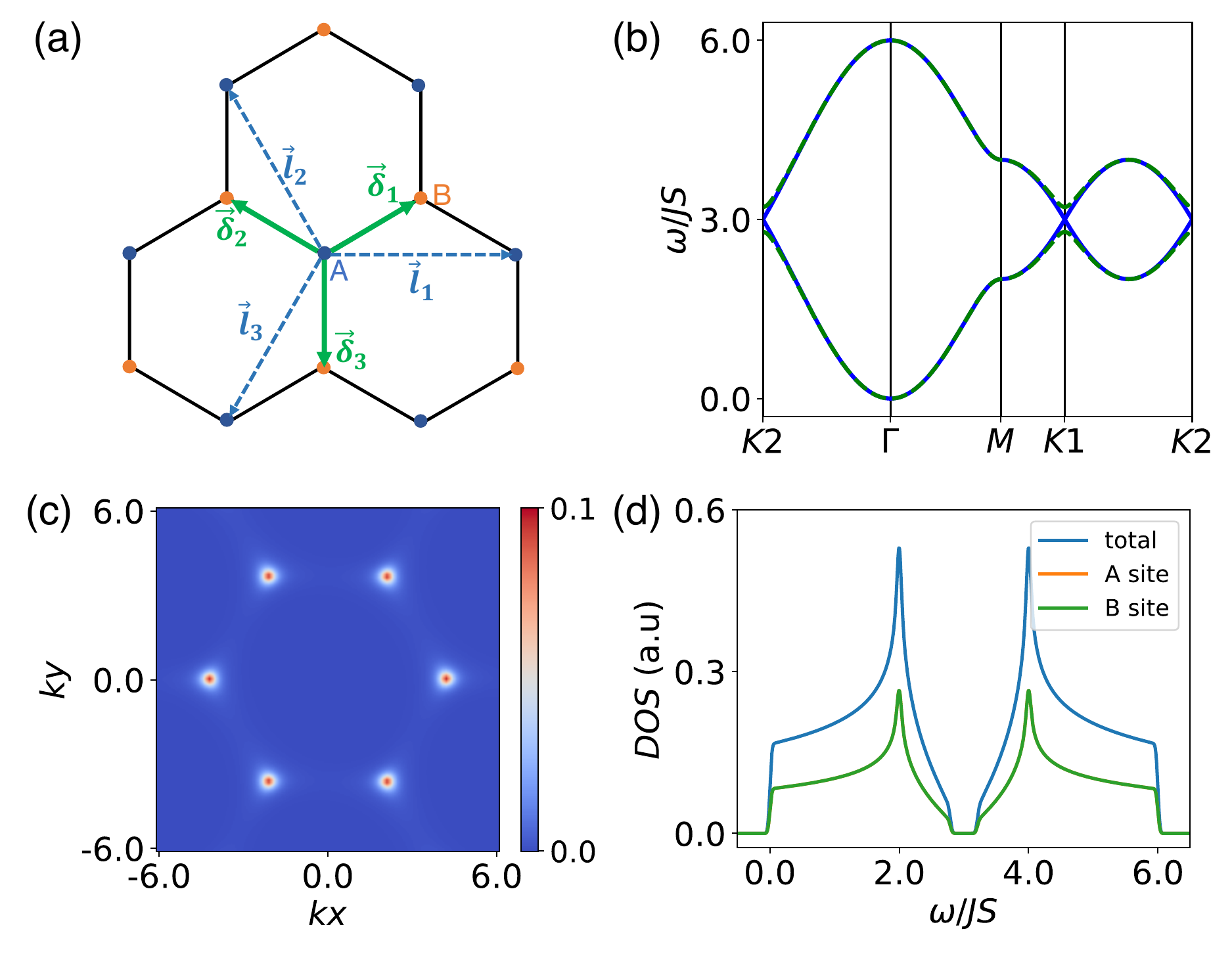}
\caption{Honeycomb structure of ferromagnet and noninteracting spectrum of magnonic excitation. (a) Geometric structure of honeycomb ferromagnets with NN bonds $\bm{\delta}_n$ and NNN bonds $\bm{\sigma}_n$. (b) Calculated magnon spectrum by single-particle Hamiltonian, dashed green curves are the topologically non-trivial bands with $D/J=0.05$. One can see the DMI-induced gaps at the Dirac points. (c) Berry curvature distribution of acoustic branch with $D/J=0.05$. (d) The density of states and partial density of states for gapped magnon bands. One can see the conserved sublattice symmetry from the equal distributions of A and B sites. }
\label{fig:1}
\end{figure}

\section{Dirac magnons in honeycomb ferromagnet}\label{sec.II}
The Heisenberg model on the honeycomb lattice is described by the Hamiltonian:
\begin{equation}
\begin{split}
H=-J\sum_{\langle ij\rangle}\bm{S}_i\cdot\bm{S}_j,
\end{split}
\end{equation}
where $J>0$ denotes the strength of ferromagnetic nearest neighbor (NN) exchange coupling and $\bm{S}_i$ denotes the local spin on the $i$-site of the lattice. The ground state is a ferromagnet, and low energy magnon excitations above the ground state are obtained through the linearized Holstein-Primakoff (HP) transformation~\cite{HP1940}:
\begin{equation}
\begin{split}
S^x_i+iS^y_i=\sqrt{2S}a_i,\quad S^x_i-iS^y_i=\sqrt{2S}a^{\dagger}_i,\quad S^z_i=S-a^{\dagger}_ia_i.
\end{split}
\end{equation}
$a~(a^{\dagger})$ is the magnon annihilation (creation) operator, and obey commutation relation $[a_i, a^{\dag}_j]=\delta_{ij}$. In the linear spin wave theory limit (neglecting any magnon-magnon interaction), the above Hamiltonian reduces to an effective tight-binding model of magnons:
\begin{equation}
\begin{split}
H=-JS\sum_{\langle ij\rangle}a^{\dagger}_ia_j+3JS\sum_{i}a^{\dagger}_ia_i.
\end{split}
\label{eq:rsHamil}
\end{equation}
Here, we have subtracted the ferromagnetic ground state energy $E_G=-3N_LJS^2$. $N_L$ is the number of unit cells\cite{Fransson2016}. 

DMI is ubiquitous in many quantum magnets and is the dominant interaction driving the topological properties of the magnons. Here we introduce a physically relevant out-of-plane next nearest neighbor (NNN) DMI,  $\bm{D}_{i'j'}\cdot(\bm{S}_{i'}\times\bm{S}_{j'})=\sum_{\alpha\beta\gamma}D\epsilon_{\alpha\beta\gamma}v^{\alpha}_{ij}S^{\beta}_iS^{\gamma}_j$, where $\bm{v}_{i'j'}=\frac{\bm{d}^1_{i'j'}\times\bm{d}^2_{i'j'}}{\abs{\bm{d}^1_{i'j'}\times\bm{d}^2_{i'j'}}}=(0,0,\pm1)$, and $D$ is the strength of the DM vector. We redefine the HP boson operator with two flavors to account for the sublattice (pseudospin) degree of freedom: $a_i\to a_i, \;\; a_j \to b_{i+\bm{\delta}_n}$. The $\bm{\delta}_{n}$ are the three NN bonds of magnon orbital with $\bm{\delta}_{1}=\left(\frac{1}{2},\frac{1}{2\sqrt{3}}\right)$, $\bm{\delta}_{2}=\left(\frac{-1}{2},\frac{1}{2\sqrt{3}}\right)$ and $\bm{\delta}_{3}=\left(0,\frac{-1}{\sqrt{3}}\right)$, shown by the green arrows shown in Fig.~\ref{fig:1}(a). Fourier transformation of Eq.(\ref{eq:rsHamil}, yields the single-magnon Hamiltonian in the momentum space: $H_0=\sum_{\bm{k}}\psi_{\bm{k}}^{\dagger}M_{\bm{k}} \psi_{\bm{k}}$ with $\psi_{\bm{k}}^{\dagger}=\left(a^{\dagger}_{\bm{k}},b^{\dagger}_{\bm{k}}\right)$. The explicit expression of the $H_0$ is given by
\begin{equation}
\begin{split}
H_0=\sum_{\bm{k}}(a^{\dagger}_{\bm{k}},b^{\dagger}_{\bm{k}})
\begin{pmatrix} 
 3J-2DS\beta_{\bm{k}}&-JS\gamma_{\bm{k}}\\ 
 -JS\gamma^*_{\bm{k}}&3JS+2DS\beta_{\bm{k}}
\end{pmatrix}
\begin{pmatrix} 
 a_{\bm{k}}\\ 
 b_{\bm{k}}
\end{pmatrix}
\end{split},
\end{equation}
where $\gamma_{\bm{k}_i}=\sum_n e^{i\bm{k}_i\cdot \bm{\delta}_n}$, $\beta_{\bm{k}}=\sum_n \Im\left(e^{i\bm{k}\cdot \bm{l}_n}\right)$, and $\bm{l}_n$ are the NNN bonds with $\bm{l}_1=\left(1,0\right)$, $\bm{l}_2=\left(\frac{-1}{2},\frac{\sqrt{3}}{2}\right)$ and $\bm{l}_3=\left(\frac{-1}{2},\frac{-\sqrt{3}}{2}\right)$, shown by the dashed blue arrows, as shown in Fig.~\ref{fig:1}(a). The spectrum of the $H_0$ is readily obtained from the diagonalization of $M_{\bm{k}}$:
\begin{equation}
\begin{split}
\omega_{\bm{k}}=3JS\pm\sqrt{J^2S^2\abs{\gamma_{\bm{k}}}^2+4D^2S^2\beta^2_{\bm{k}}}.
\end{split}
\end{equation}
When $\bm{k}\to 0$, the lowest energy reduces to gapless magnon excitation, $\omega_{\bm{k}}\approx\frac{1}{2}JS\bm{k}^2\to0$ . This gapless Goldstone mode is the consequence of the spontaneously broken symmetry in the ferromagnetic ground state. In the honeycomb lattice, magnonic excitations appear in two flavors, exhibiting a graphene-like band structure. The two bands touch linearly at the Dirac point energy when $D=0$. A nonzero $D$ provides an effective Haldane mass term that opens up a topologically non-trivial band gap $\Delta g=6\sqrt{3}DS$ between the upper and lower branches at the Dirac points. The spectra are plotted in Fig.~\ref{fig:1}(b). One can see the blue curves of the gapless bands with $D=0$ and the dashed green curves of the gapped topological bands with $D/J=0.05$. 

The non-trivial topology of the DMI-driven gapped magnon bands is reflected in the Berry curvature distributions of the lower band shown in Fig.~\ref{fig:1}(c); the finite values with the same sign around the gapped Dirac points contribute to a non-zero integer Chern number $C=1$.
Similar to the corresponding fermionic systems, the non-zero Berry curvature provides an effective gauge field for the bosonic magnons in the k-space and dominates the transport properties. One should note that a small but non-zero magnetic field is necessary for stabilizing the ferromagnetic ground state and the magnon excitations at any non-zero temperature, because the Mermin-Wagner theorem rules out the long-distance ferromagnetic order in 2D systems at finite temperatures without the magnetic field~\cite{Mermin1966}. However, the role of the magnetic field is just to stabilize the ground state; it does not contribute to the topological properties of the bands.

\subsection{Interaction renormalized Dirac magnon}\label{sec.II_A}
Considering the first-order expansion of $\frac{1}{\sqrt{S}}$ in HP transformation, we include the magnon-magnon interaction in the real lattice space:
\begin{equation}
\begin{aligned}[b]
&H'=\frac{J}{4}\sum_{i,n}a^{\dagger}_i b^{\dagger}_{i+\bm{\delta}_n}b_{i+\bm{\delta}_n}b_{i+\bm{\delta}_n}+b^{\dagger}_{i+\bm{\delta}_n} b^{\dagger}_{i+\bm{\delta}_n}a_ib_{i+\bm{\delta}_n}+a^{\dagger}_i b^{\dagger}_{i+\bm{\delta}_n}a_ia_i\\
&+a^{\dagger}_i a^{\dagger}_ia_ib_{i+\bm{\delta}_n}-J\sum_{i,n}a^{\dagger}_ia_ib^{\dagger}_{i+\bm{\delta}_n}b_{i+\bm{\delta}_n}.
\end{aligned}
\end{equation}
The interacting Hamiltonian in the momentum space can be obtained by using a Fourier transformation of the above real space interaction Hamiltonian:
\begin{equation}\label{eq:5}
\begin{aligned}[b]
H'=&\frac{J}{4N_L}\sum_{\{\bm{k}_i\}}\gamma_{\bm{k}_1}a^{\dagger}_{\bm{k}_1}b^{\dagger}_{\bm{k}_2}b_{\bm{k}_3}b_{\bm{k}_4}+\frac{J}{4N_L}\sum_{\{\bm{k}_i\}}\gamma^{*}_{\bm{k}_4}b^{\dagger}_{\bm{k}_1}b^{\dagger}_{\bm{k}_2}b_{\bm{k}_3}a_{\bm{k}_4}\\
+&\frac{J}{4N_L}\sum_{\{\bm{k}_i\}}\gamma^{*}_{\bm{k}_1}b^{\dagger}_{\bm{k}_1}a^{\dagger}_{\bm{k}_2}a_{\bm{k}_3}a_{\bm{k}_4}+\frac{J}{4N_L}\sum_{\{\bm{k}_i\}}\gamma_{\bm{k}_4}a^{\dagger}_{\bm{k}_1}a^{\dagger}_{\bm{k}_2}a_{\bm{k}_3}b_{\bm{k}_4}\\
-&\frac{J}{N_L}\sum_{\{\bm{k}_i\}}\gamma_{\bm{k}_4-\bm{k}_2}a^{\dagger}_{\bm{k}_1}b^{\dagger}_{\bm{k}_2}a_{\bm{k}_3}b_{\bm{k}_4},
\end{aligned}
\end{equation}
where $\sum_{\{\bm{k}_i\}}$ denotes summation over all ${\bf k}_i$. The momentum is conserved in all of the above interaction terms as $\frac{1}{N_L}\sum_{i}e^{i(\bm{k}_1+\bm{k}_2-\bm{k}_3-\bm{k}_4)\cdot\bm{r}_i}=\delta_{\bm{k}_1+\bm{k}_2,\bm{k}_3+\bm{k}_4}$. All the interaction terms can be expressed in a more compact form: $H'=\sum_{\{\bm{k}_i\}}V^{\bm{k}_1,\bm{k}_2}_{\bm{k}_3,\bm{k}_4}\psi^{\dagger}_{\bm{k}_1}\psi^{\dagger}_{\bm{k}_2}\psi_{\bm{k}_3}\psi_{\bm{k}_4}$, where the summation runs over all $(\bm{k}_1,\bm{k}_2,\bm{k}_3,\bm{k}_4)$ combinations constrained by the momentum conservation. 

We employ a retarded Green's function formalism to study the renormalization of the magnon bands due to magnon-magnon interactions.  In the real space-time domain, a retarded Green's function is defined as $\langle \psi(\bm{r},t);\psi^{\dagger}(\bm{r}',t')\rangle=-i\theta(t-t')\langle [\psi(\bm{r},t),\psi^{\dagger}(\bm{r}',t') ]\rangle$. Here $\langle\cdots \rangle$ denotes the ensemble average, $\theta(t-t')$ is the step function, and $\psi^{\dag}(\bm{r},t)$ is the field operator of the spinor which can be written as $\left(\sum_{i}\phi^{*}_{a,i}(\bm{r})a^{\dag}_{i}(t),\sum_{i}\phi^{*}_{b,i}(\bm{r})b^{\dag}_{i}(t)\right)$, where $\phi^{*}_{a,i}(\bm{r})$ and $\phi^{*}_{b,i}(\bm{r})$ are the single-particle orbital functions of (A, B)-site. The HP boson operators of (A, B)-site in the bipartite honeycomb lattice are given by $a^{\dag}_i(t)$ and $b^{\dag}_i(t)$. The frequency-dependent retarded Green's function in $k-$space is given by $G_R(\bm{k},\bm{k}';\omega)=\langle \psi_{\bm{k}};\psi^{\dagger}_{\bm{k}'}\rangle_{\omega}$. Using the Heisenberg picture, we write down the equation of motion of Green's function in the frequency domain: 
\begin{equation}\label{eq:eom}
\begin{split}
\omega G_R(\bm{k},\bm{k}';\omega)=\delta_{\bm{k},\bm{k}'}+\langle [\psi_{\bm{k}},H_0];\psi^{\dagger}_{\bm{k}'} \rangle_{\omega}+\langle [\psi_{\bm{k}},H'];\psi^{\dagger}_{\bm{k}'} \rangle_{\omega},
\end{split} 
\end{equation}
where  $[\psi_{\bm{k}},H']=\sum_{\{\bm{k}_i\}}V^{(\bm{k},\bm{k}_2)}_{\bm{k}_3,\bm{k}_4}\psi^{\dagger}_{\bm{k}_2}\psi_{\bm{k}_3}\psi_{\bm{k}_4}$, which gives rise to many-body effects. In order to solve the above equation, we employ the mean-field approximation (MFA) and random phase approximation (RPA): $b^{\dag}_{\bm{k}_2}a_{\bm{k}_3}b_{\bm{k}_4}\approx \delta_{\bm{k}_2,\bm{k}_3}\langle b^{\dag}_{\bm{k}_2}a_{\bm{k}_3}\rangle b_{\bm{k}_4}+\delta_{\bm{k}_2,\bm{k}_4}\langle b^{\dag}_{\bm{k}_2}b_{\bm{k}_4}\rangle a_{\bm{k}_3}$. The $\delta-$function is for the RPA, and a short explanation of its validity is in order here. Generally, the expectation value $\langle b^{\dag}_{\bm{k}_2}a_{\bm{k}_3}\rangle$ has a dominant time dependence $\langle b^{\dag}_{\bm{k}_2}a_{\bm{k}_3}\rangle\propto e^{i(\omega_{\bm{k}_2}-\omega_{\bm{k}_3})t}$. When considering summation over all $\bm{k}_2$ and $\bm{k}_3$, we neglect the terms with $\bm{k}_2 \neq \bm{k}_3$, because these are averaged to zero due to the rapid oscillations in the time domain, and therefore we only retain the dominant contribution from $\bm{k}_2 = \bm{k}_3$ term. 

From the perturbation theory, we can derive the spinor Dyson equation for the interacting magnons up to the second order, which is written as
\begin{equation}
\begin{aligned}[b]
G_R(\bm{k},\bm{k}';\omega)=&G^{(0)}_R(\bm{k},\bm{k}';\omega)+G^{(0)}_R(\bm{k},\bm{k}';\omega)\Sigma^{(1)}_{\bm{k}}G^{(0)}_R(\bm{k},\bm{k}';\omega)\\
&+G^{(0)}_R(\bm{k},\bm{k}';\omega)\Sigma^{(2)}_{\bm{k}}(\omega)G^{(0)}_R(\bm{k},\bm{k}';\omega),
\end{aligned}
\end{equation}
where $G_R(\bm{k},\bm{k}';\omega)$ is the interacting spinor Green's function of the Dirac magnon, $G^{(0)}_R(\bm{k},\bm{k}';\omega)=\frac{\delta_{\bm{k},\bm{k}'}}{\omega-M(\bm{k}')}$ is the free particle Green's function without magnon-magnon interactions. $\Sigma^{(1)}_{\bm{k}}$ and $\Sigma^{(2)}_{\bm{k}}$ are the first-order and second-order self-energies, respectively. We will show the explicit forms in the following sections. 

Using MFA and RPA, we can simply expand $[\psi_{\bm{k}},H']$ in Eq.~(\ref{eq:eom}), and obtain the Hartree-type self-energy, which is the first-order renormalization given by
\begin{equation}
\begin{aligned}[b]
\Sigma^{(1)}_{H'}(\bm{k})=\sum_{\bm{k}_2}V^{(\bm{k},\bm{k}_2)}_{(\bm{k}_2,\bm{k})}\langle\psi^{\dagger}_{\bm{k}_2}\psi_{\bm{k}_2}\rangle^{(0)}, 
\end{aligned}
\end{equation}
where $V^{(\bm{k},\bm{k}_2)}_{(\bm{k}_2,\bm{k})}=2V^{\bm{k},\bm{k}_2}_{\bm{k}_2,\bm{k}}+2V^{\bm{k},\bm{k}_2}_{\bm{k},\bm{k}_2}$ is the interaction coefficient matrix, $\langle\psi^{\dagger}_{\bm{k}_2}\psi_{\bm{k}_2}\rangle^{(0)}=\begin{pmatrix} 
\langle a^{\dagger}_{\bm{k}_2}a_{\bm{k}_2}\rangle^{(0)}&\langle a^{\dagger}_{\bm{k}_2}b_{\bm{k}_2}\rangle^{(0)}\\ 
\langle b^{\dagger}_{\bm{k}_2}a_{\bm{k}_2}\rangle^{(0)}&\langle b^{\dagger}_{\bm{k}_2}b_{\bm{k}_2}\rangle^{(0)}
\end{pmatrix}$ is the zeroth order population matrix, which is determined by the thermal excitation or the magnon parametric amplifications~\cite{Kamimaki2020, Malz2019}. 
\begin{figure}[t!]
\includegraphics[width=8.5 cm]{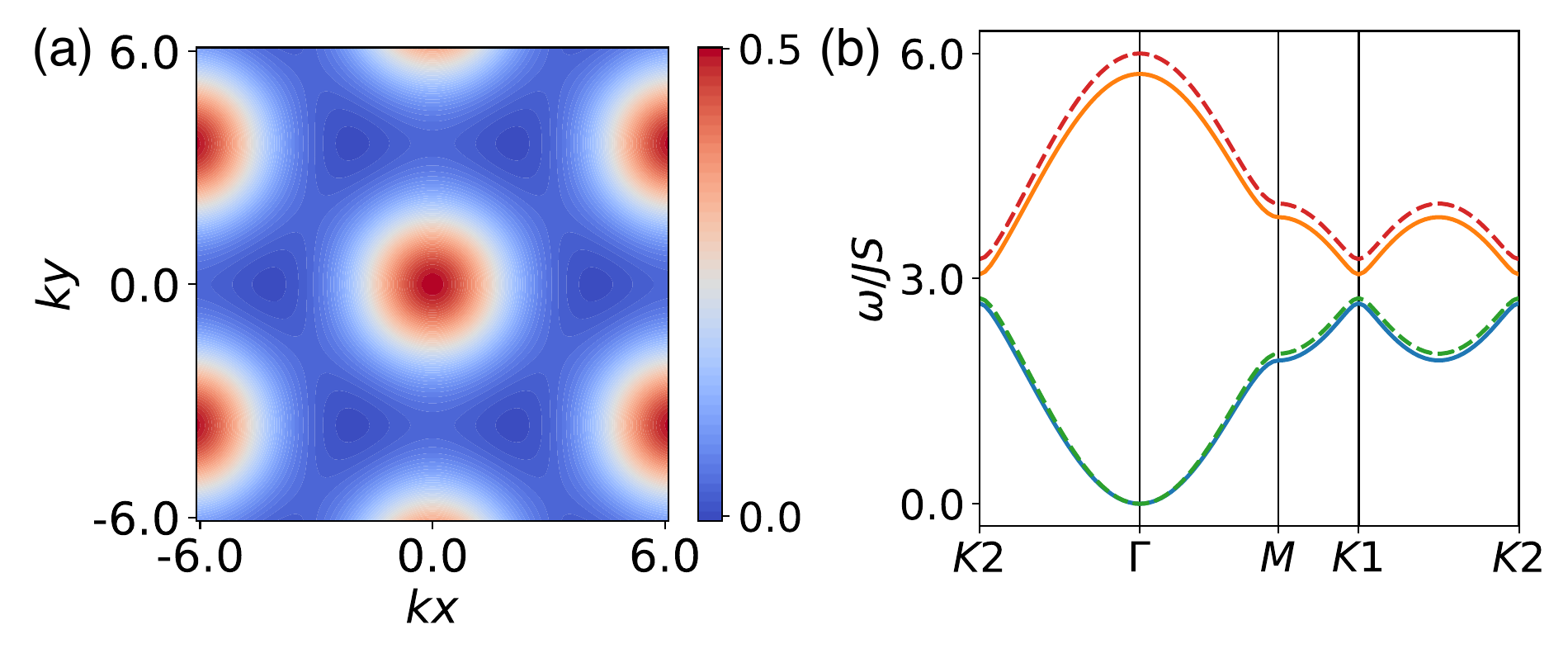}
\caption{The first order renormalization on magnon band from thermal excitation. (a) The density of renormalization factor $a^T(\bm{k})$, where $\int_{BZ}d^{2}\bm{k}a^T(\bm{k})=\alpha(T)$. (b) Renormalized magnon bands at $k_BT/JS=0.8$ are shown by solid curves, which have smaller bandwidths compared to the pristine magnon bands in dashed curves.}
\label{fig:2}
\end{figure}

\subsection{Low temperature approximation}\label{sec.II_B}
For thermally excited magnons, we usually use a low-temperature approximation, which means that we only include finite density magnonic excitations around the lowest energy state at $\Gamma$ point in the lower magnon band. The first-order self-energy $\Sigma^{(1)}_{\bm{k}}$ of the thermal excitation is given by~\cite{Pershoguba2018}:
\begin{equation}
\begin{aligned}[b]
\Sigma^{(1)}_{H'}(\bm{k})\approx-\frac{J}{2N_L}\sum_{\bm{k}_1}\frac{\omega_{\bm{k}_1}}{3JS}f(\omega_{\bm{k}_1})
\begin{pmatrix} 
3&-\gamma_{\bm{k}}\\ 
-\gamma^*_{\bm{k}}&3
\end{pmatrix},
\end{aligned}
\end{equation}
where $f(\omega_{\bm{k}_1})=\frac{1}{e^{\omega_{\bm{k}_1}\beta}-1}$ is the Bose-Einstein distribution of magnons at finite temperature $T$, and $\beta=\frac{1}{k_BT}$. We approximate the acoustic (lower) magnon band as $\omega_{\bm{q}}\approx c_2 q^2$  around the $\Gamma$ point. The summation $\frac{1}{N_L}\sum_q \omega_{q}f(\omega_{q})$ in the above equation can be easily evaluated, and we obtain $\frac{1}{N_L}\sum_q c_2 q^2\frac{1}{e^{\omega_{q}\beta}-1}=\frac{A k^2_B}{4\pi c_2}\zeta(2)T^2$, where $A$ is the area of the primitive unit cell, and $\zeta(2)$ is the Riemann zeta function. Finally, we obtain a renormalized magnon band modified by the first-order self-energy:
\begin{equation}
\begin{aligned}[b]
\omega_{\bm{k}}=3JS(1-\alpha(T))\pm\sqrt{(1-\alpha(T))^2J^2S^2\abs{\gamma_{\bm{k}}}^2+4D^2S^2\beta^2_{\bm{k}}},
\end{aligned}
\label{eq:w_lowT}
\end{equation}
where $\alpha(T)=\frac{A \pi k^2_B}{24 J^2S^3}T^2\propto T^2$ is defined as the renormalization factor, which is consistent with the results in \cite{Pershoguba2018} (also see Appendix~\ref{appen:B}). We plot the distribution of the renormalization factor from the thermal excitation at $k_BT/JS=0.8$ in Fig.~\ref{fig:2}(a), and the modified magnon band in Fig.~\ref{fig:2}(b). The first-order renormalization at a low temperature ($T \ll J$) does not change the non-trivial magnon gap but does reduce the magnon bandwidth by a factor $1-\alpha(T)$. 

\subsection{Beyond the low temperature approximation}\label{sec.II_C}
The above discussion is only valid at low temperatures. The thermally generated magnons are concentrated at the bottom of the band where the Berry curvature is minimum. As a result, the effects of non-trivial band topology are not manifested in any physical observable. Fortunately, recent studies have shown that magnons can be excited not only by thermal energy but also at any energy-momentum point of the spectrum by an external electromagnetic field~\cite{Malz2019}.  Hence one can controllably generate magnons at specific points in the magnetic Brillouin zone with a large concentration of Berry curvature so that the simultaneous effects of non-trivial band topology and interactions on magnons can be studied systematically. This is the approach taken in the following.  To encompass all possible situations, we shall focus on the full expression of the first-order self-energy from the interacting Hamiltonian:
\begin{equation}\label{eq:self}
\begin{aligned}[b]
&\Sigma^{(1)}_{H'}(\bm{k})=\frac{J}{2N_L}\sum_{\bm{k}_1}\\
&\begin{pmatrix} 
\abs{\gamma_{\bm{k}_1}}\Delta_2(\bm{k}_1)-\gamma_0\Theta_b(\bm{k}_1)&\gamma_{\bm{k}}\Theta^+(\bm{k}_1)-\theta(\bm{k},\bm{k}_1)\\ 
\gamma^*_{\bm{k}}\Theta^+(\bm{k}_1)-\theta^*(\bm{k},\bm{k}_1)&\abs{\gamma_{\bm{k}_1}}\Delta_2(\bm{k}_1)-\gamma_0\Theta_a(\bm{k}_1)
\end{pmatrix},
\end{aligned}
\end{equation}
where we have the terms:
\begin{equation}\label{eq:func}
\begin{aligned}[b]
\Delta_2(\bm{k}_1)&=\sqrt{1-\frac{B^2}{A^2}}\Theta^{-}(\bm{k}_1),\\
\Theta_b(\bm{k}_1)&=\Theta^{+}(\bm{k}_1)-\frac{B}{A}\Theta^{-}(\bm{k}_1),\\
\Theta_a(\bm{k}_1)&=\Theta^{+}(\bm{k}_1)+\frac{B}{A}\Theta^{-}(\bm{k}_1),\\
\theta(\bm{k},\bm{k}_1)&=\Delta_2(\bm{k}_1)\gamma_{\bm{k}-\bm{k}_1}e^{i\phi_{\bm{k}_1}},
\end{aligned}
\end{equation}
and $A=\sqrt{(\frac{2D\beta_{\bm{k}_1}}{J})^2+\abs{\gamma_{\bm{k}_1}}^2}$, $B=\frac{2D\beta_{\bm{k}_1}}{J}$, $\Theta^{+}(\bm{k})=f(\omega_{d_{\bm{k}}})+f(\omega_{u_{\bm{k}}})$, $\Theta^{-}(\bm{k})=f(\omega_{d_{\bm{k}}})-f(\omega_{u_{\bm{k}}})$, $f(\omega_{d_{\bm{k}}})$ and $f(\omega_{u_{\bm{k}}})$ are the magnon populations at point $\bm{k}$ of lower and upper bands, respectively. The quartic DM interactions are also considered, leading to a self-energy term given by $\Sigma^{(1)}_{DM}(\bm{k})=\begin{pmatrix} 
\sigma'_{11}&0\\ 
0&\sigma'_{22}
\end{pmatrix}$.  The explicit form of each term is (more details in Appendix~\ref{appen:B})
\begin{equation}
\begin{aligned}[b]
\sigma'_{11}&=\frac{D\beta_{\bm{k}}}{N_L}\sum_{\bm{k}_1}\Theta^+(\bm{k}_1)+\frac{D}{N_L}\sum_{\bm{k}_1}\frac{\beta_{\bm{k}_1}B}{A}\Theta^-(\bm{k}_1)\\
\sigma'_{22}&=-\frac{D\beta_{\bm{k}}}{N_L}\sum_{\bm{k}_1}\Theta^+(\bm{k}_1)+\frac{D}{N_L}\sum_{\bm{k}_1}\frac{\beta_{\bm{k}_1}B}{A}\Theta^-(\bm{k}_1).
\end{aligned}
\end{equation}
Hence the renormalized Hamiltonian at the first-order level is given by
\begin{equation}\label{eq:re_H}
\begin{aligned}[b]
&H_1=H_0+\Sigma^{(1)}_{H'}(\bm{k})+\Sigma^{(1)}_{DM}(\bm{k})\\
&=\begin{pmatrix} 
 3JS-2DS\beta_{\bm{k}}+P+m_H&-(JS-Q)\gamma_{\bm{k}}-g(\bm{k})\\ 
 -(JS-Q)\gamma^*_{\bm{k}}-g^*(\bm{k})&3JS+2DS\beta_{\bm{k}}+P-m_H
\end{pmatrix},
\end{aligned}
\end{equation}
where $P=\frac{J}{2N_L}\sum_{\bm{k}_1}(\Delta_1(\bm{k}_1)-\gamma_0\Theta^+(\bm{k}_1))$, $\Delta_1(\bm{k}_1)=\abs{\gamma_{\bm{k}_1}}\Delta_2(\bm{k}_1)+\frac{B^2}{A}\Theta^{-}(\bm{k}_1)$, and $Q=\frac{J}{2N_L}\sum_{\bm{k}_1}\Theta^+(\bm{k}_1)$. These are all overall factors and $\bm{k}$-independent. It is easy to see that the diagonal term $P$ mimics the scalar potential, and the interaction-induced off-diagonal term $g(\bm{k})=\frac{J}{2N_L}\sum_{\bm{k}_1}\theta(\bm{k},\bm{k}_1)$ mimics the vector potential in the vicinity of the magnon Dirac point using minimal coupling with $\gamma_{\bm{k}}+g(\bm{k})\to \bm{k}+g(\bm{k}_D)$. 

The most interesting term is the interaction-induced Haldane term $m_H=\frac{J}{2N_L}B(\bm{k})\sum_{\bm{k}_1}\Theta^{+}(\bm{k}_1)$. Thus the interaction effect can be seen as the emergent gauge field in the weakly interacting magnon system. The summation of the value $\bm{k} _1$ is limited to the first Brillouin zone. We diagonalize the new Hamiltonian to obtain the renormalized magnon spectrum as 
\begin{equation}\label{eq:re_e}
\begin{aligned}[b]
\omega_{\bm{k}}=3JS+P\pm\sqrt{\abs{(JS-Q)\gamma_{\bm{k}}+g(\bm{k})}^2+(2DS\beta_{\bm{k}}-m_H)^2}
\end{aligned}
\end{equation}
which reduces to Eq.(\ref{eq:w_lowT}) when using a low temperature approximation. 

\begin{figure}[t!]
\includegraphics[width=8.5 cm]{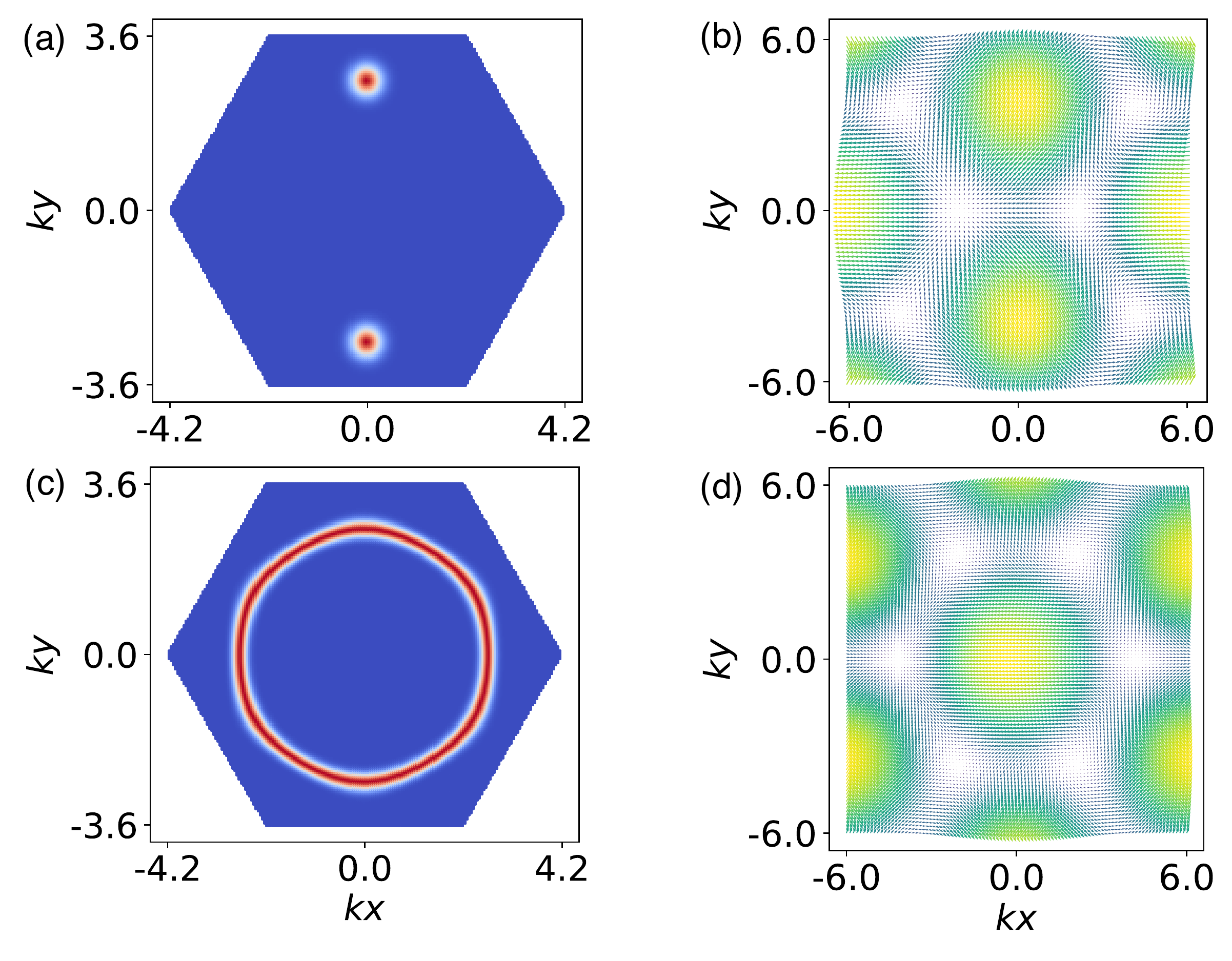}
\caption{Population of excited magnon by electromagnetic field and distribution of $g(\bm{k})$. (a) The anisotropic population of magnon by EM field amplification. Two excited magnon states with opposite $\bm{k}$ vectors ($(0,b)$ and $(0,-b)$). (b) Distribution of the in-plane components of the vector $\bm{g}$ shows the nonuniform of $g(\bm{k})$ in BZ. (c) The isotropic population of magnon by EM field amplification. (d) Distribution of the in-plane components of the vector $\bm{g}$ by isotropic populations in (c). }
\label{fig:3}
\end{figure}

\section{Band engineering from magnon-magnon interaction}\label{sec.III}
\subsection{Shifting of Dirac points and quantum geometric tensor}\label{sec.III_A} 
We use Eqs.(\ref{eq:re_H}) and (\ref{eq:re_e}) to study the many-body effect for arbitrary magnon populations, in particular for magnons amplified by the electromagnetic field. The magnon amplification will be discussed in Sec.~\ref{sec.V}. We argue that treating the interaction within a mean-field theory gives rise to an effective potential, in which the diagonal part $P$ is a scalar potential, and the off-diagonal part $g(\bm{k})$ is an effective vector potential in the vicinity of the Dirac point. The scalar potential $P$ for the energy shift will not change the physics around the Dirac point.

\begin{figure}[t!]
\includegraphics[width=9.0 cm]{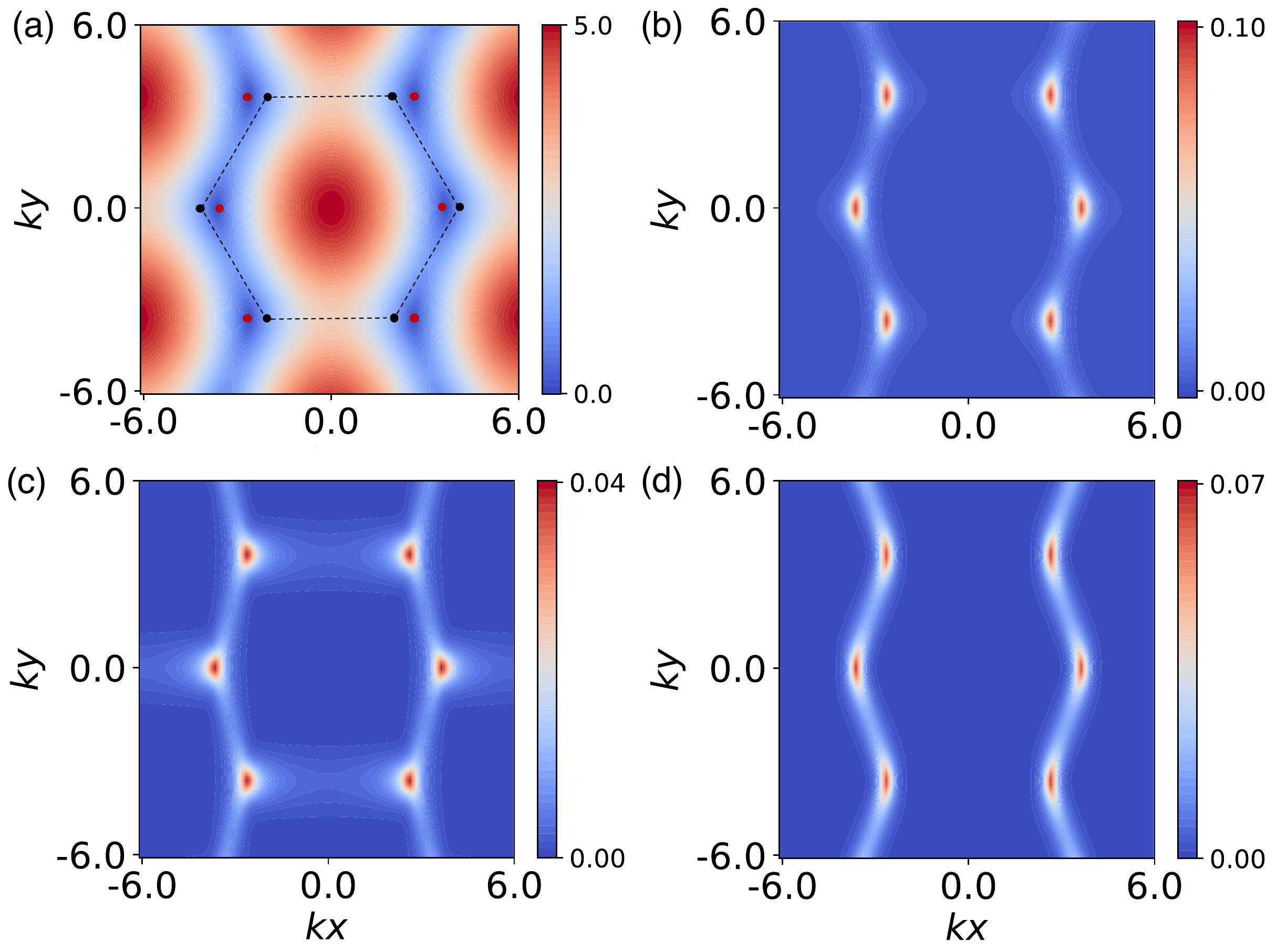}
\caption{Quantum geometric tensor in renormalized magnon bands. (a) The shift of Dirac points by anisotropic magnon amplification is shown in Fig.~\ref{fig:3}(a) without DM interaction. Black dots represent the original Dirac points, and the shifted Dirac points are shown in red dots. (b) The distribution of Berry curvature shows the broken $C_6$ symmetry by interaction effect. (c) Quantum metric $G_{xx}$ component. (d) Quantum metric $G_{yy}$ component.}
\label{fig:4}
\end{figure}

It is interesting to note that in the pseudospin space, $g(\bm{k})$ acts as an effective in-plane Zeeman field on sublattice pseudospin $\bm{\sigma}$. In Eq.~(\ref{eq:re_H}), we can rewrite the non-trivial part as $\begin{pmatrix} 
 m_H&-g(\bm{k})\\ 
 -g^*(\bm{k})&-m_H
\end{pmatrix}=\bm{g}\cdot\bm{\tau}$, where $\bm{\tau}=(\tau_x, \tau_y, \tau_z)$ is the Pauli matrix. It follows that we can reformulate the complex quantity $g(\bm{k})$ and the interaction-induced Haldane mass into a vector form given by the expression $\bm{g}=(-\Re(g(\bm{k})),\Im(g(\bm{k})),m_H)$. For an anisotropic magnon density distribution $f(\omega_{\bm{k}})$ as shown in Fig.~\ref{fig:3}(a), we have the distribution of vector $\bm{g}$, showing a nonuniform vortex feature illustrated in Fig.~\ref{fig:3}(b). $g(\bm{k})$ breaks the $C_6$ rotational symmetry of the first BZ, and as a consequence, the Dirac point will be shifted by the effective vector potential, as shown in Fig.~\ref{fig:4}(a)~\cite{Tarruell2012}. The black dots represent the original Dirac points, whereas the red dots are the new ones shifted by $\bm{g}$. However, the gapless feature of the Dirac point will still be preserved in the absence of DM interactions, since the original interaction terms in Eq~(\ref{eq:5}) do not break time-reversal or chiral symmetries.
In addition, we do not include the anomalous paring term $\langle\psi^{\dag}_{\bm{k}}\psi^{\dag}_{-\bm{k}}\rangle$ ($\langle\psi_{\bm{k}}\psi_{-\bm{k}}\rangle$) in RPA, which breaks the $U(1)$ symmetry. Dirac point shifts are observed in many cases by strain engineering~\cite{Feilhauer2015, Kim2021}; we propose that many-body effects can induce the same effect without lattice deformation.

The shift of Dirac points changes the geometric properties of the magnon dispersion. We study how the effective gauge field $g(\bm{k})$ modifies the quantum geometric tensor (QGT) that consists of the Berry curvature and the quantum metric measuring the "distance" between the eigenstates in the Hilbert space. It is defined as follows:
\begin{equation}\label{eq:qgt1}
\begin{split}
Q_{ab}=G_{ab}-\frac{i}{2}\Omega_{ab},
\end{split}
\end{equation}
where the real part $G_{ab}$ is the quantum metric, and $\Omega_{ab}$ is the Berry curvature. We already know that Berry curvature is crucial for topological phases; 
the quantum metric is associated with superfluidity in flat bands and orbital magnetic susceptibility. Finally, it should be noted that the QGT is not just an abstract mathematical construct but has been measured directly~\cite{Gianfrate2020}. Since the gauge-invariant QGT contains the structural information about the eigenstates of a parametrized Hamiltonian, we have the explicit expressions as below:
\begin{equation}\label{eq:qgt2}
\begin{aligned}[b]
G_{ab}&=\Re\left(\sum_{m\ne n}\frac{\mel{u_m}{\partial_{k_a}H}{u_n}\mel{u_n}{\partial_{k_b}H}{u_m}}{(E_m-E_n)^2}\right),\\
\Omega_{ab}&=i\left(\sum_{m\ne n}\frac{\mel{u_m}{\partial_{k_a}H}{u_n}\mel{u_n}{\partial_{k_b}H}{u_m}}{(E_m-E_n)^2}-(a\to b)\right).
\end{aligned}
\end{equation} 
It's convenient to rewrite the Hamiltonian in Eq. ~(\ref{eq:re_H}) with the pseudospin freedom $\bm{\sigma}$:
\begin{equation}\label{eq:pss}
\begin{aligned}[b]
H_1=(3JS+P)\tau_0+\bm{h}(\bm{k})\cdot\bm{\tau}+\bm{g}(\bm{k})\cdot\bm{\tau}
\end{aligned}
\end{equation}
where effective field vector $\bm{h}$ is $(-(JS-Q)\abs{\gamma_{\bm{k}}}\cos{\phi_{\bm{k}}}, (JS-Q)\abs{\gamma_{\bm{k}}}\sin{\phi_{\bm{k}}}, -2DS\beta_{\bm{k}})$,$\tau_0$ is the $2\times2$ identity, and $\bm{g}(\bm{k})=(-\abs{g_{\bm{k}}}\cos{\phi_{g_{\bm{k}}}}, \abs{g_{\bm{k}}}\sin{\phi_{g_{\bm{k}}}}, m_H)$. 
Within this representation, we put the band physics on a Bloch sphere, and all the geometric and topological properties are contained in the $\bm{h}\cdot\bm{\tau}+\bm{g}\cdot\bm{\tau}$ term. We plot the modified Berry curvature in Fig.~\ref{fig:4}(b), the $xx$ and $yy$ components of quantum metric $G$ in Fig.~\ref{fig:4}(c)-(d). We can see that the QGT is renormalized to a new distribution in BZ by interaction-induced Zeeman field $g_{\bm{k}}$. The shifted Dirac points and the renormalized bands can be detected by well-established techniques, including Brillouin light scattering, inelastic X-ray scattering, inelastic neutron scattering, and electron tunneling~\cite{Braicovich2010, Samuelsen1971, Cenker2021, Sobolev1994, Klein2018}.

\subsection{Topological bands with tunable Chern numbers}\label{sec.III_B}
From Eq.~(\ref{eq:pss}), we find that interaction gives rise to an effective mass $m_H$, which can be understood as an out-of-plane Zeeman field on the pseudospin degrees of freedom. The effective mass term determining the bandgap is essential for the geometric effect and the non-trivial band topology. The competition between the DMI term, $2DS\beta_{\bm{k}}$, and the interaction-induced Haldane term $m_H$ can lead to the different Chern numbers and thus distinct band topology. One should note that $m_H$ depends on DMI, as can be seen from the expression $m_H=\frac{JB}{2N_L}\sum_{\bm{k}_1}\Theta^{+}(\bm{k}_1)$, where the parameter $B\propto DS$. When $D=0$, $m_H$ vanishes as well.

\begin{figure}[t!]
\includegraphics[width=9.0 cm]{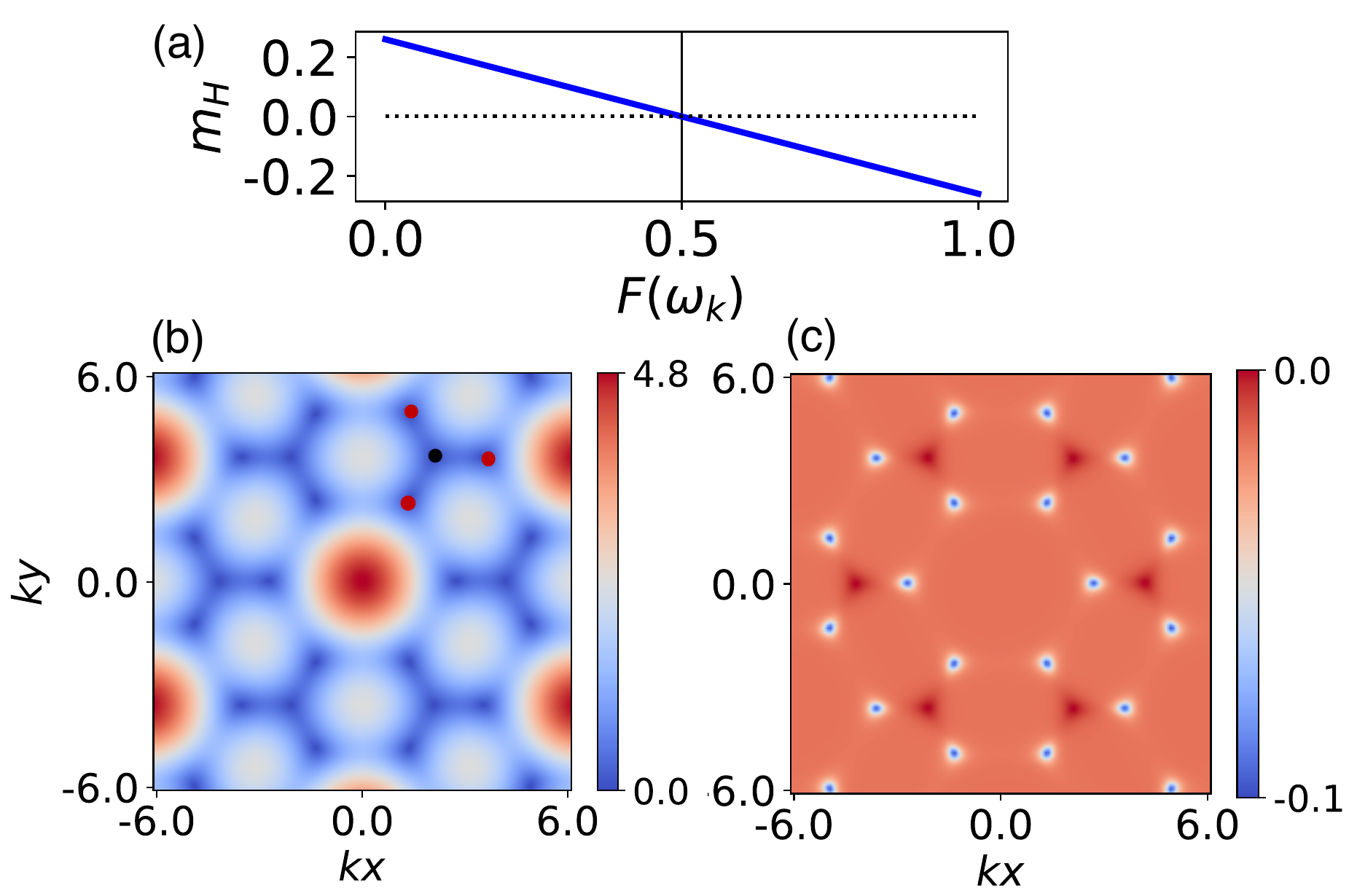}
\caption{Renormalized magnon bands with high Chern number, we use $D/J=0.05$. (a). Renormalized Haldane mass at Dirac point in magnon bands we use $D/J=0.05$, $F(\omega_{\bm{k}})=\frac{1}{4N_L}\sum_{\bm{k}_1}\left(f(\omega_{d_{\bm{k}_1}})+f(\omega_{u_{\bm{k}_1}})\right)$ is the quantity for the normalised magnon density. (b) The energy difference $\Delta\omega_{\bm{k}}/JS$ between upper and lower bands. The original Dirac points are labeled in black dots, and the three satellitic Dirac points are shown in red dots. (c) Berry curvature distributions.}
\label{fig:5}
\end{figure}

The total mass term is given by $M_H=2DS\beta_{\bm{k}}-m_H$. By tuning the magnon population, we can achieve topological phase transitions in the bottom band from $C=1$ to $C=-1$ or $C=-1$ to $C=1$. This is due to the fact that $m_H$ is dominated by the total population at point $\bm{k}_1$, which can be driven by pumping magnons in both branches or by thermal excitation. Our results show that the total mass term decreases linearly with increasing magnon population, as illustrated in Fig.~\ref{fig:5}(a). The sign-change of $m_H$ reveals a topological phase transition with reversing the Chern numbers. A recent study has proposed a similar topological phase transition by thermal excitation~\cite{Lu2021}. We argue that the band topology can be simply tuned using a magnon amplification approach, which will be discussed in Sec.~\ref{sec.V}. The parametric amplification approach offers distinct advantages over thermal excitation due to its flexibility~\cite{Malz2019}. 

By properly tuning the magnon population with the external EM field, we can also increase the number of crossing points between the upper and lower magnon branches. This, in principle, can give rise to magnon bands with higher Chern numbers. Mathematically, the interaction-induced $g(\bm{k})$ can have any form. The general expression of $g(\bm{k})$ consists of a linear combination of the $n$-th nearest geometric factors $\gamma^{(n)}_{\bm{k}}$, or $g(\bm{k})=\sum_{n}c_n\gamma^{(n)}_{\bm{k}}$, where $\gamma^{(n)}_{\bm{k}}=\sum_j e^{i\bm{k}\cdot\bm{r}^{(n)}_j}$, and $\bm{r}^{(n)}_j$ is the $n$-th nearest bond. As a specific example, let us look at the model of $g(\bm{k})=c\sum_{n}e^{i\bm{k}\cdot\bm{\eta}_n}$, where $c$ is expressed as the strength of the vector potential, and $\bm{\eta}_n$ is the geometric vector. We find that when $\eta_n$ is for the third-nearest bonds, there are three more Dirac points at each $\bm{K}$ valley, as illustrated in Fig.~\ref{fig:5}(b). We introduce a nonzero DMI to gap all the Dirac points and obtain the Berry curvature distribution of the lower band as shown in Fig.~\ref{fig:5}(c). In this case the Chern number of the lower magnon band is given by $C=\int_{BZ} \Omega_{\bm{k}}d\bm{k}=-2$. 

\subsection{Topological flat bands by parametric pumping}\label{sec.III_C} 
\begin{figure}[t!]
\includegraphics[width=9.0 cm]{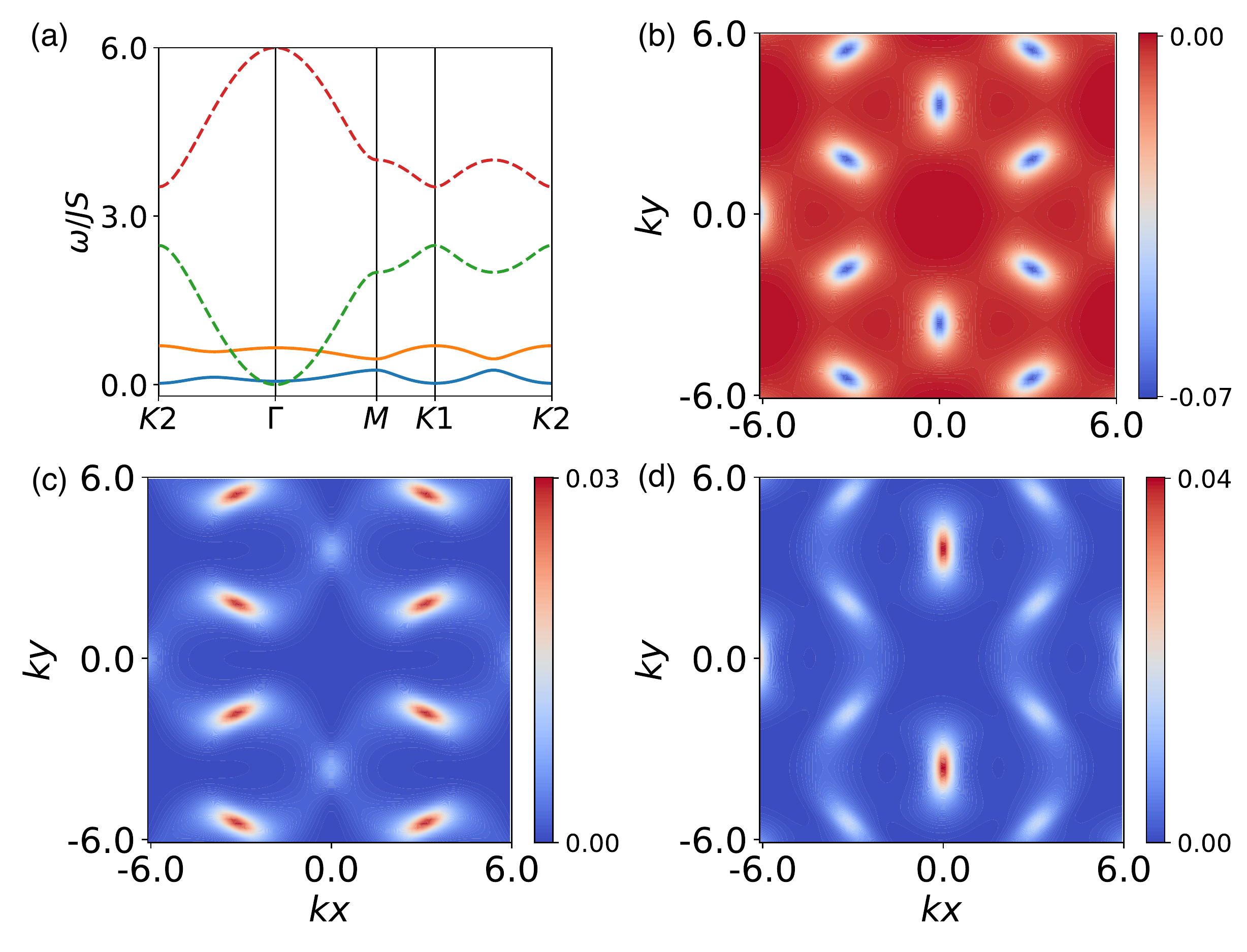}
\caption{QGT in topologically flat bands. (a) The topologically flat band by magnon-magnon interactions. Solid curves are interaction-induced flat bands, and the dashed curves are noninteracting magnon bands. (b) The Berry curvature is redistributed by the interaction effect. (c) Quantum metric $G_{xx}$ component. (d) Quantum metric $G_{yy}$ component.}
\label{fig:6}
\end{figure}

A striking effect of interactions between magnons in regions of the Brillouin zone with large Berry curvature is a strong reduction of bandwidth. Using isotropic parametric pumping (see details in Sec.~\ref{sec.V_A}), we can engineer a topologically flat magnon band with a very small bandwidth. We would like to point out that since we focus mainly on the qualitative behaviors of the magnon-magnon interaction, higher-order terms need to be considered carefully when the magnon population density is very high, where the self-energy is as large as the bare band energy. Dealing with the high magnon density case with more accurate renormalization effects is difficult, since the perturbative expansion will be extremely complicated in magnon-magnon interactions with two sublattices in the honeycomb model. To address this problem, a non-perturbative treatment of local interactions—e.g. dynamical mean-field theory (DMFT) has to be employed~\cite{Otsuki2013}. We employ Eq.~(\ref{eq:re_e}) with pumped magnons with fixed energy, but all possible momenta (shown by a circular band in  Fig.~\ref{fig:3}(c)). The pumping energy of magnon used here is $\omega(\bm{k})\approx 1.6J$, for $D/J=0.1$, and the pumped population intensity $I_p$ is about 7.2, which is determined by the scattering effect, dissipative damping of the magnons at finite temperature in the presence of disorder. The pumping dynamics are described in detail in Sec.~\ref{sec.V_A}.  
The resulting flat bands are shown in Fig.~\ref{fig:6}(a) by the solid curves. The flatness of the topological bands can be simply characterized by the ratio $r_f=\frac{w_p}{w_r}$, where $w_p$ is the bandwidth without magnon-magnon interactions, and $w_r$ is the bandwidth of the renormalized magnon spectrum. The computed value of $r_f \approx 0.09$ implies that the renormalized bands are almost dispersionless. 

The almost flat topological bands give rise to a more uniform Berry curvature and quantum metric tensors that mimic  Landau level physics of interacting particles~\cite{Wang2021}. The quantum geometric properties are illustrated in Fig.~\ref{fig:6}(b)-(d). The Berry curvature is redistributed, and the hotspots are shifted from $K$ to $M$ points in the BZ. The elements $G_{xx}$ and $G_{yy}$ of the quantum metric tensor in Fig.~\ref{fig:6}(c)-(d) are also strongly renormalized. We would also like to stress that the resulting steady state under parametric pumping is not only constrained by the energy conservation but also has a magnon density redistribution effect caused by scattering among the pumped areas~\cite{Hahn2020}. However, the renormalization effect is dominated by the magnon population in the whole k-space, thus this redistribution effect has no influence on band renormalization.

The renormalized magnon Bloch bands also conform to the ideal flat band condition that is provided by~\cite{Claassen2015}:  
\begin{equation}\label{}
\begin{aligned}[b]
\sqrt{\det G(\bm{k})}-\frac{1}{2}\abs{\Omega(\bm{k})}=0,
\end{aligned}
\end{equation}
where $\det G(\bm{k})=G_{xx}G_{yy}-G_{xy}G_{yx}$ is the determinant of quantum metric tensor. The nearly dispersionless magnon bands with ideal flat conditions allow us to investigate strongly correlated behavior since the interaction strength between the magnons is comparable to the bandwidth. Thus, we can expect the superfluid-Bose insulator transition at the partially filled lowest band~\cite{Fisher1989}. More interestingly, the band minimum is also modified, which shifts from $\Gamma$ point to $K$ point, where there is a gapped Dirac cone with finite Berry curvatures and a nonzero quantum metric. The induced flat band can potentially host a novel Bose-Einstein condensate that is stabilized by the nonzero quantum metric~\cite{Julku2021}. Furthermore, the emergent Goldstone modes from this flat band condensation reveal a quantum geometric dependency~\cite{Julku2021prl}. We argue that our interaction-induced flat magnon bands provide a realistic platform for studying this novel phenomenon, which deserves further theoretical study.

\section{Second order renormalization and magnon scattering}\label{sec.IV}
In the previous section, we employed a mean-field approximation of the term $[\psi_{\bm{k}}, H']$ and directly obtained the first-order self-energy. In order to include the higher-order renormalization, we must consider the dynamics of the commutator  $[\psi_{\bm{k}},H']$. The second-order effect from the quartic DMI is neglected since its magnitude is  ${\mathcal O}(\frac{D^2}{J^2}) \ll 1$ for realistic values of $D$. In this section, we focus on the second-order approximation by considering the equation of motion of retarded Green's function with the commutator $[\psi_{\bm{k}}, H']$:
\begin{equation}\label{}
\begin{aligned}[b]
\omega \langle [\psi_{\bm{k}},H'];\psi^{\dagger}_{\bm{k}'} \rangle_{\omega}=\langle [[\psi_{\bm{k}},H'],\psi^{\dagger}_{\bm{k}'}] \rangle+\langle [[\psi_{\bm{k}},H'],H];\psi^{\dagger}_{\bm{k}'} \rangle_{\omega}.
\end{aligned}
\end{equation}

The second-order self-energy from perturbation theory can be written as
\begin{equation}
\begin{aligned}[b]
\Sigma^{(2)}_{\bm{k}}=\frac{1}{2}\sum_{\{\bm{k}_i\}}\frac{V^{(\bm{k},\bm{k}_2)}_{(\bm{k}_3,\bm{k}_4)}V^{(\bm{k}_4,\bm{k}_3)}_{(\bm{k}_2,\bm{k})}(n_{\bm{k}_2}(1+n_{\bm{k}_3}+n_{\bm{k}_4})-n_{\bm{k}_3}n_{\bm{k}_4})}{\omega_{\bm{k}}+i\epsilon-\mathcal{M}(\bm{k}_2,\bm{k}_3,\bm{k}_4)},
\end{aligned}
\end{equation}
where $V^{(\bm{k},\bm{k}_2)}_{(\bm{k}_3,\bm{k}_4)}=V^{\bm{k},\bm{k}_2}_{\bm{k}_3,\bm{k}_4}+V^{\bm{k}_2,\bm{k}}_{\bm{k}_3,\bm{k}_4}+V^{\bm{k},\bm{k}_2}_{\bm{k}_4,\bm{k}_3}+V^{\bm{k}_2,\bm{k}}_{\bm{k}_4,\bm{k}_3}$ is the interacting coefficient matrix. The nonzero matrix elements from the interacting Hamiltonian are $V^{1,2}_{1,2}=-\frac{J}{N_L}\gamma_{\bm{k}_4-\bm{k}_2}$, $V^{1,2}_{2,2}=\frac{J}{4N_L}\gamma_{\bm{k}_1}$, $V^{2,1}_{1,1}=\frac{J}{4N_L}\gamma^*_{\bm{k}_1}$, $V^{2,2}_{2,1}=\frac{J}{4N_L}\gamma^*_{\bm{k}_4}$ and $V^{1,1}_{1,2}=\frac{J}{4N_L}\gamma_{\bm{k}_4}$. We define $V^{a,b}_{c,d}\triangleq V^{a\bm{k}_1,b\bm{k}_2}_{c\bm{k}_3,d\bm{k}_4}$, and $a$,$b$,$c$,$d$ $\in \{1,2\}$ are the labels for the two components of spinor $\psi_{\bm{k}}(\psi^{\dag}_{\bm{k}})$. The function $n_{\bm{k}}$ denotes the zeroth order population matrix $\langle\psi^{\dag}_{\bm{k}}\psi_{\bm{k}}\rangle^{(0)}$, which can be the equilibrium or non-equilibrium distribution function. $\mathcal{M}(\bm{k}_2,\bm{k}_3,\bm{k}_4)$ is a tensor dominated by the elements of $M(\bm{k}_2)$, $M(\bm{k}_3)$ and $M(\bm{k}_4)$. The definition of $\mathcal{M}(\bm{k}_2,\bm{k}_3,\bm{k}_4)$ comes from the operation:
\begin{equation}
\begin{aligned}[b]
\mathcal{M}\psi^{\dag}_{\bm{k}_2}\psi_{\bm{k}_3}\psi_{\bm{k}_4}&=\psi^{\dag}_{\bm{k}_2}M(\bm{k}_3)\psi_{\bm{k}_3}\psi_{\bm{k}_4}+\psi^{\dag}_{\bm{k}_2}\psi_{\bm{k}_3}M(\bm{k}_4)\psi_{\bm{k}_4}\\
&-\psi^{\dag}_{\bm{k}_2}M(\bm{k}_2)\psi_{\bm{k}_3}\psi_{\bm{k}_4}
\end{aligned}
\end{equation}
We simplify this operation and obtain $Q'_{a,b,c}(\bm{k}_2,\bm{k}_3,\bm{k}_4)=\sum_{d,e,f}\mathcal{M}^{a,b,c}_{d,e,f}(\bm{k}_2,\bm{k}_3,\bm{k}_4)Q_{d,e,f}(\bm{k}_2,\bm{k}_3,\bm{k}_4)$, where $Q'_{a,b,c}$ is the element in RHS of above equation, $Q_{a,b,c}$ is the element in $\psi^{\dag}_{\bm{k}_2}\psi_{\bm{k}_3}\psi_{\bm{k}_4}$. Since $\psi^{\dag}_{\bm{k}_2}\psi_{\bm{k}_3}\psi_{\bm{k}_4}$ has 8 elements, it gives rise to $2^6=64$ elements in $\mathcal{M}$. Here we list the eight diagonal elements of $\mathcal{M}$:
\begin{equation}
\begin{aligned}[b]
\mathcal{M}^{111}_{111}&=-M^{11}(\bm{k}_2)+M^{11}(\bm{k}_3)+M^{11}(\bm{k}_4),\\
\mathcal{M}^{222}_{222}&=-M^{22}(\bm{k}_2)+M^{22}(\bm{k}_3)+M^{22}(\bm{k}_4),\\
\mathcal{M}^{112}_{112}&=-M^{11}(\bm{k}_2)+M^{11}(\bm{k}_3)+M^{22}(\bm{k}_4),\\
\mathcal{M}^{121}_{121}&=-M^{11}(\bm{k}_2)+M^{22}(\bm{k}_3)+M^{11}(\bm{k}_4),\\
\mathcal{M}^{122}_{122}&=-M^{11}(\bm{k}_2)+M^{22}(\bm{k}_3)+M^{22}(\bm{k}_4),\\
\mathcal{M}^{211}_{211}&=-M^{22}(\bm{k}_2)+M^{11}(\bm{k}_3)+M^{11}(\bm{k}_4),\\
\mathcal{M}^{212}_{212}&=-M^{22}(\bm{k}_2)+M^{11}(\bm{k}_3)+M^{22}(\bm{k}_4),\\
\mathcal{M}^{221}_{221}&=-M^{22}(\bm{k}_2)+M^{22}(\bm{k}_3)+M^{11}(\bm{k}_4).\\
\end{aligned}
\end{equation}
These diagonal elements are the most important terms for determining the magnon-magnon scattering channels. It is more convenient to work with the self-energy matrix in an eigenmode basis, and transform the $\begin{pmatrix}a_{\bm{k}}\\b_{\bm{k}}\end{pmatrix}$ basis to the eigenmode basis $\begin{pmatrix}d_{\bm{k}}\\u_{\bm{k}}\end{pmatrix}$ using a unitary transformation $U$, where $d_{\bm{k}}$($u_{\bm{k}}$) is the magnon operator for the lower (upper) band. As a consequence, $\mathcal{M}$ has only diagonal components, and $M$ is a diagonal matrix of the eigenvalues. Thus we have the second-order self-energy matrix in the eigenmode basis:
\begin{equation}\label{eq:seself}
\begin{aligned}[b]
\tilde{\Sigma}^{(2)}_{\bm{k}}=\frac{1}{2}\sum_{\{\bm{k}_i\}}\frac{\tilde{V}^{(\bm{k},\bm{k}_2)}_{(\bm{k}_3,\bm{k}_4)}\tilde{V}^{(\bm{k}_4,\bm{k}_3)}_{(\bm{k}_2,\bm{k})}(\tilde{n}_{\bm{k}_2}(1+\tilde{n}_{\bm{k}_3}+\tilde{n}_{\bm{k}_4})-\tilde{n}_{\bm{k}_3}\tilde{n}_{\bm{k}_4})}{\omega_{\bm{k}}+i\epsilon+\omega_{\bm{k}_2}-\omega_{\bm{k}_3}-\omega_{\bm{k}_4}}
\end{aligned}
\end{equation}
where $\tilde{V}^{(\bm{k},\bm{k}_2)}_{(\bm{k}_3,\bm{k}_4)}=U^{\dag}_{\bm{k}}U^{\dag}_{\bm{k}_2}V^{(\bm{k},\bm{k}_2)}_{(\bm{k}_3,\bm{k}_4)}U_{\bm{k}_3}U_{\bm{k}_4}$ (see details in Appendix~\ref{appen:D}). The expression of unitary $U$ is given by
\begin{equation}\label{}
\begin{split}
\begin{pmatrix} 
a_{\bm{k}}\\
b_{\bm{k}}
\end{pmatrix}=\frac{1}{\sqrt{2}}
\begin{pmatrix} 
\sqrt{1+\frac{B}{A}}e^{i\frac{\phi_{\bm{k}}}{2}}&\sqrt{1-\frac{B}{A}}e^{i\frac{\phi_{\bm{k}}}{2}}\\ 
\sqrt{1-\frac{B}{A}}e^{-i\frac{\phi_{\bm{k}}}{2}}&-\sqrt{1+\frac{B}{A}}e^{-i\frac{\phi_{\bm{k}}}{2}}
\end{pmatrix}
\begin{pmatrix} 
d_{\bm{k}}\\u_{\bm{k}}
\end{pmatrix}
\end{split}
\end{equation}
where $A$ and $B$ factors are given in Eq.~(\ref{eq:func}), which are dominated by DM interactions. With the unitary transformations, we have all the interaction matrix elements on an eigenmode basis (see details in Appendix~\ref{appen:D}). 

The second-order self-energy term has a more complicated form, which not only modifies the bandwidth, but also leads to magnon damping. Replacing $\omega$ by $\omega_{\bm{k}}+i\epsilon$ in Eq.~(\ref{eq:seself}), where the infinitesimal imaginary part $i\epsilon$ is the argument of the retarded Green's function, the second order self energy can be written as $\tilde{\Sigma}^{(2)}_{\bm{k}}=\tilde{\Sigma}^{(2)'}_{\bm{k}}-i\tilde{\Sigma}^{(2)''}_{\bm{k}}$. The real part represents the renormalization of the magnon energy bands, and the imaginary part describes the damping or magnon-magnon scattering rate~\cite{Mook2021}. The scattering rate contributes to the broadening of the magnon band, which gives rise to an upper bound of the magnon lifetime $t_m =\hbar/2\tilde{\Sigma}^{(2)''}_{\bm{k}}$. 

\begin{figure}[t!]
\includegraphics[width=8.5 cm]{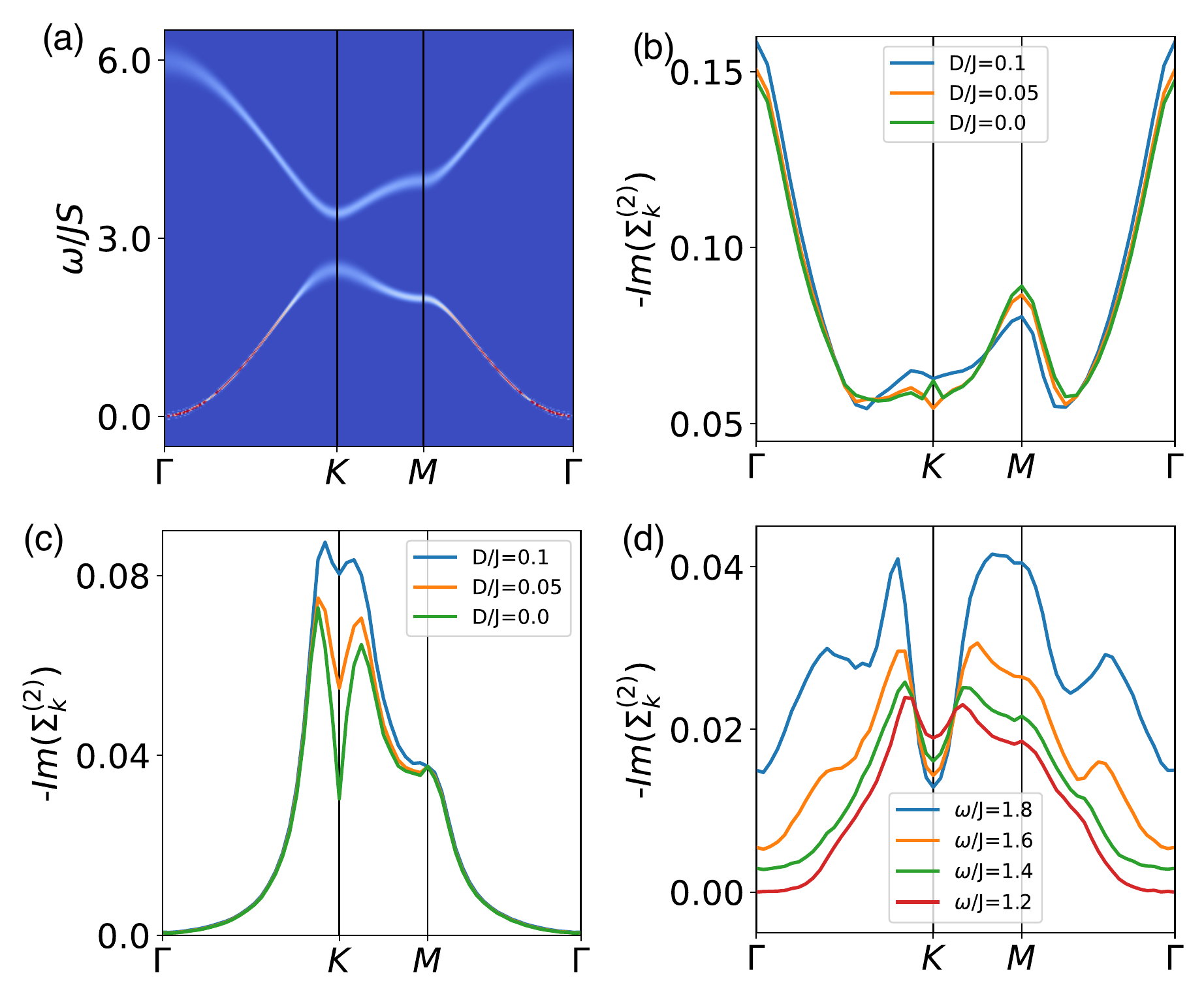}
\caption{Second-order renormalization of magnon dispersion, we use parameter $T/J$=0.5, and $D/J=0.1$. (a) Spectral function $A(\bm{k},\omega_{\bm{k}})$. In the plot, we use 1/10 of the maximal $A(\bm{k},\omega_{\bm{k}})$ for the color mapping to make a better display of damping components. (b)-(c) The magnon scattering rates with different DMI. One pronounced feature is that DMI significantly enhances the scattering rate around the Dirac point of the lower band, and gives rise to a faster decay dynamics. (d) Scattering rate in the lower band by the magnon parametric amplifications. We here employ a very small temperature $T/J=0.1$ to purely show the scattering effect from the amplified magnon population.}
\label{fig:7}
\end{figure}

We define the spectral function $A(\bm{k},\omega)$, which describes quasiparticle properties in magnon spectrum (band information) and density of states. $A(\bm{k},\omega)$ is readily obtained as the imaginary part of the retarded Green's function:
\begin{equation}\label{}
\begin{aligned}[b]
A(\bm{k},\omega)=-\frac{1}{\pi}\Im(G_R(\bm{k},\omega))
\end{aligned}
\end{equation}
The renormalized magnon dispersion at temperature $T/J=0.5$ with the second-order self-energy correction is shown in Fig.~\ref{fig:7}(a). The broadening of the bands denote magnon damping described the imaginary component $\tilde{\Sigma}^{(2)''}_{\bm{k}}$. We compare the scattering rates for the upper and lower band with and without DMI, as shown in Fig.~\ref{fig:7}(b) and (c). The DMI strongly influences the scattering rate, and the effect is momentum dependent in both magnon bands. In the upper band shown in Fig.~\ref{fig:7}(b), the rate is slightly enhanced at the $K$-point but suppressed at the $M$-point. Nevertheless, the influence of DMI on the scattering rate in the lower band is very pronounced, as shown in Fig.~\ref{fig:7}(c). The rate near the $K$-point (Dirac point) with DMI is significantly enhanced, indicating that the DMI facilitates scattering rate and results in faster magnon dynamics. The enhancement of the scattering rate at the Dirac point can be simply explained that the nonzero topological gap contributes to a finite density of states at that point, and thus provides more channels for the magnon scattering.

We also study the scattering effect from the magnon parametric amplification, which can dramatically modify the scattering rate by changing the energy position of the amplified magnons; the basic features are shown in Fig.~\ref{fig:7} (d). With a fixed magnon population intensity $I_p$, we increase the energy position from $\omega_{d_{\bm{k}}}=1.2J$ to $1.8J$ for magnons out of equilibrium. The scattering rate along the k-points is increased except for the Dirac points.  There are pronounced peaks between $\Gamma$ and $M$ points due to the obvious scattering around the pumped magnons. We employ a very small temperature of $T/J=0.1$ in the calculation to focus on the scattering effect from non-equilibrium magnon populations illustrating that the scattering rate can be flexibly tuned by the light-induced magnons. Magnon scattering rates are closely linked to microscopic dynamics, which provide powerful tools for measuring and controlling the magnetic order in the ultrafast time regime. This rapidly developing research area is relevant for magnetic memory and spintronics~\cite{Yang2020, Kang_2021}.

\section{Parametric amplifications and thermal Hall effect}\label{sec.V}
\subsection{Parametric magnon amplifications}\label{sec.V_A} 
Recently, magnon amplification with tailored electromagnetic (EM) pulses has been proposed as a way to controllably generate magnons with fixed energy~\cite{Malz2019}. In particular, the EM field can be used to create a finite density of magnons at and in the vicinity of the Dirac points, where the Berry curvature is concentrated. This, in turn, leads to an enhancement of physical manifestations of topological magnons, such as the thermal Hall effect and driven magnon Hall effect, that has eluded experimental detection in many quantum magnets known to host topological magnons. Interestingly, the same phenomenon can be used to explore the effects of magnon-magnon interaction in a systematic manner by tuning the density of magnons at fixed energy using the same physical observables to detect these effects experimentally. In the following, we show the effects of interactions on the thermal Hall effect due to a finite density of magnons near the Dirac points. We start with a brief description of the magnon amplification scheme. The coupling of the EM field to a quantum spin system can be described phenomenologically by the following effective Hamiltonian:
\begin{equation}\label{}
\begin{aligned}[b]
H_{int}=\sum_{\bm{k}}\frac{g_{\bm{k}}}{2}(d^{\dagger}_{-\bm{k}}d^{\dagger}_{\bm{k}} b+b^{\dagger}d_{\bm{k}}d_{-\bm{k}}),
\end{aligned}
\end{equation} 
where the field operator $b\approx\beta\exp(-i\Omega_0t)$ describes the external EM field or the pump photon, and $g_{\bm{k}}\approx g$ is the coupling coefficient between external field and magnon excitation that can be treated as momentum independent over a small bandwidth.  
The microscopic model for coupling coefficient $g_{\bm{k}}$ can be understood that the polarization operator couples to the external electric field via the Hamiltonian $H=H_0+\bm{E}(t)\cdot\bm{P}$, where $\bm{P}$ is the polarization operator that is given by~\cite{Moriya1968,Malz2019}
\begin{equation}\label{}
\begin{aligned}[b]
\bm{P}&=\sum_{\beta\gamma}\bm{K}^{\beta\gamma}_iS^{\beta}_{i}S^{\gamma}_{i}+\sum_{kl}\left(\bm{\pi}_{ij}\delta^{kl}+\bm{\Gamma}^{(kl)}_{ij}+\bm{D}^{[kl]}_{ij}\right)S^{k}_iS^{l}_j\\
&\approx K_ia_ia_i+K_ia^{\dag}_ia^{\dag}_i+Q_{ij}a_ia_j+Q^{*}_{ij}a^{\dag}_ia^{\dag}_j.
\end{aligned}
\end{equation}
where $\bm{K}^{\beta\gamma}_i$ is the coefficient tensor of single spin terms, $\bm{\pi}$ is an isotropic tensor, $\bm{D}$ and $\bm{\Gamma}$ are anisotropic traceless antisymmetric and symmetric tensors of paired spin terms, respectively~\cite{Malz2019}.Translating spin operators into the spin-wave picture, we have the anomalous magnon pairing terms while absorbing a photon, and the Fourier components of $Q_{ij}$ and $k_{ij}$ give rise to the coupling coefficient $g_{\bm{k}}$. We would like a breaking inversion symmetry to have a more considerable coupling strength due to the symmetry constraints (see Appendix~\ref{appen:E}). We note that our two-dimensional honeycomb flake on a substrate or some stacking (A-B-A type) of the layers will give rise to viable options for breaking the inversion symmetry. 

\begin{figure}[t!]
\includegraphics[width=8.5 cm]{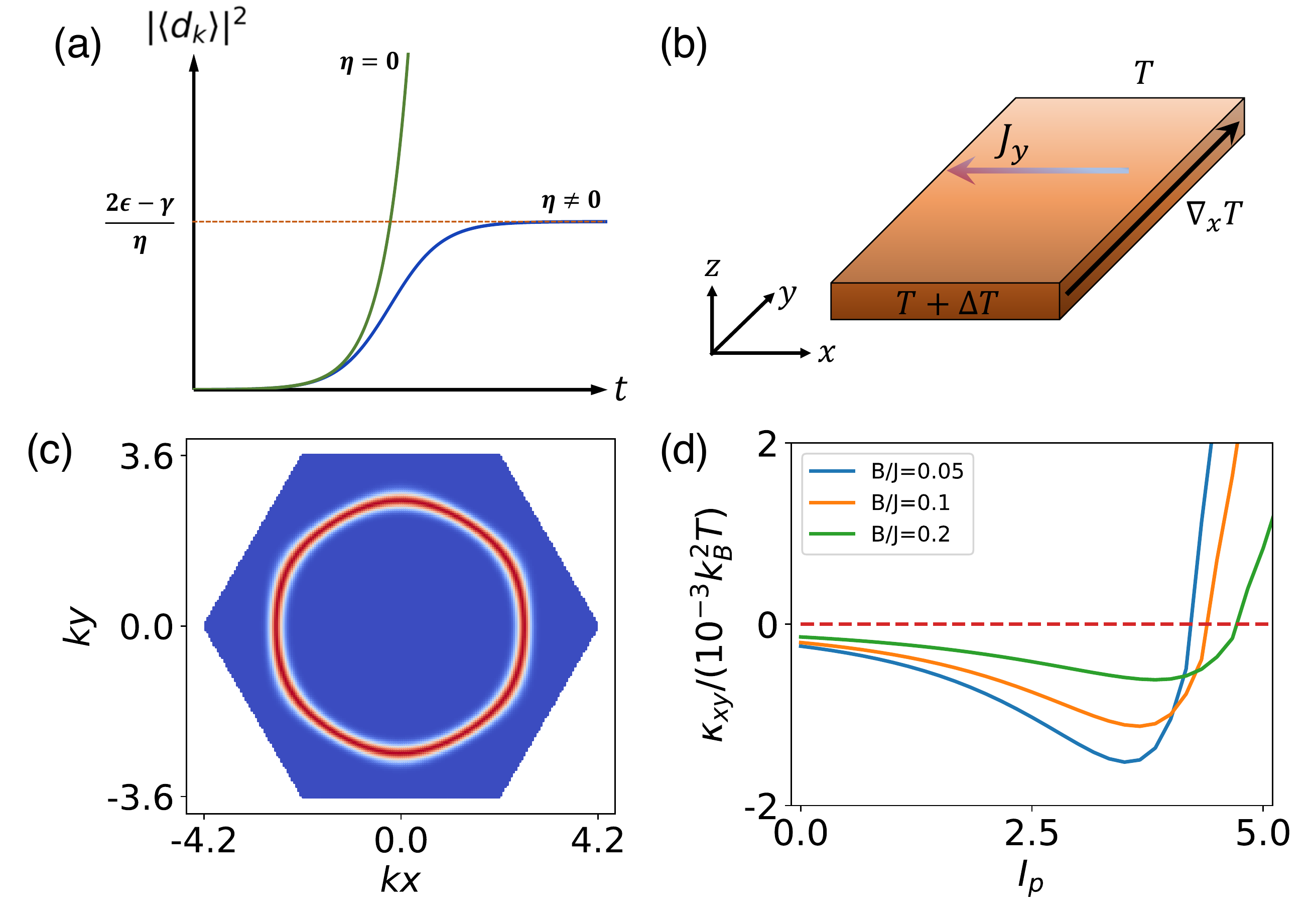}
\caption{Thermal Hall effect with parametric amplifications  (a) Magnon intensity is a function of time by parametric amplification. (b) The schematic plot for thermal Hall measurements. (c) The position of amplified magnon in Brillouin zone. (d) The thermal Hall conductivity with different magnetic fields. We set $T/J=0.2$ in the calculations.}
\label{fig:8}
\end{figure}

$H_{int}$ describes the process of creating a pair of magnons by absorption of a photon (and, by hermiticity of the coupling Hamiltonian, the reverse process of annihilation of a pair of magnons accompanied by the emission of a photon). Magnons must be created (or annihilated) in pairs with equal and opposite momenta to satisfy the conservation of momentum during the process (photons carry zero momentum). For the single-particle Hamiltonian, we have to include the scattering terms $H_{imp}$ caused by impurities and disorders. The magnon density is determined by the time-dependent Heisenberg equation of motion as follows~\cite{Malz2019}:
\begin{equation}\label{}
\begin{aligned}[b]
i\frac{d T_{\bm{k}}(t)}{dt}=\langle[\hat{T}_{\bm{k}},H]\rangle=\tilde{\Omega}_{\bm{k}} T_{\bm{k}}(t),\quad H=H_1+H_{int} + H_{imp},
\end{aligned}
\end{equation} 
where operator $\hat{T}_{\bm{k}}(t)=( d_{\bm{k}}, d^{\dagger}_{-\bm{k}})$, and $T_{\bm{k}}(t)=(\langle d_{\bm{k}} \rangle, \langle d^{\dagger}_{-\bm{k}} \rangle)$ are the classical amplitudes of the magnon fields, and $\tilde{\Omega}_{\bm{k}}$ is the dynamical matrix, which has the eigenvalues:
\begin{equation}\label{}
\begin{aligned}[b]
\omega_{\bm{k},\pm}=\frac{\omega_{\bm{k}}-\omega_{-\bm{k}}}{2}-\frac{i\gamma}{2}\pm\sqrt{\frac{(\omega_{\bm{k}}+\omega_{-\bm{k}}-\Omega_0)^2}{2}-\epsilon^2},
\end{aligned}
\end{equation} 
where $\gamma$ denotes linear dissipative damping, which is dominated by $H_{imp}$, and $\epsilon=g\beta$ is the overall coupling strength. When the detuning term $\omega_{\bm{k}}+\omega_{-\bm{k}}-\Omega_0\approx0$, the imaginary part of $\omega_{\bm{k},+}$ becomes $\epsilon-\gamma/2$, and the magnon density at $\bm{k}$ is $\langle d_{\bm{k}} \rangle \propto e^{(\epsilon-\gamma/2)t}$. When the coupling strength $\epsilon$ exceeds the dissipation $\gamma$, there is an exponential growth of the magnon density in mode $\bm{k}$, and resonant amplification is achieved. The eventual limit in the growth in amplified magnon density (due to decoherence, scattering, etc.) can be modeled by a phenomenological nonlinear damping constant $\eta$. Thus, the dynamical equation of magnon population $\langle d_{\bm{k}} \rangle$ is given by
\begin{equation}
\begin{aligned}[b]
&i\frac{dI_p}{dt}=\\
&\left(\tilde{\omega}_{\bm{k}}-\tilde{\omega}_{-\bm{k}}-i(\gamma+\eta I_p)+i\sqrt{4\epsilon^2-(\tilde{\omega}_{\bm{k}}+\tilde{\omega}_{-\bm{k}})^2}\right)I_p,
\end{aligned}
\end{equation}
where $I_p=\abs{\langle d_{\bm{k}} \rangle}^2$ is the magnon population intensity. When the resonance condition $\omega_{\bm{k}}+\omega_{-\bm{k}}=\Omega_0$ is reached, we can readily solve the dynamical equation and obtain the typical solution as
\begin{equation}
\begin{aligned}[b]
I_p=\frac{1}{\frac{\eta}{2\epsilon-\gamma}+c_0e^{-2(\epsilon-\gamma)t}}.
\end{aligned}
\end{equation}
When $t\to\infty$, we have the steady-state magnon intensity $I_p=\frac{2\epsilon-\gamma}{\eta}$, which is determined by the magnon-photon coupling strength and the dissipation coefficients $\gamma$ and $\eta$. The amplification dynamics is shown in Fig.~\ref{fig:8}(a). Thus, we can employ the amplification scheme to populate the magnon at specific momenta in the BZ, giving rise to a tunable many-body effect on magnon quasiparticles and transport properties. 

\subsection{Thermal Hall effect for Dirac magnon}\label{sec.V_B}
One direct application of the parametric amplification is the modification of the magnon thermal Hall effect and the use of this effect as an experimental tool for probing the topological character of the magnon bands and their renormalization due to interaction effects. The schematic plot of thermal Hall measurement is shown in Fig.~\ref{fig:8}(b). With the thermal gradient along the x-direction, one can measure the thermal Hall conductance along the y-direction. The calculation of the thermal conductivity $\kappa^{xy}$ in a slab geometry yields~\cite{Romhanyi2015, Matsumoto2011, Sun2021}: 
\begin{equation}\label{}
\begin{aligned}[b]
\kappa^{xy}=-\frac{k_{B}^2T}{(2\pi)^2}\sum_{n}\int d^2\bm{k} c_{2}\left(f(\omega_n(\bm{k}))\right)\Omega^{xy}_n(\bm{k}),
\end{aligned}
\end{equation}   
where $k_B$ is Boltzmann constant,$f(\omega_n(\bm{k}))$ is the magnon density distribution that, in the present case, has contributions from both thermally excited magnons -- whose population is given by the Bose-Einstein (BE) distribution function, $f(\omega_n(\bm{k}))=\frac{1}{e^{\omega_n(\bm{k})\beta}-1}$ -- as well as those generated by EM radiation. The function $c_2(u)$ is given by~\cite{Matsumoto2011}: 
\begin{equation}\label{eq:}
\begin{aligned}[b]
c_2(u)=(1+u)\left[\ln\left(1+\frac{1}{u}\right)\right]^2-\left[\ln(u)\right]^2-2\text{Li}_2(-u),
\end{aligned}
\end{equation} 
with $Li_2(u)=\sum^{\infty}_{k=1}\frac{u^k}{k^2}$ as the polylogarithm function, and $\Omega^{xy}_n(\bm{k})$ is the Berry curvature of $n$-th band. 

The interaction between the thermally generated and light-induced magnons is delicate in thermal Hall transport with the magnon amplifications. To fully solve this transport problem, we need to employ a self-consistent solution. However, this is beyond the scope of this paper. For simplicity, we consider the system at a low temperature so that the band renormalization effect from thermally excited magnons is negligible. Furthermore, due to the low temperature, we ignore the second-order scattering processes from magnon-magnon interactions in the transport measurement.

 With these simplifications, we focus on the thermal transport in the mean-field renormalized band by amplified magnons. We apply a monochromatic external EM field with energy $\Omega_0=3.2J$ that can only produce magnon populations at the energy $\omega_{\bm{k}}=\frac{\Omega_0}{2}$ in the lower band, as shown in Fig.~\ref{fig:8} (c). The advantage of this scheme is that the amplified magnons with the energy $\Omega_0=1.6J$ do not contribute to the thermal Hall signal due to the zero Berry curvature at their locations; thus, we can only have the thermal Hall response from the thermally excited magnons. 
We use the magnon intensity $I_p$ as the controllable variable, and the calculated thermal Hall conductivity is shown in Fig.~\ref{fig:8}(d). One can see that each curve has a transition point $I^c_p$ where the thermal Hall signal is going to reverse the signs. This originates from the topological phase transitions that are discussed in Sec.~\ref{sec.III_B}. The thermal Hall signals change the sign when the Chern number of the magnon band is reversed. An increase in the thermal Hall signal is expected to occur before the transition point due to the bandwidth reduction. When $I_p$ is close to the transition point, the topological gap is closed; thus, the signal becomes zero. The thermal Hall signal sharply increases after the reopening of the topological gap.

We also show that the transition points are magnetic field-dependent. The increase of the magnetic field gives rise to a larger critical point $I^c_p$ as shown in Fig.~\ref{fig:8}(d). Thus with a larger magnetic field, one may need a higher magnon intensity to achieve the topological phase transition. Therefore the thermal Hall measurement can be a useful experimental tool for exploring the interaction-induced topological phase transition.

\section{Conclusions}\label{con}
To summarize, we show much interesting physics arises from the magnon systems with non-trivial DM interaction, when the effect of magnon-magnon interaction is correctly accounted for. Our perturbative calculation shows the possibility of magnon band engineering exploiting such interaction, especially when the magnon density can be flexibly induced in different parts of the BZ with parametric amplification by the electromagnetic field. Such band engineering allows us to experimentally tune the bandwidth and the Berry curvature distribution. With properly designed amplification schemes, we can even induce band inversions leading to topological phase transitions. The transport properties of magnons under either a magnetic field gradient or a thermal gradient are very sensitive to the geometric and topological properties of the band. This allows us to propose a number of scenarios to experimentally probe the interplay between geometry, topology, and many-body effects in magnon systems.

The interaction-induced reduction of bandwidth and shift of the Dirac points can be detected by well-established techniques, including Brillouin light scattering, inelastic X-ray scattering, inelastic neutron scattering, and electron tunneling. By combining the computed quantum geometric tensor and the interaction-induced Haldane mass term, the thermal Hall conductivity of the magnons provides a direct diagnosis of the geometric and topological properties of the magnonic Chern bands by the external pumping field. More importantly, we can achieve flat bands with non-trivial topology, that provides a realistic platform for the study of strongly correlated bosonic systems such as superfluid-insulator transitions and flat band Bose-Einstein condensation. 

By applying a second-order perturbation theory, we also reveal that the magnon scattering rate is significantly enhanced near the Dirac point with the DMI, which has not been studied before. Such scattering from magnon-magnon interaction can play an essential role in the stability of the Dirac magnons and their decay dynamics. This is especially the case for systems with sizeable intrinsic interaction and small bandwidth. Our study highlights that the magnonic Dirac materials could serve as valuable platforms for many-body physics arising from lattice bosons with non-trivial band topologies.

\begin{acknowledgements}
B.Y. would like to acknowledge the support from the Singapore National Research Foundation (NRF) under NRF fellowship award NRF-NRFF12-2020-0005, and a Nanyang Technological University start-up grant (NTU-SUG). P.S. acknowledges financial support from the Ministry of Education, Singapore through MOE2019-T2-2-119.
\end{acknowledgements}

\appendix
\section{Effective tight-binding model of magnon by Holstein-Primakov transformation}\label{appen:A}
Generally, by HP transformation, the spin operators can be expressed as the nonlinear functions of magnon operators:
\begin{equation}\label{seq:1}
\begin{split}
S^{+}_i=\sqrt{2S}\sqrt{1-\frac{a^{\dagger}_ia_i}{2S}}a_i,\quad S^{-}_i=\sqrt{2S}a^{\dagger}_i\sqrt{1-\frac{a^{\dagger}_ia_i}{2S}}
\end{split}
\end{equation}
Up to the first-order Taylor expansion, we have
\begin{equation}\label{seq:2}
\begin{split}
S^{+}_i=\sqrt{2S}\left(a_i-\frac{a^{\dagger}_ia_ia_i}{4S}\right),\quad S^{-}_i=\sqrt{2S}\left(a^{\dagger}_i-\frac{a^{\dagger}_ia^{\dagger}_ia_i}{4S}\right)
\end{split}
\end{equation}
Then we have the Hamiltonian with magnon-magnon interaction terms:
\begin{equation}
\begin{split}
H&=\left(-JS\sum_{ij}a^{\dagger}_ia_j+h.c.\right)+3JS\sum_{i}a^{\dagger}_ia_i-3N_LJS^2
-J\sum_{ij}a^{\dagger}_ia_ia^{\dagger}_ja_j\\
&+\frac{J}{4}\sum_{ij}\left(a^{\dagger}_i a^{\dagger}_ja_ja_j +a^{\dagger}_j a^{\dagger}_ja_ia_j+a^{\dagger}_i a^{\dagger}_ja_ia_i+a^{\dagger}_i a^{\dagger}_ia_ia_j\right).
\end{split}
\end{equation}
Further, we have:
\begin{widetext}
\begin{equation}
\begin{split}
&\sum_{\alpha\beta\gamma}\epsilon_{\alpha\beta\gamma}v^{\alpha}_{ij}S^{\beta}_iS^{\gamma}_j=v^{z}_{ij}\left(S^{x}_iS^{y}_j-S^{y}_iS^{x}_j\right)\\
&=iv^{z}_{ij}\frac{S}{2}\left(a^{\dagger}_i+a_i-\frac{a^{\dagger}_ia_ia_i+a^{\dagger}_ia^{\dagger}_ia_i}{4S}\right)\left(a^{\dagger}_j-a_j-\frac{a^{\dagger}_ja^{\dagger}_ja_j-a^{\dagger}_ja_ja_j}{4S}\right)-iv^{z}_{ij}\frac{S}{2}\left(a^{\dagger}_i-a_i-\frac{a^{\dagger}_ia^{\dagger}_ia_i-a^{\dagger}_ia_ia_i}{4S}\right)\left(a^{\dagger}_j+a_j-\frac{a^{\dagger}_ja_ja_j+a^{\dagger}_ja^{\dagger}_ja_j}{4S}\right)\\
&=-iv^z_{ij}Sa^{\dagger}_ia_j+H.c.+\boxed{i\frac{v^z_{ij}}{4}a^{\dagger}_ia^{\dagger}_ia_ia_j+i\frac{v^z_{ij}}{4}a^{\dagger}_ia^{\dagger}_ja_ja_j+H.c.}.
\end{split}
\end{equation}
\end{widetext}
The terms in the box are quartic DM interactions, which are very important for the topological phase transitions. The quartic DM interactions in k-space are written as:
\begin{equation}
\begin{aligned}[b]
H'_{DM}=\sum_{\bm{k}_i}D^{\bm{k}_1,\bm{k}_2}_{\bm{k}_3,\bm{k}_4}\left(a^{\dagger}_{\bm{k}_1}a^{\dagger}_{\bm{k}_2}a_{\bm{k}_3}a_{\bm{k}_4}-b^{\dagger}_{\bm{k}_1}b^{\dagger}_{\bm{k}_2}b_{\bm{k}_3}b_{\bm{k}_4}\right),
\end{aligned}
\end{equation}
where $D^{\bm{k}_1,\bm{k}_2}_{\bm{k}_3,\bm{k}_4}=\frac{D}{2N_L}(\beta_{\bm{k}_4}+\beta_{\bm{k}_1})$.

If there is a NNN spin coupling, we then have the following spin Hamiltonian:
\begin{equation}\label{eq:the}
\begin{aligned}[b]
H_{nnn}=\sum_{i,j}J'\bm{S}_i\cdot\bm{S}_j,
\end{aligned}
\end{equation}
where $J'$ is the next nearest neighbor exchange parameter. Using HP transformation, we have the effective tight-binding model, which is written as
\begin{equation}\label{eq:the}
\begin{aligned}[b]
&H_{nnn}=J'S\sum_{\langle i,j\rangle}(a_ia^{\dag}_j+a^{\dag}_ia_j)+\\
&\frac{J'}{4}\sum_{\langle i,j\rangle}(4a^{\dagger}_ia_ia^{\dagger}_ja_j-a^{\dagger}_i a^{\dagger}_ja_ja_j -a^{\dagger}_j a^{\dagger}_ja_ia_j-a^{\dagger}_i a^{\dagger}_ja_ia_i-a^{\dagger}_i a^{\dagger}_ia_ia_j).
\end{aligned}
\end{equation}
The first term in RHS is the NNN single-particle term. Employing the Fourier transformation, we have the interacting Hamiltonian in k-space:
\begin{equation}
\begin{aligned}[b]
H'_{nnn}=\sum_{\bm{k}_i}G^{\bm{k}_1,\bm{k}_2}_{\bm{k}_3,\bm{k}_4}\left(a^{\dagger}_{\bm{k}_1}a^{\dagger}_{\bm{k}_2}a_{\bm{k}_3}a_{\bm{k}_4}+b^{\dagger}_{\bm{k}_1}b^{\dagger}_{\bm{k}_2}b_{\bm{k}_3}b_{\bm{k}_4}\right).
\end{aligned}
\end{equation}
 where the $G^{\bm{k}_1,\bm{k}_2}_{\bm{k}_3,\bm{k}_4}=\frac{J'}{2N_L}(2-\sum_{n}\Re(e^{i\bm{k}_1\cdot\bm{\sigma}_n})-\sum_{n}\Re(e^{i\bm{k}_4\cdot\bm{\sigma}_n}))$ is the interaction coefficient. 
 
\section{First-order renormalization of magnon band}\label{appen:B}

In this section, we give a detailed derivation of the first-order renormalization. We follow Zubarev and employ the retarded Green's function formalism~\cite{Zubarev_1960}:
\begin{equation}
\begin{split}
&\langle \psi(\bm{r},t);\psi^{\dagger}(\bm{r}',t')\rangle=-i\theta(t-t')\langle [\psi(\bm{r},t),\psi^{\dagger}(\bm{r}',t') ]\rangle \\
&=-i\theta(t-t')\left(\langle\psi(\bm{r},t)\psi^{\dagger}(\bm{r}',t')\rangle-\langle\psi^{\dagger}(\bm{r}',t')\psi(\bm{r},t)\rangle\right) \\
\end{split}
\end{equation}
where $\theta(t-t')$ is the step function, and $\psi^{\dag}(\bm{r},t)$ is the field operator of spinor which can be written as $(\sum_{i}\phi^{*}_{a,i}(\bm{r})a^{\dag}_{i}(t),\sum_{i}\phi^{*}_{b,i}(\bm{r})b^{\dag}_{i}(t))$, where $a^{\dag}_{i}(t)$ and $b^{\dag}_{i}(t)$ are the magnon creation operators of (A, B)-site in bipartite honeycomb lattice. The equation of motion for retarded Green’s function is
\begin{equation}\label{seq:6}
\begin{split}
i\frac{\partial}{\partial t}\langle \psi(\bm{r},t);&\psi^{\dagger}(\bm{r}',t')\rangle\\
&=\delta(\bm{r}-\bm{r}')\delta(t-t')+\langle [\psi(\bm{r},t),H(t)];\psi^{\dagger}(\bm{r}',t') \rangle
\end{split}
\end{equation}
Using Fourier transformation, we study the above dynamical equation in the frequency domain and k-space. Then we have
\begin{equation}
\begin{split}
\omega \langle \psi_{\bm{k}};&\psi^{\dagger}_{\bm{k}'}\rangle_{\omega}=\langle [\psi_{\bm{k}},\psi^{\dagger}_{\bm{k}'}] \rangle+\langle [\psi_{\bm{k}},H];\psi^{\dagger}_{\bm{k}'} \rangle_{\omega}\\
&=\langle [\psi_{\bm{k}},\psi^{\dagger}_{\bm{k}'}] \rangle+\langle [\psi_{\bm{k}},H_0];\psi^{\dagger}_{\bm{k}'} \rangle_{\omega}+\langle [\psi_{\bm{k}},H'];\psi^{\dagger}_{\bm{k}'} \rangle_{\omega}.
\end{split} 
\end{equation}
Fist we check the zero-order Green's function with $H=H_0$, which is for the noninteracting magnon particle: $\omega \langle \psi_{\bm{k}};\psi^{\dagger}_{\bm{k}'}\rangle_{\omega}=\delta_{\bm{k},\bm{k}'}+\langle [\psi_{\bm{k}},H_0];\psi^{\dagger}_{\bm{k}'} \rangle_{\omega}$:
\begin{equation}
\begin{split}
\langle \psi_{\bm{k}};\psi^{\dagger}_{\bm{k}'}\rangle^{0}_{\omega}=\frac{\delta_{\bm{k},\bm{k}'}}{\omega-M_{\bm{k}}}
\end{split}
\end{equation}
where $[\psi_{\bm{k}},H_0]=[\psi_{\bm{k}},\sum_{\bm{k}''}\psi_{\bm{k}''}^{\dagger}M_{\bm{k}''} \psi_{\bm{k}''}]=M_{\bm{k}} \psi_{\bm{k}}$. The equation of motion of Green's function now reads
\begin{equation}\label{}
\begin{aligned}[b]
\langle \psi_{\bm{k}};\psi^{\dagger}_{\bm{k}'}\rangle_{\omega}&=\frac{\langle [\psi_{\bm{k}},\psi^{\dagger}_{\bm{k}'}] \rangle}{\omega-M(\bm{k})}+(\omega-M(\bm{k}))^{-1}\langle [\psi_{\bm{k}},H'];\psi^{\dagger}_{\bm{k}'} \rangle_{\omega}\\
&=\langle \psi_{\bm{k}};\psi^{\dagger}_{\bm{k}'}\rangle^{0}_{\omega}+\langle \psi_{\bm{k}};\psi^{\dagger}_{\bm{k}'}\rangle^{0}_{\omega}\langle [\psi_{\bm{k}},H'];\psi^{\dagger}_{\bm{k}'} \rangle_{\omega}.
\end{aligned}
\end{equation}
The poles of Green's function give rise to the 2x2 noninteracting magnon bands. We can have a general interaction terms with the form: $H'=\sum_{\{\bm{k}_i\}}V^{\bm{k}_1,\bm{k}_2}_{\bm{k}_3,\bm{k}_4}\psi^{\dagger}_{\bm{k}_1}\psi^{\dagger}_{\bm{k}_2}\psi_{\bm{k}_3}\psi_{\bm{k}_4}$. Each wave vector index in interaction matrix element has two components due to the spinor property of $\psi^{\dagger}_{\bm{k}}$: 
\begin{equation}\label{}
\begin{aligned}[b]
\sum_{\bm{k}_1,\bm{k}_3}V^{\bm{k}_1,\bm{k}_2}_{\bm{k}_3,\bm{k}_4}\psi^{\dagger}_{\bm{k}_1}\psi_{\bm{k}_3}&=\sum_{\bm{k}_1,\bm{k}_3}V^{1\bm{k}_1,\bm{k}_2}_{1\bm{k}_3,\bm{k}_4}a^{\dagger}_{\bm{k}_1}a_{\bm{k}_3}+\sum_{\bm{k}_1,\bm{k}_3}V^{1\bm{k}_1,\bm{k}_2}_{2\bm{k}_3,\bm{k}_4}a^{\dagger}_{\bm{k}_1}b_{\bm{k}_3}\\
&+\sum_{\bm{k}_1,\bm{k}_3}V^{2\bm{k}_1,\bm{k}_2}_{1\bm{k}_3,\bm{k}_4}b^{\dagger}_{\bm{k}_1}a_{\bm{k}_3}+\sum_{\bm{k}_1,\bm{k}_3}V^{2\bm{k}_1,\bm{k}_2}_{2\bm{k}_3,\bm{k}_4}b^{\dagger}_{\bm{k}_1}b_{\bm{k}_3}
\end{aligned}
\end{equation}
We have the commutator term: 
\begin{equation}\label{seq:8}
\begin{split}
[\psi_{\bm{k}},H']&=\sum_{\{\bm{k}_i\}}V^{\bm{k}_1,\bm{k}_2}_{\bm{k}_3,\bm{k}_4}[\psi_{\bm{k}},\psi^{\dagger}_{\bm{k}_1}\psi^{\dagger}_{\bm{k}_2}\psi_{\bm{k}_3}\psi_{\bm{k}_4}]\\
&=\sum_{\{\bm{k}_i\}}V^{(\bm{k},\bm{k}_2)}_{\bm{k}_3,\bm{k}_4}\psi^{\dagger}_{\bm{k}_2}\psi_{\bm{k}_3}\psi_{\bm{k}_4}
\end{split}
\end{equation}
which is very important and will also be used in second-order renormalization. The first-order renormalization can be obtained by the mean-field approximation:
\begin{equation}
\begin{split}
\langle[\psi_{\bm{k}},H'];\psi^{\dag}_{\bm{k}'}\rangle_{\omega}
&\approx\sum_{\{\bm{k}_i\}}(V^{(\bm{k},\bm{k}_2)}_{\bm{k}_2,\bm{k}}+V^{(\bm{k},\bm{k}_2)}_{\bm{k},\bm{k}_2})\langle\psi^{\dagger}_{\bm{k}_2}\psi_{\bm{k}_2}\rangle\langle\psi_{\bm{k}};\psi^{\dag}_{\bm{k}'}\rangle_{\omega}\\
&\approx \Sigma^{(1)}_{H'}({\bm{k}})\langle\psi_{\bm{k}};\psi^{\dag}_{\bm{k}'}\rangle_{\omega}
\end{split}
\end{equation}
where $\Sigma^{(1)}_{H'}({\bm{k}})=\sum_{\bm{k}_2}V^{(\bm{k},\bm{k}_2)}_{(\bm{k}_2,\bm{k})}\langle\psi^{\dagger}_{\bm{k}_2}\psi_{\bm{k}_2}\rangle^0$ is the first-order self-energy, $\langle\psi^{\dagger}_{\bm{k}_2}\psi_{\bm{k}_2}\rangle^0$ is the zeroth order population matrix. $\Sigma^{(1)}_{H'}({\bm{k}})$ is a $2\times2$ matrix can be written by
$\begin{pmatrix} 
\sigma_{11}&\sigma_{12}\\ 
\sigma_{21}&\sigma_{22}
\end{pmatrix}$.
Each element term has an explicit form:
\begin{equation}\label{seq:10}
\begin{split}
\sigma_{11}&=\frac{-J}{N_L}\sum_{\bm{k}_1}(\gamma_0\langle b^{\dag}_{\bm{k}_1}b_{\bm{k}_1}\rangle-\frac{1}{2}\gamma^*_{\bm{k}_1}\langle b^{\dag}_{\bm{k}_1}a_{\bm{k}_1}\rangle-\frac{1}{2}\gamma_{\bm{k}_1}\langle a^{\dag}_{\bm{k}_1}b_{\bm{k}_1}\rangle), \\
\sigma_{12}&=\frac{-J}{N_L}\sum_{\bm{k}_1}(\gamma_{\bm{k}-\bm{k}_1}\langle b^{\dag}_{\bm{k}_1}a_{\bm{k}_1}\rangle-\frac{1}{2}\gamma_{\bm{k}}\langle b^{\dag}_{\bm{k}_1}b_{\bm{k}_1}\rangle-\frac{1}{2}\gamma_{\bm{k}}\langle a^{\dag}_{\bm{k}_1}a_{\bm{k}_1}\rangle),\\
\sigma_{21}&=\frac{-J}{N_L}\sum_{\bm{k}_1}(\gamma_{\bm{k}_1-\bm{k}}\langle a^{\dag}_{\bm{k}_1}b_{\bm{k}_1}\rangle-\frac{1}{2}\gamma^*_{\bm{k}}\langle a^{\dag}_{\bm{k}_1}a_{\bm{k}_1}\rangle-\frac{1}{2}\gamma^*_{\bm{k}}\langle b^{\dag}_{\bm{k}_1}b_{\bm{k}_1}\rangle),\\
\sigma_{22}&=\frac{-J}{N_L}\sum_{\bm{k}_1}(\gamma_0\langle a^{\dag}_{\bm{k}_1}a_{\bm{k}_1}\rangle-\frac{1}{2}\gamma_{\bm{k}_1}\langle a^{\dag}_{\bm{k}_1}b_{\bm{k}_1}\rangle-\frac{1}{2}\gamma^*_{\bm{k}_1}\langle b^{\dag}_{\bm{k}_1}a_{\bm{k}_1}\rangle).
\end{split}
\end{equation}
The expectation value can be calculated by magnon statistics. The grand partition function has the form $Z=e^{-\beta E_G}Tr(e^{-\beta H_0})$, where $E_G=-3JS^2N_L$ is the energy of ferromagnetic ground state, $H_0$ is the noninteracting magnon Hamiltonian. In eigenmode basis, $H_0=\omega_{d_{\bm{k}}}d^{\dagger}_{\bm{k}}d_{\bm{k}}+\omega_{u_{\bm{k}}}u^{\dagger}_{\bm{k}}u_{\bm{k}}$, and $\psi_{\bm{k}}=Ud_{\bm{k}}$,where $U$ is the unitary matrix of basis transformation. We have  
\begin{equation}\label{}
\begin{aligned}[b]
\begin{pmatrix} 
a_{\bm{k}}\\
b_{\bm{k}}
\end{pmatrix}=\frac{1}{\sqrt{2}}
\begin{pmatrix} 
\sqrt{1+\frac{B}{A}}e^{i\frac{\phi_{\bm{k}}}{2}}&\sqrt{1-\frac{B}{A}}e^{i\frac{\phi_{\bm{k}}}{2}}\\ 
\sqrt{1-\frac{B}{A}}e^{-i\frac{\phi_{\bm{k}}}{2}}&-\sqrt{1+\frac{B}{A}}e^{-i\frac{\phi_{\bm{k}}}{2}}
\end{pmatrix}
\begin{pmatrix} 
d_{\bm{k}}\\ 
u_{\bm{k}}
\end{pmatrix}
\end{aligned}
\end{equation}
where $A=\sqrt{(\frac{2D\beta_{\bm{k}}}{J})^2+\abs{\gamma_{\bm{k}}}^2}$,  and $B=\frac{2D\beta_{\bm{k}}}{J}$ are functions of wave vector $\bm{k}$. We obtain the averages of population operators from the grand canonical ensemble:
\begin{equation}\label{}
\begin{aligned}[b]
\langle a^{\dag}_{\bm{k}}a_{\bm{k}}\rangle&=\frac{1}{2}\left(1+\frac{B}{A}\right)\langle d^{\dagger}_{\bm{k}}d_{\bm{k}}\rangle+\frac{1}{2}\left(1-\frac{B}{A}\right)\langle u^{\dagger}_{\bm{k}}u_{\bm{k}}\rangle\\
&=\frac{1}{2}\left(1+\frac{B}{A}\right)f(\omega_{d_{\bm{k}}})+\frac{1}{2}\left(1-\frac{B}{A}\right)f(\omega_{u_{\bm{k}}}),\\ 
\langle b^{\dag}_{\bm{k}}b_{\bm{k}}\rangle&=\frac{1}{2}\left(1-\frac{B}{A}\right)f(\omega_{d_{\bm{k}}})+\frac{1}{2}\left(1+\frac{B}{A}\right)f(\omega_{u_{\bm{k}}}),\\
\langle a^{\dag}_{\bm{k}}b_{\bm{k}}\rangle&=\langle b^{\dag}_{\bm{k}}a_{\bm{k}}\rangle^*=\frac{1}{2}e^{-i\phi_{\bm{k}}}\sqrt{1-\frac{B^2}{A^2}}\langle d^{\dagger}_{\bm{k}}d_{\bm{k}}-u^{\dagger}_{\bm{k}}u_{\bm{k}}\rangle\\
&=\frac{1}{2}e^{-i\phi_{\bm{k}}}\sqrt{1-\frac{B^2}{A^2}}(f(\omega_{d_{\bm{k}}})-f(\omega_{u_{\bm{k}}})). 
\end{aligned}
\end{equation}
In first-order renormalization, we have to include the contributions from quartic DM interactions. We conclude a diagonal $2\times2$ self-energy matrix $\Sigma^{(1)}_{DM}(\bm{k})=\begin{pmatrix} 
\sigma'_{11}&0\\ 
0&\sigma'_{22}
\end{pmatrix}$. The diagonal matrix elements have the explicit forms $\sigma'_{11}=\sum_{\bm{k}_1}D^{(\bm{k}_1,\bm{k})}_{(\bm{k},\bm{k}_1)}\langle a^{\dag}_{\bm{k}_1}a_{\bm{k}_1}\rangle$, and $\sigma'_{22}=-\sum_{\bm{k}_1}D^{(\bm{k}_1,\bm{k})}_{(\bm{k},\bm{k}_1)}\langle b^{\dag}_{\bm{k}_1}b_{\bm{k}_1}\rangle$, where $D^{(\bm{k}_1,\bm{k})}_{(\bm{k},\bm{k}_1)}=D^{\bm{k}_1,\bm{k}}_{\bm{k},\bm{k}_1}+D^{\bm{k},\bm{k}_1}_{\bm{k},\bm{k}_1}+D^{\bm{k}_1,\bm{k}}_{\bm{k}_1,\bm{k}}+D^{\bm{k},\bm{k}_1}_{\bm{k}_1,\bm{k}}$ is symmetrized coefficient. We can have its explicit form $D^{(\bm{k}_1,\bm{k})}_{(\bm{k},\bm{k}_1)}=\frac{2D\beta_{\bm{k}}}{N_L}+\frac{2D\beta_{\bm{k}_1}}{N_L}$. Thus, we obtain
\begin{equation}
\begin{aligned}[b]
\sigma'_{11}&=\frac{D\beta_{\bm{k}}}{N_L}\sum_{\bm{k}_1}(f(\omega_{d_{\bm{k}_1}})+f(\omega_{u_{\bm{k}_1}}))\\
&+\frac{D}{N_L}\sum_{\bm{k}_1}\frac{\beta_{\bm{k}_1}B}{A}(f(\omega_{d_{\bm{k}_1}})-f(\omega_{u_{\bm{k}_1}}))\\
\sigma'_{22}&=-\frac{D\beta_{\bm{k}}}{N_L}\sum_{\bm{k}_1}(f(\omega_{d_{\bm{k}_1}})+f(\omega_{u_{\bm{k}_1}}))\\
&+\frac{D}{N_L}\sum_{\bm{k}_1}\frac{\beta_{\bm{k}_1}B}{A}(f(\omega_{d_{\bm{k}_1}})-f(\omega_{u_{\bm{k}_1}})).
\end{aligned}
\end{equation}
We can see that $\sigma'_{11}\neq \sigma'_{22}$, this term will contribute the effective Haldane mass. Finally, we obtain the total first-order self-energy, which is given by
\begin{widetext}
\begin{equation}
\begin{aligned}[b]
\Sigma^{(1)}_{H'}({\bm{k}})+\Sigma^{(1)}_{DM}(\bm{k})=\frac{J}{2N_L}\sum_{\bm{k}_1}
\begin{pmatrix} 
\Delta_1(\bm{k}_1)-\gamma_0\Theta_b(\bm{k}_1)+m_H&\gamma_{\bm{k}}\Theta^{+}(\bm{k}_1)-\gamma_{\bm{k}-\bm{k}_1}e^{i\phi_{{\bm{k}}_1}}\Delta_2(\bm{k}_1)\\ 
\gamma^*_{\bm{k}}\Theta^{+}(\bm{k}_1)-\gamma_{\bm{k}_1-\bm{k}}e^{-i\phi_{\bm{k}_1}}\Delta_2(\bm{k}_1)&\Delta_1(\bm{k}_1)-\gamma_0\Theta_a(\bm{k}_1)-m_H
\end{pmatrix}.
\end{aligned}
\end{equation}
\end{widetext}
Where we have the terms:
\begin{equation}
\begin{aligned}[b]
\Delta_1(\bm{k}_1)&=\left(\abs{\gamma_{\bm{k}_1}}\sqrt{1-\frac{B^2}{A^2}}+\frac{B^2}{A}\right)\Theta^{-}(\bm{k}_1),\\
\Delta_2(\bm{k}_1)&=\sqrt{1-\frac{B^2}{A^2}}\Theta^{-}(\bm{k}_1),\\
\Theta_b(\bm{k}_1)&=\Theta^{+}(\bm{k}_1)-\frac{B}{A}\Theta^{-}(\bm{k}_1),\\
\Theta_a(\bm{k}_1)&=\Theta^{+}(\bm{k}_1)+\frac{B}{A}\Theta^{-}(\bm{k}_1),\\
\end{aligned}
\end{equation}
where $\Theta^{+}(\bm{k})=f(\omega_{d_{\bm{k}}})+f(\omega_{u_{\bm{k}}})$, $\Theta^{-}(\bm{k})=f(\omega_{d_{\bm{k}}})-f(\omega_{u_{\bm{k}}})$, and $m_H=\frac{2D\beta_{\bm{k}}}{J}\Theta^{+}(\bm{k}_1)$ is the interaction-induced Haldane mass term. The DMI changes the population balances between $\langle a^{\dag}_{\bm{k}}a_{\bm{k}}\rangle$ and $\langle b^{\dag}_{\bm{k}}b_{\bm{k}}\rangle$, which can not be neglect around the Dirac point ($\frac{B}{A}=1$). When DMI parameter $D$ is zero or near the $\Gamma$ point ($B\approx 0$), we have a simpler unitary transform matrix, which is given by
\begin{equation}\label{seq:11}
\begin{split}
\begin{pmatrix} 
a_{\bm{k}}\\
b_{\bm{k}}
\end{pmatrix}\approx
\frac{1}{\sqrt{2}}\begin{pmatrix} 
e^{i\frac{\phi_{\bm{k}}}{2}}&e^{i\frac{\phi_{\bm{k}}}{2}}\\ 
e^{-i\frac{\phi_{\bm{k}}}{2}}&-e^{-i\frac{\phi_{\bm{k}}}{2}}
\end{pmatrix}
\begin{pmatrix} 
d_{\bm{k}}\\ 
u_{\bm{k}}
\end{pmatrix}.
\end{split}
\end{equation}
In the low temperature approximation, all thermal excited magnons mainly populate around the $\Gamma$ point of the lower band, where we have $B\approx 0$, $f(\omega_{u_{\bm{k}}})=0$, $m_H\approx 0$ and $\Theta^{+}(\bm{k}_1)=\Delta(\bm{k}_1)=f(\omega_{d_{\bm{k}}})$. The first-order self-energy at low temperature is thus given by
\begin{widetext}
\begin{equation}
\begin{split}
\Sigma^{(1)}_{H'}({\bm{k}})=
\frac{J}{2N_L}\sum_{\bm{k}_1}f(\omega_{\bm{k}_1})
\begin{pmatrix} 
Re(\gamma_{\bm{k}_1}e^{-i\phi_{\bm{k}_1}})-\gamma_0&\gamma_{\bm{k}}-\gamma_{\bm{k}-\bm{k}_1}e^{i\phi_{{\bm{k}}_1}}\\ 
\gamma^*_{\bm{k}}-\gamma_{\bm{k}_1-\bm{k}}e^{-i\phi_{\bm{k}_1}}&Re(\gamma_{\bm{k}_1}e^{-i\phi_{\bm{k}_1}})-\gamma_0
\end{pmatrix}
\approx-\frac{J}{2N_L}\sum_{\bm{k}_1}\frac{\omega_{\bm{k}_1}}{3JS}f(\omega_{\bm{k}_1})
\begin{pmatrix} 
3&-\gamma_{\bm{k}}\\ 
-\gamma^*_{\bm{k}}&3
\end{pmatrix}
\end{split}
\end{equation}
\end{widetext}
where $Re(\gamma_{k_1}e^{-i\phi_{k_1}})=\abs{\gamma_{k_1}}$, and $1-\gamma_{k-k_1}e^{i\phi_{k_1}}/\gamma_k\approx 1-\gamma_{-k_1}e^{i\phi_{k_1}}/\gamma_0=1-\abs{\gamma_{k_1}}/\gamma_0$ are employed.
In low temperatures, the magnon excitations only exist around the zone center. We can use isotropic approximation, and the magnon band has quadratic expression $\omega_{\bm{k}}=c_2 k^2$. The summation can be written as
\begin{equation}
\begin{split}
&\frac{1}{N_L}\sum_{\bm{k}_1}\omega_{\bm{k}_1}f(\omega_{\bm{k}_1})=\frac{1}{N_L}\sum_{\bm{k}_1}\frac{c_2 k_1^2}{e^{\omega_{\bm{k}-1}\beta}-1}\\
&=\frac{A}{(2\pi)^2}\sum^{+\infty}_{n=1}\int d^2\bm{k}_1c_2 k_1^2 e^{-n\beta c_2 k_1^2}\\
&=\frac{A}{4\pi\beta^2c_2}\sum^{+\infty}_{n=1}\frac{1}{n^2}=\frac{A k^2_B}{4\pi c_2}\zeta(2)T^2,
\end{split}
\end{equation}
where $\frac{1}{N_L}\sum_{\bm{k}_1}=\frac{A}{(2\pi)^2}\int_{BZ}d^2\bm{k}_1$ with $A$ is the area of the primitive unit cell, $\zeta(p)=\sum_{n=1}\frac{1}{n^p}$ is Riemann zeta function. When $p=2$, $\zeta(2)=\frac{\pi^2}{6}$, and self-energy $\Sigma^{(1)}_{\bm{k}}=-\frac{A \pi k^2_B}{24 J^2S^3}T^2H_0(\bm{k})$. One can see that the renormalization magnitude is proportional to $T^2$ at a low-temperature approximation.

\onecolumngrid
\section{Second-order renormalization of magnon band}\label{appen:C}
We here to derive the second-order self-energy in a weekly interacting magnon system. By the Eq.~(\ref{seq:6}), we have the Green's functions equation of motion:
\begin{equation}\label{seq:15}
\begin{aligned}[b]
\omega \langle [\psi_{\bm{k}},H'];\psi^{\dagger}_{\bm{k}'} \rangle_{\omega}=\langle [[\psi_{\bm{k}},H'],\psi^{\dagger}_{\bm{k}'}] \rangle+\langle [[\psi_{\bm{k}},H'],H];\psi^{\dagger}_{\bm{k}'} \rangle_{\omega}.
\end{aligned}
\end{equation}
We have $[\psi_{\bm{k}},H']=\sum_{\{\bm{k}_i\}}V^{\bm{k}_1,\bm{k}_2}_{\bm{k}_3,\bm{k}_4}[\psi_{\bm{k}},\psi^{\dagger}_{\bm{k}_1}\psi^{\dagger}_{\bm{k}_2}\psi_{\bm{k}_3}\psi_{\bm{k}_4}]=\sum_{\{\bm{k}_i\}}V^{(\bm{k},\bm{k}_2)}_{\bm{k}_3,\bm{k}_4}\psi^{\dagger}_{\bm{k}_2}\psi_{\bm{k}_3}\psi_{\bm{k}_4}=\frac{1}{2}\sum_{\{\bm{k}_i\}}V^{(\bm{k},\bm{k}_2)}_{(\bm{k}_3,\bm{k}_4)}\psi^{\dagger}_{\bm{k}_2}\psi_{\bm{k}_3}\psi_{\bm{k}_4}$. For the last expression, we split the original form and relabel the indices for the symmetric interaction matrix $V^{(\bm{k},\bm{k}_2)}_{(\bm{k}_3,\bm{k}_4)}$. We now transform the above equation to
\begin{equation}\label{seq:}
\begin{aligned}[b]
\frac{1}{2}\sum_{\{\bm{k}_i\}}V^{(\bm{k},\bm{k}_2)}_{(\bm{k}_3,\bm{k}_4)}\omega \langle \psi^{\dagger}_{\bm{k}_2}\psi_{\bm{k}_3}\psi_{\bm{k}_4};\psi^{\dagger}_{\bm{k}'} \rangle_{\omega}=\frac{1}{2}\sum_{\{\bm{k}_i\}}V^{(\bm{k},\bm{k}_2)}_{(\bm{k}_3,\bm{k}_4)}\langle [\psi^{\dagger}_{\bm{k}_2}\psi_{\bm{k}_3}\psi_{\bm{k}_4},\psi^{\dagger}_{\bm{k}'}] \rangle+\frac{1}{2}\sum_{\{\bm{k}_i\}}V^{(\bm{k},\bm{k}_2)}_{(\bm{k}_3,\bm{k}_4)}\langle [\psi^{\dagger}_{\bm{k}_2}\psi_{\bm{k}_3}\psi_{\bm{k}_4},H];\psi^{\dagger}_{\bm{k}'} \rangle_{\omega}.
\end{aligned}
\end{equation}
We can neglect the summation first and focus on equation of motion of $\psi^{\dagger}_{\bm{k}_2}\psi_{\bm{k}_3}\psi_{\bm{k}_4}$:
\begin{equation}\label{seq:17}
\begin{aligned}[b]
\omega \langle \psi^{\dagger}_{\bm{k}_2}\psi_{\bm{k}_3}\psi_{\bm{k}_4};\psi^{\dagger}_{\bm{k}'} \rangle_{\omega}=\langle [\psi^{\dagger}_{\bm{k}_2}\psi_{\bm{k}_3}\psi_{\bm{k}_4},\psi^{\dagger}_{\bm{k}'}] \rangle+\langle [\psi^{\dagger}_{\bm{k}_2}\psi_{\bm{k}_3}\psi_{\bm{k}_4},H];\psi^{\dagger}_{\bm{k}'} \rangle_{\omega}.
\end{aligned}
\end{equation}
We expand the first term in the RHS of the above equation, which is the commutator. From the Eq.~(\ref{seq:8})
\begin{equation}\label{seq:18}
\begin{aligned}[b]
\langle[\psi^{\dagger}_{\bm{k}_2}\psi_{\bm{k}_3}\psi_{\bm{k}_4},\psi^{\dagger}_{\bm{k}'}]\rangle\approx\langle\psi^{\dagger}_{\bm{k}_2}\psi_{\bm{k}_4}\rangle\delta_{\bm{k}_2,\bm{k}_4}\delta_{\bm{k}_3,\bm{k}'}+\langle\psi^{\dagger}_{\bm{k}_2}\psi_{\bm{k}_3}\rangle\delta_{\bm{k}_2,\bm{k}_3}\delta_{\bm{k}_4,\bm{k}'},
\end{aligned}
\end{equation}
The second term in Eq.~({\ref{seq:17}}) is little bit complicated. We spilt the commutator $[\psi^{\dagger}_{\bm{k}_2}\psi_{\bm{k}_3}\psi_{\bm{k}_4},H]$ into two parts by separate the $H=H_0+H'$, and obtain
\begin{equation}
\begin{aligned}[b]
[\psi^{\dagger}_{\bm{k}_2}\psi_{\bm{k}_3}\psi_{\bm{k}_4},H_0]=\psi^{\dagger}_{\bm{k}_2}\psi_{\bm{k}_3}M_{\bm{k}_4}\psi_{\bm{k}_4}+\psi^{\dagger}_{\bm{k}_2}M_{\bm{k}_3}\psi_{\bm{k}_3}\psi_{\bm{k}_4}-\psi^{\dagger}_{\bm{k}_2}M_{\bm{k}_2}\psi_{\bm{k}_3}\psi_{\bm{k}_4}
\end{aligned}
\end{equation}
where the $M_{\bm{k}}\psi_{\bm{k}}$ gives rise to the factors related to the magnon spectrum. We define a tensor with the symbol $\mathcal{M}$, which is the function $\mathcal{M}=\mathcal{M}(\bm{k_2},\bm{k}_3,\bm{k}_4)$. We has the transformation $Q'_{a,bc}=\sum_{b,e,f}\mathcal{M}^{a,bc}_{b,ef}Q_{a,bc}$. Where $Q'=\psi^{\dagger}_{\bm{k}_2}\psi_{\bm{k}_3}M_{\bm{k}_4}\psi_{\bm{k}_4}+\psi^{\dagger}_{\bm{k}_2}M_{\bm{k}_3}\psi_{\bm{k}_3}\psi_{\bm{k}_4}-\psi^{\dagger}_{\bm{k}_2}M_{\bm{k}_2}\psi_{\bm{k}_3}\psi_{\bm{k}_4}$, $Q_{a,bc}=\psi^{\dagger}_{\bm{k}_2}\psi_{\bm{k}_3}\psi_{\bm{k}_4}$, and $a$,$b$ and $c$ are integers in $\{1,2\}$ indexed the components of the quantity. For instance, $Q_{1,11}=a^{\dagger}_{\bm{k}_2}a_{\bm{k}_3}a_{\bm{k}_4}$, $Q_{1,21}=a^{\dagger}_{\bm{k}_2}b_{\bm{k}_3}a_{\bm{k}_4}$. Thus we can have a neat expression which is
\begin{equation}
\begin{aligned}[b]
[[\psi_{\bm{k}},H'],H_0]=\frac{1}{2}\sum_{\{\bm{k}_i\}}V^{(\bm{k},\bm{k}_2)}_{(\bm{k}_3,\bm{k}_4)}\sum_{b,e,f}\mathcal{M}^{a,bc}_{b,ef}Q_{a,bc}=\frac{1}{2}\sum_{\{\bm{k}_i\}}V^{(\bm{k},\bm{k}_2)}_{(\bm{k}_3,\bm{k}_4)}\mathcal{M}\psi^{\dagger}_{\bm{k}_2}\psi_{\bm{k}_3}\psi_{\bm{k}_4}
\end{aligned}
\end{equation}
Thus we have $\langle[\psi^{\dagger}_{\bm{k}_2}\psi_{\bm{k}_3}\psi_{\bm{k}_4},H_0];\psi^{\dag}_{\bm{k}'}\rangle_{\omega}=\mathcal{M}\langle\psi^{\dagger}_{\bm{k}_2}\psi_{\bm{k}_3}\psi_{\bm{k}_4};\psi^{\dag}_{\bm{k}'}\rangle_{\omega}$. Now we focus on the term $[\psi^{\dagger}_{\bm{k}_2}\psi_{\bm{k}_3}\psi_{\bm{k}_4},H']$, It gives rise more complicate terms written as:
\begin{equation}\label{}
\begin{aligned}[b]
[\psi^{\dagger}_{\bm{k}_2}\psi_{\bm{k}_3}\psi_{\bm{k}_4},H']=&-\sum_{\bm{k}'_i}V^{\bm{k}'_1,\bm{k}'_2}_{(\bm{k}'_3,\bm{k}_2)}\psi^{\dagger}_{\bm{k}'_1}\psi^{\dagger}_{\bm{k}'_2}\psi_{\bm{k}'_3}\psi_{\bm{k}_3}\psi_{\bm{k}_4}+\sum_{\bm{k}'_i}V^{(\bm{k}_3,\bm{k}'_2)}_{\bm{k}'_3,\bm{k}'_4}\psi^{\dagger}_{\bm{k}_2}\psi^{\dagger}_{\bm{k}'_2}\psi_{\bm{k}'_3}\psi_{\bm{k}'_4}\psi_{\bm{k}_4}\\
+&\sum_{\bm{k}'_i}V^{(\bm{k}_4,\bm{k}'_2)}_{\bm{k}'_3,\bm{k}'_4}\psi^{\dagger}_{\bm{k}_2}\psi^{\dagger}_{\bm{k}'_2}\psi_{\bm{k}_3}\psi_{\bm{k}'_3}\psi_{\bm{k}'_4}+\sum_{\bm{k}'_i}V^{(\bm{k}_4,\bm{k}_3)}_{\bm{k}'_3,\bm{k}'_4}\psi^{\dagger}_{\bm{k}_2}\psi_{\bm{k}'_3}\psi_{\bm{k}'_4}.
\end{aligned}
\end{equation}
We still employ the RPA and consider all possible contractions. We have 
\begin{equation}\label{seq:19}
\begin{aligned}[b]
\sum_{\bm{k}'_i}V^{\bm{k}'_1,\bm{k}'_2}_{(\bm{k}'_3,\bm{k}_2)}\langle\psi^{\dagger}_{\bm{k}'_1}\psi^{\dagger}_{\bm{k}'_2}\psi_{\bm{k}'_3}\psi_{\bm{k}_3}\psi_{\bm{k}_4};\psi^{\dagger}_{\bm{k}'}\rangle
\approx&
V^{(\bm{k}_4,\bm{k}_3)}_{(\bm{k},\bm{k}_2)}\langle\psi^{\dag}_{\bm{k}_3}\psi_{\bm{k}_3}\rangle\langle\psi^{\dag}_{\bm{k}_4}\psi_{\bm{k}_4}\rangle\langle\psi_{\bm{k}};\psi^{\dagger}_{\bm{k}'}\rangle+\sum_{\bm{k}'_3}V^{(\bm{k}_4,\bm{k}'_3)}_{(\bm{k}'_3,\bm{k}_2)}\delta_{\bm{k}_3,\bm{k}}\langle\psi^{\dag}_{\bm{k}_4}\psi_{\bm{k}_4}\rangle\langle\psi^{\dag}_{\bm{k}'_3}\psi_{\bm{k}'_3}\rangle\langle\psi_{\bm{k}};\psi^{\dagger}_{\bm{k}'}\rangle\\
&+\sum_{\bm{k}'_3}V^{(\bm{k}_3,\bm{k}'_3)}_{(\bm{k}'_3,\bm{k}_2)}\delta_{\bm{k}_4,\bm{k}}\langle\psi^{\dag}_{\bm{k}_3}\psi_{\bm{k}_3}\rangle\langle\psi^{\dag}_{\bm{k}'_3}\psi_{\bm{k}'_3}\rangle\langle\psi_{\bm{k}};\psi^{\dagger}_{\bm{k}'}\rangle
\end{aligned}
\end{equation}
For the second and third terms in Eq~(\ref{seq:18}), we apply the same procedure:
\begin{equation}\label{seq:20}
\begin{aligned}[b]
\sum_{\bm{k}'_i}V^{(\bm{k}_3,\bm{k}'_2)}_{\bm{k}'_3,\bm{k}'_4}&\langle\psi^{\dagger}_{\bm{k}_2}\psi^{\dagger}_{\bm{k}'_2}\psi_{\bm{k}'_3}\psi_{\bm{k}'_4}\psi_{\bm{k}_4};\psi^{\dagger}_{\bm{k}'}\rangle+\sum_{\bm{k}'_i}V^{(\bm{k}_4,\bm{k}'_2)}_{\bm{k}'_3,\bm{k}'_4}\langle\psi^{\dagger}_{\bm{k}_2}\psi^{\dagger}_{\bm{k}'_2}\psi_{\bm{k}_3}\psi_{\bm{k}'_3}\psi_{\bm{k}'_4};\psi^{\dagger}_{\bm{k}'}\rangle\\
=&\sum_{\bm{k}'_2}V^{(\bm{k}_3,\bm{k}'_2)}_{(\bm{k}'_2,\bm{k}_2)}\delta_{\bm{k}_4,\bm{k}}\langle\psi^{\dag}_{\bm{k}_2}\psi_{\bm{k}_2}\rangle\langle\psi^{\dag}_{\bm{k}'_2}\psi_{\bm{k}'_2}\rangle\langle\psi_{\bm{k}};\psi^{\dagger}_{\bm{k}'}\rangle+\sum_{\bm{k}'_2}V^{(\bm{k}_3,\bm{k}'_2)}_{(\bm{k}'_2,\bm{k})}\delta_{\bm{k}_4,\bm{k}_2}\langle\psi^{\dag}_{\bm{k}_2}\psi_{\bm{k}_2}\rangle\langle\psi^{\dag}_{\bm{k}'_2}\psi_{\bm{k}'_2}\rangle\langle\psi_{\bm{k}};\psi^{\dagger}_{\bm{k}'}\rangle\\
=&V^{(\bm{k}_3,\bm{k}_4)}_{(\bm{k}_2,\bm{k})}\langle\psi^{\dag}_{\bm{k}_2}\psi_{\bm{k}_2}\rangle\langle\psi^{\dag}_{\bm{k}_4}\psi_{\bm{k}_4}\rangle\langle\psi_{\bm{k}};\psi^{\dagger}_{\bm{k}'}\rangle+\sum_{\bm{k}'_2}V^{(\bm{k}_4,\bm{k}'_2)}_{(\bm{k}'_2,\bm{k}_2)}\delta_{\bm{k}_3,\bm{k}}\langle\psi^{\dag}_{\bm{k}_2}\psi_{\bm{k}_2}\rangle\langle\psi^{\dag}_{\bm{k}'_2}\psi_{\bm{k}'_2}\rangle\langle\psi_{\bm{k}};\psi^{\dagger}_{\bm{k}'}\rangle\\
=&V^{(\bm{k}_3,\bm{k}_4)}_{(\bm{k}_2,\bm{k})}\langle\psi^{\dag}_{\bm{k}_2}\psi_{\bm{k}_2}\rangle\langle\psi^{\dag}_{\bm{k}_3}\psi_{\bm{k}_3}\rangle\langle\psi_{\bm{k}};\psi^{\dagger}_{\bm{k}'}\rangle+\sum_{\bm{k}'_2}V^{(\bm{k}_4,\bm{k}'_2)}_{(\bm{k}'_2,\bm{k})}\delta_{\bm{k}_3,\bm{k}_2}\langle\psi^{\dag}_{\bm{k}_2}\psi_{\bm{k}_2}\rangle\langle\psi^{\dag}_{\bm{k}'_2}\psi_{\bm{k}'_2}\rangle\langle\psi_{\bm{k}};\psi^{\dagger}_{\bm{k}'}\rangle
\end{aligned}
\end{equation}
The first and fourth terms in the bracket cancel the last two terms in Eq.~(\ref{seq:19}), the second term is
\begin{equation}\label{}
\begin{aligned}[b]
 \sum_{\bm{k}'_2}V^{(\bm{k}_3,\bm{k}'_2)}_{(\bm{k}'_2,\bm{k})}\delta_{\bm{k}_4,\bm{k}_2}\langle\psi^{\dag}_{\bm{k}_2}\psi_{\bm{k}_2}\rangle\langle\psi^{\dag}_{\bm{k}'_2}\psi_{\bm{k}'_2}\rangle\langle\psi_{\bm{k}};\psi^{\dagger}_{\bm{k}'}\rangle=\delta_{\bm{k}_4,\bm{k}_2}\langle\psi^{\dag}_{\bm{k}_2}\psi_{\bm{k}_2}\rangle\Sigma^{(1)}_{\bm{k}}\langle\psi_{\bm{k}};\psi^{\dagger}_{\bm{k}'}\rangle
\end{aligned}
\end{equation}
and we obtain the similar result for the last term $\delta_{\bm{k}_3,\bm{k}_2}\langle\psi^{\dag}_{\bm{k}_2}\psi_{\bm{k}_2}\rangle\Sigma^{(1)}_{\bm{k}}\langle\psi_{\bm{k}};\psi^{\dagger}_{\bm{k}'}\rangle$. Finally, we calculate the last term in Eq.~(\ref{seq:18}), which can be written as
\begin{equation}\label{seq:21}
\begin{aligned}[b]
\sum_{\bm{k}'_i}V^{(\bm{k}_4,\bm{k}_3)}_{\bm{k}'_3,\bm{k}'_4}\langle\psi^{\dagger}_{\bm{k}_2}\psi_{\bm{k}'_3}\psi_{\bm{k}'_4};\psi^{\dagger}_{\bm{k}'}\rangle
=V^{(\bm{k}_4,\bm{k}_3)}_{(\bm{k}_2,\bm{k})}\langle\psi^{\dag}_{\bm{k}_2}\psi_{\bm{k}_2}\rangle\langle\psi_{\bm{k}};\psi^{\dagger}_{\bm{k}'}\rangle_{\omega}
\end{aligned}
\end{equation}
Combining the Eqs~(\ref{seq:19}--\ref{seq:21}), we obtain
\begin{equation}\label{seq:22}
\begin{aligned}[b]
\langle [\psi^{\dag}_{\bm{k}_2}\psi_{\bm{k}_3}\psi_{\bm{k}_4},H'];\psi^{\dagger}_{\bm{k}'} \rangle_{\omega}=
\left(V^{(\bm{k}_4,\bm{k}_3)}_{(\bm{k}_2,\bm{k})}(n_{\bm{k}_2}(1+n_{\bm{k}_3}+n_{\bm{k}_4})-n_{\bm{k}_3}n_{\bm{k}_4})+\left(\delta_{\bm{k}_4,\bm{k}_2}+\delta_{\bm{k}_3,\bm{k}_2}\right)n_{\bm{k}_2}\Sigma^{(1)}_{\bm{k}}\right)\langle\psi_{\bm{k}};\psi^{\dagger}_{\bm{k}'}\rangle_{\omega}
\end{aligned}
\end{equation}
where $n_{\bm{k}_i}$ is short for $\langle\psi^{\dag}_{\bm{k}_i}\psi_{\bm{k}_i}\rangle$. We can replace the Green's function $\langle\psi_{\bm{k}};\psi^{\dagger}_{\bm{k}'}\rangle_{\omega}$ with free particle one $\langle\psi_{\bm{k}};\psi^{\dagger}_{\bm{k}'}\rangle^0_{\omega}$ in this order. We note that it is very convenient to write the full expression in eigenmode space for the second-order renormalization. We will do it later using a unitary transformation.
We have the expression:
\begin{equation}
\begin{aligned}[b]
\left(\omega-\mathcal{M}\right)&\langle\psi^{\dagger}_{\bm{k}_2}\psi_{\bm{k}_3}\psi_{\bm{k}_4};\psi^{\dag}_{\bm{k}'}\rangle_{\omega}=
n_{\bm{k}_4}\delta_{\bm{k}_2,\bm{k}_4}\delta_{\bm{k}_3,\bm{k}'}+n_{\bm{k}_3}\delta_{\bm{k}_2,\bm{k}_3}\delta_{\bm{k}_4,\bm{k}'}+\langle [\psi^{\dag}_{\bm{k}_2}\psi_{\bm{k}_3}\psi_{\bm{k}_4},H'];\psi^{\dagger}_{\bm{k}'} \rangle_{\omega}\\
&=n_{\bm{k}_2}(\delta_{\bm{k}_2,\bm{k}_4}\delta_{\bm{k}_3,\bm{k}'}+\delta_{\bm{k}_2,\bm{k}_3}\delta_{\bm{k}_4,\bm{k}'})+\left(V^{(\bm{k}_4,\bm{k}_3)}_{(\bm{k}_2,\bm{k})}(n_{\bm{k}_2}(1+n_{\bm{k}_3}+n_{\bm{k}_4})-n_{\bm{k}_3}n_{\bm{k}_4})+\left(\delta_{\bm{k}_4,\bm{k}_2}+\delta_{\bm{k}_3,\bm{k}_2}\right)n_{\bm{k}_2}\Sigma^{(1)}_{\bm{k}}\right)\langle\psi_{\bm{k}};\psi^{\dagger}_{\bm{k}'}\rangle^0_{\omega}
\end{aligned}
\end{equation}
Thus the high-order renormalization effect can be summarized as
\begin{equation}
\begin{aligned}[b]
\frac{1}{2}\sum_{\bm{k}_i}V^{(\bm{k},\bm{k}_2)}_{(\bm{k}_3,\bm{k}_4)}&\langle\psi^{\dagger}_{\bm{k}_2}\psi_{\bm{k}_3}\psi_{\bm{k}_4};\psi^{\dag}_{\bm{k}'}\rangle_{\omega}=\frac{1}{2}\sum_{\bm{k}_i}V^{(\bm{k},\bm{k}_2)}_{(\bm{k}_3,\bm{k}_4)}(\omega-\mathcal{M})^{-1}n_{\bm{k}_2}(\delta_{\bm{k}_2,\bm{k}_4}\delta_{\bm{k}_3,\bm{k}'}+\delta_{\bm{k}_2,\bm{k}_3}\delta_{\bm{k}_4,\bm{k}'})\\
&+\frac{1}{2}\sum_{\bm{k}_i}V^{(\bm{k},\bm{k}_2)}_{(\bm{k}_3,\bm{k}_4)}(\omega-\mathcal{M})^{-1}\left(V^{(\bm{k}_4,\bm{k}_3)}_{(\bm{k}_2,\bm{k})}(n_{\bm{k}_2}(1+n_{\bm{k}_3}+n_{\bm{k}_4})-n_{\bm{k}_3}n_{\bm{k}_4})+\left(\delta_{\bm{k}_4,\bm{k}_2}+\delta_{\bm{k}_3,\bm{k}_2}\right)n_{\bm{k}_2}\Sigma^{(1)}_{\bm{k}}\right)\langle\psi_{\bm{k}};\psi^{\dagger}_{\bm{k}'}\rangle^0_{\omega}
\end{aligned}
\end{equation}
Where the first term in RHS can be simplified
\begin{equation}
\begin{aligned}[b]
\frac{1}{2}\sum_{\bm{k}_i}V^{(\bm{k},\bm{k}_2)}_{(\bm{k}_3,\bm{k}_4)}(\omega-\mathcal{M})^{-1}n_{\bm{k}_2}(\delta_{\bm{k}_2,\bm{k}_4}\delta_{\bm{k}_3,\bm{k}'}+\delta_{\bm{k}_2,\bm{k}_3}\delta_{\bm{k}_4,\bm{k}'})\approx\Sigma^{(1)}_{\bm{k}}\langle\psi_{\bm{k}};\psi^{\dagger}_{\bm{k}'}\rangle^0_{\omega}
\end{aligned}
\end{equation} 
using $(\omega-\mathcal{M})^{-1}n_{\bm{k}_2}(\delta_{\bm{k}_2,\bm{k}_4}\delta_{\bm{k}_3,\bm{k}'}+\delta_{\bm{k}_2,\bm{k}_3}\delta_{\bm{k}_4,\bm{k}'})=\frac{2n_{\bm{k}_2}}{\omega-M(\bm{k}')}$ and $\sum_{\bm{k}_2}V^{(\bm{k},\bm{k}_2)}_{(\bm{k},\bm{k}_2)}n_{\bm{k}_2}=\Sigma^{(1)}_{\bm{k}}$. The second term in RHS: 
\begin{equation}
\begin{aligned}[b]
&\frac{1}{2}\sum_{\bm{k}_i}V^{(\bm{k},\bm{k}_2)}_{(\bm{k}_3,\bm{k}_4)}(\omega-\mathcal{M})^{-1}\left(V^{(\bm{k}_4,\bm{k}_3)}_{(\bm{k}_2,\bm{k})}(n_{\bm{k}_2}(1+n_{\bm{k}_3}+n_{\bm{k}_4})-n_{\bm{k}_3}n_{\bm{k}_4})+\left(\delta_{\bm{k}_4,\bm{k}_2}+\delta_{\bm{k}_3,\bm{k}_2}\right)n_{\bm{k}_2}\Sigma^{(1)}_{\bm{k}}\right)\langle\psi_{\bm{k}};\psi^{\dagger}_{\bm{k}'}\rangle^0_{\omega}\\
&=\frac{1}{2}\sum_{\bm{k}_i}\frac{V^{(\bm{k},\bm{k}_2)}_{(\bm{k}_3,\bm{k}_4)}V^{(\bm{k}_4,\bm{k}_3)}_{(\bm{k}_2,\bm{k})}(n_{\bm{k}_2}(1+n_{\bm{k}_3}+n_{\bm{k}_4})-n_{\bm{k}_3}n_{\bm{k}_4})}{\omega-\mathcal{M}}\langle\psi_{\bm{k}};\psi^{\dagger}_{\bm{k}'}\rangle^0_{\omega}+\langle\psi_{\bm{k}};\psi^{\dagger}_{\bm{k}'}\rangle^0_{\omega}(\Sigma^{(1)}_{\bm{k}})^2\langle\psi_{\bm{k}};\psi^{\dagger}_{\bm{k}'}\rangle^0_{\omega}.
\end{aligned}
\end{equation}
We drop $\left(\Sigma^{(1)}_{\bm{k}}\right)^2$ term, since it is just 2 loops of first-order renormalization and can be reducible. The second-order self-energy now has the expression:
\begin{equation}
\begin{aligned}[b]
\Sigma^{(2)}_{\bm{k}}=\frac{1}{2}\sum_{\{\bm{k}_i\}}\frac{V^{(\bm{k},\bm{k}_2)}_{(\bm{k}_3,\bm{k}_4)}V^{(\bm{k}_4,\bm{k}_3)}_{(\bm{k}_2,\bm{k})}(n_{\bm{k}_2}(1+n_{\bm{k}_3}+n_{\bm{k}_4})-n_{\bm{k}_3}n_{\bm{k}_4})}{\omega_{\bm{k}}+i\epsilon-\mathcal{M}(\bm{k}_2,\bm{k}_3,\bm{k}_4)}
\end{aligned}
\end{equation}
with replacing the $\omega$ by $\omega_{\bm{k}}+i\epsilon$ where the infinitesimal imaginary part is the argument of the retarded Green's function, we can separate $\Sigma^{(2)}_{\bm{k}}=\Sigma^{(2)'}_{\bm{k}}-i\Sigma^{(2)''}_{\bm{k}}$, where the real part is the renormalization term and the imaginary part is the damping term.

Thus, we have the final Green's functions with second-order renormalizations:
\begin{equation}
\begin{aligned}[b]
\langle \psi_{\bm{k}};\psi^{\dagger}_{\bm{k}'}\rangle_{\omega+i\epsilon}&=\frac{\langle [\psi_{\bm{k}},\psi^{\dagger}_{\bm{k}'}] \rangle}{\omega+i\epsilon-M_{\bm{k}}}+\frac{1}{\omega+i\epsilon-M_{\bm{k}}}\Sigma^{(1)}_{\bm{k}}\langle\psi_{\bm{k}};\psi^{\dagger}_{\bm{k}'}\rangle^{(0)}+\frac{1}{\omega+i\epsilon-M_{\bm{k}}}\Sigma^{(2)}_{\bm{k}}\langle\psi_{\bm{k}};\psi^{\dagger}_{\bm{k}'}\rangle^{(0)}\\
&=\langle\psi_{\bm{k}};\psi^{\dagger}_{\bm{k}'}\rangle^{(0)}_{\omega+i\epsilon}+\langle\psi_{\bm{k}};\psi^{\dagger}_{\bm{k}'}\rangle^{(0)}_{\omega+i\epsilon}\Sigma^{(1)}_{\bm{k}}\langle\psi_{\bm{k}};\psi^{\dagger}_{\bm{k}'}\rangle^{(0)}_{\omega+i\epsilon}+\langle\psi_{\bm{k}};\psi^{\dagger}_{\bm{k}'}\rangle^{(0)}_{\omega+i\epsilon}\Sigma^{(2)}_{\bm{k}}(\omega_{\bm{k}}+i\epsilon)\langle\psi_{\bm{k}};\psi^{\dagger}_{\bm{k}'}\rangle^{(0)}_{\omega+i\epsilon},
\end{aligned}
\end{equation}
which follows the Dyson equation in spinor form. We have built the formalism for the many-body renormalization of a spinor. In order to calculate the spectrum structure, it is convenient to transform the sublattice basis to the eigenmode basis by the unitary in Eq.~(\ref{seq:11}):
\begin{equation}
\begin{aligned}[b]
\langle d_{\bm{k}};d^{\dagger}_{\bm{k}'}\rangle&=U^{\dag}\langle \psi_{\bm{k}};\psi^{\dagger}_{\bm{k}'}\rangle U\\
&=U^{\dag}\langle\psi_{\bm{k}};\psi^{\dagger}_{\bm{k}'}\rangle^{(0)}U+U^{\dag}\langle\psi_{\bm{k}};\psi^{\dagger}_{\bm{k}'}\rangle^{(0)}UU^{\dag}\Sigma^{(1)}_{\bm{k}}UU^{\dag}\langle\psi_{\bm{k}};\psi^{\dagger}_{\bm{k}'}\rangle^{(0)}U+U^{\dag}\langle\psi_{\bm{k}};\psi^{\dagger}_{\bm{k}'}\rangle^{(0)}UU^{\dag}\Sigma^{(2)}_{\bm{k}}UU^{\dag}\langle\psi_{\bm{k}};\psi^{\dagger}_{\bm{k}'}\rangle^{(0)}U\\
&=\langle d_{\bm{k}};d^{\dagger}_{\bm{k}'}\rangle^{0}+\langle d_{\bm{k}};d^{\dagger}_{\bm{k}'}\rangle^{0}U^{\dag}\Sigma^{(1)}_{\bm{k}}U\langle d_{\bm{k}};d^{\dagger}_{\bm{k}'}\rangle^{0}+\langle d_{\bm{k}};d^{\dagger}_{\bm{k}'}\rangle^{0}U^{\dag}\Sigma^{(2)}_{\bm{k}}U\langle d_{\bm{k}};d^{\dagger}_{\bm{k}'}\rangle^{0}.
\end{aligned}
\end{equation}
{\em Calculation rules and interaction matrix}--We notify the calculation rules since there are plenty of matrices and scalars in the formulae. First, we define the commutator $[\psi_{\bm{k}},\psi^{\dag}_{\bm{k}'}]$ of spinors:
\begin{equation}
\begin{aligned}[b]
&[\psi_{\bm{k}},\psi^{\dag}_{\bm{k}'}]=
\left[
\begin{pmatrix} 
 a_{\bm{k}}\\ 
 b_{\bm{k}}
\end{pmatrix}
,
\begin{pmatrix} 
 a^{\dag}_{\bm{k}'}&b^{\dag}_{\bm{k}'}\\ 
\end{pmatrix}
\right]=
\begin{pmatrix} 
 [a_{\bm{k}},a^{\dag}_{\bm{k}'}] &[a_{\bm{k}},b^{\dag}_{\bm{k}'}]\\ 
 [b_{\bm{k}},a^{\dag}_{\bm{k}'}] &[b_{\bm{k}},b^{\dag}_{\bm{k}'}]
\end{pmatrix}
=\delta_{\bm{k},\bm{k}'}\bm{I}\\
&\sum_{\bm{k}}\langle\psi^{\dag}_{\bm{k}}\psi_{\bm{k}}\rangle=\sum_{\bm{k}}\left(\langle a^{\dag}_{\bm{k}}a_{\bm{k}}\rangle +\langle a^{\dag}_{\bm{k}}b_{\bm{k}}\rangle+\langle b^{\dag}_{\bm{k}}a_{\bm{k}}\rangle + \langle b^{\dag}_{\bm{k}}b_{\bm{k}}\rangle \right)\\
&\sum_{\bm{k}}\psi^{\dag}_{\bm{k}}M_{\bm{k}}\psi_{\bm{k}}=\sum_{\bm{k}}\left(M^{11}_{\bm{k}}a^{\dag}_{\bm{k}}a_{\bm{k}} +M^{12}_{\bm{k}}a^{\dag}_{\bm{k}}b_{\bm{k}}+M^{21}_{\bm{k}}b^{\dag}_{\bm{k}}a_{\bm{k}} + M^{22}_{\bm{k}}b^{\dag}_{\bm{k}}b_{\bm{k}}\right)
\end{aligned}
\end{equation}
We can see that in $\sum_{\bm{k}}\psi^{\dag}_{\bm{k}}M_{\bm{k}}\psi_{\bm{k}}$, $M^{ab}_{\bm{k}}$ is always following the $\psi^{\dag}_{a,\bm{k}}\psi_{b,\bm{k}}$, which is nothing but matrix multiplication. Thus, if we keep in mind that we always have this expansion rule in expressions which have the form $\psi^{\dag}_{\bm{k}}A_{\bm{k}}\psi_{\bm{k}}$, then we can change the order to the form $A_{\bm{k}}\psi^{\dag}_{\bm{k}}\psi_{\bm{k}}$. This is why the interaction Hamiltonian can be written in the form: $\sum_{\{\bm{k}_i\}}V^{\bm{k}_1,\bm{k}_2}_{\bm{k}_3,\bm{k}_4}\psi^{\dagger}_{\bm{k}_1}\psi^{\dagger}_{\bm{k}_2}\psi_{\bm{k}_3}\psi_{\bm{k}_4}$. The reordering can bring a lot of conveniences in the formalism.

The interacting matrix $V^{\bm{k}_1,\bm{k}_2}_{\bm{k}_3,\bm{k}_3}$ is 4-dimensional matrix. The nonzero matrix elements from the interacting Hamiltonian are $V^{1,2}_{1,2}=-\frac{J}{N_L}\gamma_{\bm{k}_4-\bm{k}_2}$, $V^{1,2}_{2,2}=\frac{J}{4N_L}\gamma_{\bm{k}_1}$, $V^{2,1}_{1,1}=\frac{J}{4N_L}\gamma^*_{\bm{k}_1}$, $V^{2,2}_{2,1}=\frac{J}{4N_L}\gamma^*_{\bm{k}_4}$ and $V^{1,1}_{1,2}=\frac{J}{4N_L}\gamma_{\bm{k}_4}$. We use a $V^{a,b}_{c,d}$ short for $V^{a\bm{k}_1,b\bm{k}_2}_{c\bm{k}_3,d\bm{k}_4}$, and $a$,$b$,$c$,$d$ $\in \{1,2\}$ are the labels for the two components of spinor $\psi_{\bm{k}}(\psi^{\dag}_{\bm{k}})$. One can check the first-order renormalization $\Sigma^{(1)}_{\bm{k}}=\sum_{\bm{k}_2}V^{(\bm{k},\bm{k}_2)}_{(\bm{k}_2,\bm{k})}\langle\psi^{\dagger}_{\bm{k}_2}\psi_{\bm{k}_2}\rangle^0$. Since $\bm{k}_2$ is a dummy index, and the summation (contraction) of $\bm{k}_2$ leaves a 2x2 matrix. Inserting each nonzero interacting element gives rise to the explicit form in Eq.~(\ref{seq:10}).

\section{Interaction matrix element in eigenmode basis}\label{appen:D}
To calculate the second-order self-energy, we need to calculate each matrix element. We have the retarded Green's function:
\begin{equation}\label{}
\begin{aligned}[b]
G_R(\bm{k},\bm{k}';\omega)=\langle \psi_{\bm{k}};\psi^{\dagger}_{\bm{k}'}\rangle_{\omega}=\begin{pmatrix} 
 \langle a_{\bm{k}};a^{\dag}_{\bm{k}'}\rangle_{\omega} & \langle a_{\bm{k}}; b^{\dag}_{\bm{k}'} \rangle_{\omega}\\ 
 \langle b_{\bm{k}};a^{\dag}_{\bm{k}'}\rangle_{\omega} &\langle b_{\bm{k}}; b^{\dag}_{\bm{k}'}\rangle_{\omega}
\end{pmatrix}
\end{aligned}
\end{equation}
The equation of motion of Green's function can be written as 
\begin{equation}\label{}
\begin{aligned}[b]
\omega \langle \psi_{\bm{k}};\psi^{\dagger}_{\bm{k}'}\rangle_{\omega}&=\langle [\psi_{\bm{k}},\psi^{\dagger}_{\bm{k}'}] \rangle+\langle [\psi_{\bm{k}},H];\psi^{\dagger}_{\bm{k}'} \rangle_{\omega}\\
&=\langle [\psi_{\bm{k}},\psi^{\dagger}_{\bm{k}'}] \rangle+\langle [\psi_{\bm{k}},H_0];\psi^{\dagger}_{\bm{k}'} \rangle_{\omega}+\langle [\psi_{\bm{k}},H'];\psi^{\dagger}_{\bm{k}'} \rangle_{\omega}.
\end{aligned}
\end{equation}
We have the $[\psi_{\bm{k}},H']$, which can be written as
\begin{equation}
\begin{aligned}[b]
\Phi_{\bm{k}}=[\psi_k,H']=
-\frac{J}{4N_L}\sum_{\bm{k}_2,\bm{k}_3,\bm{k}_4}\begin{pmatrix} 
 4\gamma_{\bm{k}_4-\bm{k}_2}b^{\dag}_{\bm{k}_2}a_{\bm{k}_3}b_{\bm{k}_4}-\gamma_{\bm{k}}b^{\dag}_{\bm{k}_2}b_{\bm{k}_3}b_{\bm{k}_4}
 -\gamma^*_{\bm{k}_2}b^{\dag}_{\bm{k}_2}a_{\bm{k}_3}a_{\bm{k}_4}-2\gamma_{\bm{k}_4}a^{\dag}_{\bm{k}_2}a_{\bm{k}_3}b_{\bm{k}_4}\\ 
 4\gamma_{\bm{k}_4-\bm{k}}a^{\dag}_{\bm{k}_2}a_{\bm{k}_3}b_{\bm{k}_4}-\gamma_{\bm{k}_2}a^{\dag}_{\bm{k}_2}b_{\bm{k}_3}b_{\bm{k}_4}
 -\gamma^*_{\bm{k}}a^{\dag}_{\bm{k}_2}a_{\bm{k}_3}a_{\bm{k}_4}-2\gamma^*_{\bm{k}_4}b^{\dag}_{\bm{k}_2}b_{\bm{k}_3}a_{\bm{k}_4}
\end{pmatrix}.
\end{aligned}
\end{equation}
We have the equation motion of $\Phi_{\bm{k}}$:
\begin{equation}\label{}
\begin{aligned}[b]
\omega \langle \Phi_{\bm{k}};\psi^{\dagger}_{\bm{k}'} \rangle_{\omega}=\langle [\Phi_{\bm{k}},\psi^{\dagger}_{\bm{k}'}] \rangle+\langle [\Phi_{\bm{k}},H];\psi^{\dagger}_{\bm{k}'} \rangle_{\omega}=\omega
\begin{pmatrix} 
 \langle \Phi^1_{\bm{k}};a^{\dagger}_{\bm{k}'} \rangle_{\omega}&\langle \Phi^1_{\bm{k}};b^{\dagger}_{\bm{k}'} \rangle_{\omega}\\ 
\langle \Phi^2_{\bm{k}};a^{\dagger}_{\bm{k}'} \rangle_{\omega}&\langle \Phi^2_{\bm{k}};b^{\dagger}_{\bm{k}'} \rangle_{\omega}
\end{pmatrix}
\end{aligned}
\end{equation}
Where $\langle [\Phi_{\bm{k}},\psi^{\dagger}_{\bm{k}'}] \rangle$ has
\begin{equation}\label{}
\begin{aligned}[b]
\langle \cdots \rangle_{11}&=-\frac{J}{4N_L}\sum_{\bm{k}_2,\bm{k}_3,\bm{k}_4}
 4\gamma_{\bm{k}_4-\bm{k}_2}\langle b^{\dag}_{\bm{k}_2}b_{\bm{k}_4}\rangle\delta_{\bm{k}_3,\bm{k}'}
 -\gamma^*_{\bm{k}_2}(\langle b^{\dag}_{\bm{k}_2}a_{\bm{k}_3}\rangle\delta_{\bm{k}_4,\bm{k}'}+\langle b^{\dag}_{\bm{k}_2}a_{\bm{k}_4}\rangle\delta_{\bm{k}_3,\bm{k}'})-2\gamma_{\bm{k}_4}\langle a^{\dag}_{\bm{k}_2}b_{\bm{k}_4}\rangle\delta_{\bm{k}_3,\bm{k}'}\\
 \langle \cdots \rangle_{12}&=-\frac{J}{4N_L}\sum_{\bm{k}_2,\bm{k}_3,\bm{k}_4}
 4\gamma_{\bm{k}_4-\bm{k}_2}\langle b^{\dag}_{\bm{k}_2}a_{\bm{k}_3}\rangle\delta_{\bm{k}_4,\bm{k}'}
 -\gamma_{\bm{k}}(\langle b^{\dag}_{\bm{k}_2}b_{\bm{k}_3}\rangle\delta_{\bm{k}_4,\bm{k}'}+\langle b^{\dag}_{\bm{k}_2}b_{\bm{k}_4}\rangle\delta_{\bm{k}_3,\bm{k}'})-2\gamma_{\bm{k}_4}\langle a^{\dag}_{\bm{k}_2}a_{\bm{k}_3}\rangle\delta_{\bm{k}_4,\bm{k}'}\\
\langle \cdots \rangle_{21}&=-\frac{J}{4N_L}\sum_{\bm{k}_2,\bm{k}_3,\bm{k}_4}
4\gamma_{\bm{k}_4-\bm{k}}\langle a^{\dag}_{\bm{k}_2}b_{\bm{k}_4}\rangle\delta_{\bm{k}_3,\bm{k}'}
-\gamma^*_{\bm{k}}(\langle a^{\dag}_{\bm{k}_2}a_{\bm{k}_3}\rangle\delta_{\bm{k}_4,\bm{k}'}+\langle a^{\dag}_{\bm{k}_2}a_{\bm{k}_4}\rangle\delta_{\bm{k}_3,\bm{k}'})-2\gamma^*_{\bm{k}_4}\langle b^{\dag}_{\bm{k}_2}b_{\bm{k}_3}\rangle\delta_{\bm{k}_4,\bm{k}'}\\
 \langle \cdots \rangle_{22}&=-\frac{J}{4N_L}\sum_{\bm{k}_2,\bm{k}_3,\bm{k}_4}
 4\gamma_{\bm{k}_4-\bm{k}}\langle a^{\dag}_{\bm{k}_2}a_{\bm{k}_3}\rangle\delta_{\bm{k}_4,\bm{k}'}
 -\gamma_{\bm{k}_2}(\langle a^{\dag}_{\bm{k}_2}b_{\bm{k}_3}\rangle\delta_{\bm{k}_4,\bm{k}'}+\langle a^{\dag}_{\bm{k}_2}b_{\bm{k}_4}\rangle\delta_{\bm{k}_3,\bm{k}'})-2\gamma^*_{\bm{k}_4}\langle b^{\dag}_{\bm{k}_2}a_{\bm{k}_4}\rangle\delta_{\bm{k}_3,\bm{k}'}.
\end{aligned}
\end{equation}
Next, we have
\begin{equation}\label{}
\begin{aligned}[b]
[\Phi_{\bm{k}},H_0]=\begin{pmatrix}[\Phi^1_{\bm{k}},H_0]\\ [\Phi^2_{\bm{k}},H_0]\end{pmatrix}=-\frac{J}{4N_L}\sum_{\bm{k}_2,\bm{k}_3,\bm{k}_4}\begin{pmatrix} 
 [4\gamma_{\bm{k}_4-\bm{k}_2}b^{\dag}_{\bm{k}_2}a_{\bm{k}_3}b_{\bm{k}_4}-\gamma_{\bm{k}}b^{\dag}_{\bm{k}_2}b_{\bm{k}_3}b_{\bm{k}_4}
 -\gamma^*_{\bm{k}_2}b^{\dag}_{\bm{k}_2}a_{\bm{k}_3}a_{\bm{k}_4}-2\gamma_{\bm{k}_4}a^{\dag}_{\bm{k}_2}a_{\bm{k}_3}b_{\bm{k}_4},H_0]\\ 
 [4\gamma_{\bm{k}_4-\bm{k}}a^{\dag}_{\bm{k}_2}a_{\bm{k}_3}b_{\bm{k}_4}-\gamma_{\bm{k}_2}a^{\dag}_{\bm{k}_2}b_{\bm{k}_3}b_{\bm{k}_4}
 -\gamma^*_{\bm{k}}a^{\dag}_{\bm{k}_2}a_{\bm{k}_3}a_{\bm{k}_4}-2\gamma^*_{\bm{k}_4}b^{\dag}_{\bm{k}_2}b_{\bm{k}_3}a_{\bm{k}_4},H_0]
\end{pmatrix}.
\end{aligned}
\end{equation}
Inside, we have
\begin{equation}\label{}
\begin{aligned}[b]
[b^{\dag}_{\bm{k}_2}a_{\bm{k}_3}b_{\bm{k}_4},H_0]&=\left(M^{11}_{\bm{k}_3}+M^{22}_{\bm{k}_4}-M^{22}_{\bm{k}_2}\right)b^{\dag}_{\bm{k}_2}a_{\bm{k}_3}b_{\bm{k}_4}+M^{12}_{\bm{k}_3}b^{\dag}_{\bm{k}_2}b_{\bm{k}_3}b_{\bm{k}_4}
+M^{21}_{\bm{k}_4}b^{\dag}_{\bm{k}_2}a_{\bm{k}_3}a_{\bm{k}_4}-M^{12}_{\bm{k}_2}a^{\dag}_{\bm{k}_2}a_{\bm{k}_3}b_{\bm{k}_4}\\
[b^{\dag}_{\bm{k}_2}b_{\bm{k}_3}b_{\bm{k}_4},H_0]&=\left(M^{22}_{\bm{k}_3}+M^{22}_{\bm{k}_4}-M^{22}_{\bm{k}_2}\right)b^{\dag}_{\bm{k}_2}b_{\bm{k}_3}b_{\bm{k}_4}+M^{21}_{\bm{k}_3}b^{\dag}_{\bm{k}_2}a_{\bm{k}_3}b_{\bm{k}_4}+M^{21}_{\bm{k}_4}b^{\dag}_{\bm{k}_2}b_{\bm{k}_3}a_{\bm{k}_4}-M^{12}_{\bm{k}_2}a^{\dag}_{\bm{k}_2}b_{\bm{k}_3}b_{\bm{k}_4}\\
[b^{\dag}_{\bm{k}_2}a_{\bm{k}_3}a_{\bm{k}_4},H_0]&=\left(M^{11}_{\bm{k}_3}+M^{11}_{\bm{k}_4}-M^{22}_{\bm{k}_2}\right)b^{\dag}_{\bm{k}_2}a_{\bm{k}_3}a_{\bm{k}_4}+M^{12}_{\bm{k}_3}b^{\dag}_{\bm{k}_2}b_{\bm{k}_3}a_{\bm{k}_4}+M^{12}_{\bm{k}_4}b^{\dag}_{\bm{k}_2}a_{\bm{k}_3}b_{\bm{k}_4}-M^{12}_{\bm{k}_2}a^{\dag}_{\bm{k}_2}a_{\bm{k}_3}a_{\bm{k}_4}\\
[a^{\dag}_{\bm{k}_2}a_{\bm{k}_3}b_{\bm{k}_4},H_0]&=\left(M^{11}_{\bm{k}_3}+M^{22}_{\bm{k}_4}-M^{11}_{\bm{k}_2}\right)a^{\dag}_{\bm{k}_2}a_{\bm{k}_3}b_{\bm{k}_4}+M^{12}_{\bm{k}_3}a^{\dag}_{\bm{k}_2}b_{\bm{k}_3}b_{\bm{k}_4}+M^{21}_{\bm{k}_4}a^{\dag}_{\bm{k}_2}a_{\bm{k}_3}a_{\bm{k}_4}-M^{21}_{\bm{k}_2}b^{\dag}_{\bm{k}_2}a_{\bm{k}_3}b_{\bm{k}_4}\\
[a^{\dag}_{\bm{k}_2}b_{\bm{k}_3}b_{\bm{k}_4},H_0]&=\left(M^{22}_{\bm{k}_3}+M^{22}_{\bm{k}_4}-M^{11}_{\bm{k}_2}\right)a^{\dag}_{\bm{k}_2}b_{\bm{k}_3}b_{\bm{k}_4}+M^{21}_{\bm{k}_3}a^{\dag}_{\bm{k}_2}a_{\bm{k}_3}b_{\bm{k}_4}+M^{21}_{\bm{k}_4}a^{\dag}_{\bm{k}_2}b_{\bm{k}_3}a_{\bm{k}_4}-M^{21}_{\bm{k}_2}b^{\dag}_{\bm{k}_2}b_{\bm{k}_3}b_{\bm{k}_4}\\
[a^{\dag}_{\bm{k}_2}a_{\bm{k}_3}a_{\bm{k}_4},H_0]&=\left(M^{11}_{\bm{k}_3}+M^{11}_{\bm{k}_4}-M^{11}_{\bm{k}_2}\right)a^{\dag}_{\bm{k}_2}a_{\bm{k}_3}a_{\bm{k}_4}+M^{12}_{\bm{k}_3}a^{\dag}_{\bm{k}_2}b_{\bm{k}_3}a_{\bm{k}_4}+M^{12}_{\bm{k}_4}a^{\dag}_{\bm{k}_2}a_{\bm{k}_3}b_{\bm{k}_4}-M^{21}_{\bm{k}_2}b^{\dag}_{\bm{k}_2}a_{\bm{k}_3}a_{\bm{k}_4}\\
[b^{\dag}_{\bm{k}_2}b_{\bm{k}_3}a_{\bm{k}_4},H_0]&=\left(M^{22}_{\bm{k}_3}+M^{11}_{\bm{k}_4}-M^{22}_{\bm{k}_2}\right)b^{\dag}_{\bm{k}_2}b_{\bm{k}_3}a_{\bm{k}_4}+M^{21}_{\bm{k}_3}b^{\dag}_{\bm{k}_2}a_{\bm{k}_3}a_{\bm{k}_4}+M^{12}_{\bm{k}_4}b^{\dag}_{\bm{k}_2}b_{\bm{k}_3}b_{\bm{k}_4}-M^{12}_{\bm{k}_2}a^{\dag}_{\bm{k}_2}b_{\bm{k}_3}a_{\bm{k}_4}.
\end{aligned}
\end{equation}
Insert these terms into $[\Phi^1_{\bm{k}},H_0]$, we obtain
\begin{equation}\label{}
\begin{aligned}[b]
[\Phi^1_{\bm{k}},H_0]=&4\gamma_{\bm{k}_4-\bm{k}_2}[b^{\dag}_{\bm{k}_2}a_{\bm{k}_3}b_{\bm{k}_4},H_0]-\gamma_{\bm{k}}[b^{\dag}_{\bm{k}_2}b_{\bm{k}_3}b_{\bm{k}_4},H_0]-\gamma^*_{\bm{k}_2}[b^{\dag}_{\bm{k}_2}a_{\bm{k}_3}a_{\bm{k}_4},H_0]-2\gamma_{\bm{k}_4}[a^{\dag}_{\bm{k}_2}a_{\bm{k}_3}b_{\bm{k}_4},H_0]\\
=&4\left(\left(M^{11}_{\bm{k}_3}+M^{22}_{\bm{k}_4}-M^{22}_{\bm{k}_2}\right)b^{\dag}_{\bm{k}_2}a_{\bm{k}_3}b_{\bm{k}_4}+M^{12}_{\bm{k}_3}b^{\dag}_{\bm{k}_2}b_{\bm{k}_3}b_{\bm{k}_4}
+M^{21}_{\bm{k}_4}b^{\dag}_{\bm{k}_2}a_{\bm{k}_3}a_{\bm{k}_4}-M^{12}_{\bm{k}_2}a^{\dag}_{\bm{k}_2}a_{\bm{k}_3}b_{\bm{k}_4}\right)\gamma_{\bm{k}_4-\bm{k}_2}\\
&-\left(\left(M^{22}_{\bm{k}_3}+M^{22}_{\bm{k}_4}-M^{22}_{\bm{k}_2}\right)b^{\dag}_{\bm{k}_2}b_{\bm{k}_3}b_{\bm{k}_4}+M^{21}_{\bm{k}_3}b^{\dag}_{\bm{k}_2}a_{\bm{k}_3}b_{\bm{k}_4}+M^{21}_{\bm{k}_4}b^{\dag}_{\bm{k}_2}b_{\bm{k}_3}a_{\bm{k}_4}-M^{12}_{\bm{k}_2}a^{\dag}_{\bm{k}_2}b_{\bm{k}_3}b_{\bm{k}_4}\right)\gamma_{\bm{k}}\\
&-\left(\left(M^{11}_{\bm{k}_3}+M^{11}_{\bm{k}_4}-M^{22}_{\bm{k}_2}\right)b^{\dag}_{\bm{k}_2}a_{\bm{k}_3}a_{\bm{k}_4}+M^{12}_{\bm{k}_3}b^{\dag}_{\bm{k}_2}b_{\bm{k}_3}a_{\bm{k}_4}+M^{12}_{\bm{k}_4}b^{\dag}_{\bm{k}_2}a_{\bm{k}_3}b_{\bm{k}_4}-M^{12}_{\bm{k}_2}a^{\dag}_{\bm{k}_2}a_{\bm{k}_3}a_{\bm{k}_4}\right)\gamma^*_{\bm{k}_2}\\
&-2\left(\left(M^{11}_{\bm{k}_3}+M^{22}_{\bm{k}_4}-M^{11}_{\bm{k}_2}\right)a^{\dag}_{\bm{k}_2}a_{\bm{k}_3}b_{\bm{k}_4}+M^{12}_{\bm{k}_3}a^{\dag}_{\bm{k}_2}b_{\bm{k}_3}b_{\bm{k}_4}+M^{21}_{\bm{k}_4}a^{\dag}_{\bm{k}_2}a_{\bm{k}_3}a_{\bm{k}_4}-M^{12}_{\bm{k}_2}b^{\dag}_{\bm{k}_2}a_{\bm{k}_3}b_{\bm{k}_4}\right)\gamma_{\bm{k}_4}\\
=&\left(4\gamma_{\bm{k}_4-\bm{k}_2}\left(M^{11}_{\bm{k}_3}+M^{22}_{\bm{k}_4}-M^{22}_{\bm{k}_2}\right)-\gamma_{\bm{k}}M^{21}_{\bm{k}_3}-\gamma^{*}_{\bm{k}_2}M^{12}_{\bm{k}_4}+2\gamma_{\bm{k}_4}M^{12}_{\bm{k}_2}\right)b^{\dag}_{\bm{k}_2}a_{\bm{k}_3}b_{\bm{k}_4}\\
&+\left(4\gamma_{\bm{k}_4-\bm{k}_2}M^{12}_{\bm{k}_3}-\gamma_{\bm{k}}\left(M^{22}_{\bm{k}_3}+M^{22}_{\bm{k}_4}-M^{22}_{\bm{k}_2}\right)\right)b^{\dag}_{\bm{k}_2}b_{\bm{k}_3}b_{\bm{k}_4}+\left(4\gamma_{\bm{k}_4-\bm{k}_2}M^{21}_{\bm{k}_4}-\gamma^*_{\bm{k}_2}\left(M^{11}_{\bm{k}_3}+M^{11}_{\bm{k}_4}-M^{22}_{\bm{k}_2}\right)\right)b^{\dag}_{\bm{k}_2}a_{\bm{k}_3}a_{\bm{k}_4}\\
&-\left(4\gamma_{\bm{k}_4-\bm{k}_2}M^{12}_{\bm{k}_2}+2\gamma_{\bm{k}_4}\left(M^{11}_{\bm{k}_3}+M^{22}_{\bm{k}_4}-M^{11}_{\bm{k}_2}\right)\right)a^{\dag}_{\bm{k}_2}a_{\bm{k}_3}b_{\bm{k}_4}-\left(\gamma_{\bm{k}}M^{21}_{\bm{k}_4}+\gamma^*_{\bm{k}_2}M^{12}_{\bm{k}_3}\right)b^{\dag}_{\bm{k}_2}b_{\bm{k}_3}a_{\bm{k}_4}\\
&+\left(\gamma_{\bm{k}}M^{12}_{\bm{k}_2}-2\gamma_{\bm{k}_4}M^{12}_{\bm{k}_3}\right)a^{\dag}_{\bm{k}_2}b_{\bm{k}_3}b_{\bm{k}_4}+\left(\gamma^*_{\bm{k}_2}M^{12}_{\bm{k}_2}-2\gamma_{\bm{k}_4}M^{21}_{\bm{k}_4}\right)a^{\dag}_{\bm{k}_2}a_{\bm{k}_3}a_{\bm{k}_4}\\
=&\left(4\gamma_{\bm{k}_4-\bm{k}_2}\left(M^{11}_{\bm{k}_3}+M^{22}_{\bm{k}_4}-M^{22}_{\bm{k}_2}\right)-2\gamma_{\bm{k}}M^{21}_{\bm{k}_3}-2\gamma^{*}_{\bm{k}_2}M^{12}_{\bm{k}_4}+2\gamma_{\bm{k}_4}M^{12}_{\bm{k}_2}\right)b^{\dag}_{\bm{k}_2}a_{\bm{k}_3}b_{\bm{k}_4}\\
&+\left(4\gamma_{\bm{k}_4-\bm{k}_2}M^{12}_{\bm{k}_3}-\gamma_{\bm{k}}\left(M^{22}_{\bm{k}_3}+M^{22}_{\bm{k}_4}-M^{22}_{\bm{k}_2}\right)\right)b^{\dag}_{\bm{k}_2}b_{\bm{k}_3}b_{\bm{k}_4}+\left(4\gamma_{\bm{k}_4-\bm{k}_2}M^{21}_{\bm{k}_4}-\gamma^*_{\bm{k}_2}\left(M^{11}_{\bm{k}_3}+M^{11}_{\bm{k}_4}-M^{22}_{\bm{k}_2}\right)\right)b^{\dag}_{\bm{k}_2}a_{\bm{k}_3}a_{\bm{k}_4}\\
&-\left(4\gamma_{\bm{k}_4-\bm{k}_2}M^{12}_{\bm{k}_2}+2\gamma_{\bm{k}_4}\left(M^{11}_{\bm{k}_3}+M^{22}_{\bm{k}_4}-M^{11}_{\bm{k}_2}\right)\right)a^{\dag}_{\bm{k}_2}a_{\bm{k}_3}b_{\bm{k}_4}\\
&+\left(\gamma_{\bm{k}}M^{12}_{\bm{k}_2}-2\gamma_{\bm{k}_4}M^{12}_{\bm{k}_3}\right)a^{\dag}_{\bm{k}_2}b_{\bm{k}_3}b_{\bm{k}_4}+\left(\gamma^*_{\bm{k}_2}M^{12}_{\bm{k}_2}-2\gamma_{\bm{k}_4}M^{21}_{\bm{k}_4}\right)a^{\dag}_{\bm{k}_2}a_{\bm{k}_3}a_{\bm{k}_4}.
\end{aligned}
\end{equation}

The same for $[\Phi^2_{\bm{k}},H_0]$, we have
\begin{equation}\label{}
\begin{aligned}[b]
[\Phi^2_{\bm{k}},H_0]=&\left(4\gamma_{\bm{k}_4-\bm{k}}\left(M^{11}_{\bm{k}_3}+M^{22}_{\bm{k}_4}-M^{11}_{\bm{k}_2}\right)-\gamma_{\bm{k}_2}M^{21}_{\bm{k}_3}-\gamma^{*}_{\bm{k}}M^{12}_{\bm{k}_4}+2\gamma^*_{\bm{k}_4}M^{12}_{\bm{k}_2}\right)a^{\dag}_{\bm{k}_2}a_{\bm{k}_3}b_{\bm{k}_4}\\
&+\left(4\gamma_{\bm{k}_4-\bm{k}}M^{12}_{\bm{k}_3}-\gamma_{\bm{k}_2}\left(M^{22}_{\bm{k}_3}+M^{22}_{\bm{k}_4}-M^{11}_{\bm{k}_2}\right)\right)a^{\dag}_{\bm{k}_2}b_{\bm{k}_3}b_{\bm{k}_4}+\left(4\gamma_{\bm{k}_4-\bm{k}}M^{21}_{\bm{k}_4}-\gamma^*_{\bm{k}}\left(M^{11}_{\bm{k}_3}+M^{11}_{\bm{k}_4}-M^{11}_{\bm{k}_2}\right)\right)a^{\dag}_{\bm{k}_2}a_{\bm{k}_3}a_{\bm{k}_4}\\
&-\left(4\gamma_{\bm{k}_4-\bm{k}}M^{12}_{\bm{k}_2}+2\gamma^*_{\bm{k}_4}\left(M^{11}_{\bm{k}_3}+M^{22}_{\bm{k}_4}-M^{22}_{\bm{k}_2}\right)\right)b^{\dag}_{\bm{k}_2}a_{\bm{k}_3}b_{\bm{k}_4}-\left(\gamma_{\bm{k}_2}M^{21}_{\bm{k}_4}+\gamma^*_{\bm{k}}M^{12}_{\bm{k}_3}\right)a^{\dag}_{\bm{k}_2}b_{\bm{k}_3}a_{\bm{k}_4}\\
&+\left(\gamma_{\bm{k}_2}M^{21}_{\bm{k}_2}-2\gamma^*_{\bm{k}_4}M^{12}_{\bm{k}_4}\right)b^{\dag}_{\bm{k}_2}b_{\bm{k}_3}b_{\bm{k}_4}+\left(\gamma^*_{\bm{k}}M^{21}_{\bm{k}_2}-2\gamma^*_{\bm{k}_4}M^{21}_{\bm{k}_3}\right)b^{\dag}_{\bm{k}_2}a_{\bm{k}_3}a_{\bm{k}_4}
\end{aligned}
\end{equation}
As we can see, above terms are all included in $\sum_{abc}\sum_{def}V^{(\bm{k},a\bm{k}_2)}_{(b\bm{k}_3,c\bm{k}_4)}\mathcal{M}^{abs}_{def}\psi^{\dag d}_{\bm{k}_2}\psi^{e}_{\bm{k}_3}\psi^{f}_{\bm{k}_4}$.

We transform the interaction Hamiltonian $H'$ in eigenmode basis and show it with different scattering channels:
\begin{equation}\label{}
\begin{aligned}[b]
H'=&\sum_{\{\bm{k}_i\}}V^{(1)}d^{\dag}_{\bm{k}_1}d^{\dag}_{\bm{k}_2}d_{\bm{k}_3}d_{\bm{k}_4}+V^{(2)}u^{\dag}_{\bm{k}_1}u^{\dag}_{\bm{k}_2}u_{\bm{k}_3}u_{\bm{k}_4}+V^{(3)}u^{\dag}_{\bm{k}_1}u^{\dag}_{\bm{k}_2}d_{\bm{k}_3}u_{\bm{k}_4}+V^{(4)}d^{\dag}_{\bm{k}_1}d^{\dag}_{\bm{k}_2}d_{\bm{k}_3}u_{\bm{k}_4}+V^{(5)}u^{\dag}_{\bm{k}_1}d^{\dag}_{\bm{k}_2}u_{\bm{k}_3}d_{\bm{k}_4}\\
&+V^{(6)}u^{\dag}_{\bm{k}_1}u^{\dag}_{\bm{k}_2}d_{\bm{k}_3}d_{\bm{k}_4}+V^{(7)}d^{\dag}_{\bm{k}_1}d^{\dag}_{\bm{k}_2}u_{\bm{k}_3}u_{\bm{k}_4}+V^{(8)}u^{\dag}_{\bm{k}_1}d^{\dag}_{\bm{k}_2}u_{\bm{k}_3}u_{\bm{k}_4}+V^{(9)}u^{\dag}_{\bm{k}_1}d^{\dag}_{\bm{k}_2}d_{\bm{k}_3}d_{\bm{k}_4}.
\end{aligned}
\end{equation}
One can see that each band has eight scattering channels. We list the interaction matrix elements of each channel:
\begin{equation}\label{}
\begin{aligned}[b]
V^{(1)}&=\frac{-J}{4N_L}(4\gamma_{\bm{k}_4-\bm{k}_2}U^{11*}_{\bm{k}_1}U^{21*}_{\bm{k}_2}U^{11}_{\bm{k}_3}U^{21}_{\bm{k}_4}-\gamma_{\bm{k}_1}U^{11*}_{\bm{k}_1}U^{21*}_{\bm{k}_2}U^{21}_{\bm{k}_3}U^{21}_{\bm{k}_4}-\gamma^*_{\bm{k}_1}U^{21*}_{\bm{k}_1}U^{11*}_{\bm{k}_2}U^{11}_{\bm{k}_3}U^{11}_{\bm{k}_4}-\gamma^*_{\bm{k}_4}U^{21*}_{\bm{k}_1}U^{21*}_{\bm{k}_2}U^{21}_{\bm{k}_3}U^{11}_{\bm{k}_4}-\gamma_{\bm{k}_4}U^{11*}_{\bm{k}_1}U^{11*}_{\bm{k}_2}U^{11}_{\bm{k}_3}U^{21}_{\bm{k}_4})\\
V^{(2)}&=\frac{-J}{4N_L}(4\gamma_{\bm{k}_4-\bm{k}_2}U^{12*}_{\bm{k}_1}U^{22*}_{\bm{k}_2}U^{12}_{\bm{k}_3}U^{22}_{\bm{k}_4}-\gamma_{\bm{k}_1}U^{12*}_{\bm{k}_1}U^{22*}_{\bm{k}_2}U^{22}_{\bm{k}_3}U^{22}_{\bm{k}_4}-\gamma^*_{\bm{k}_1}U^{22*}_{\bm{k}_1}U^{12*}_{\bm{k}_2}U^{12}_{\bm{k}_3}U^{12}_{\bm{k}_4}-\gamma^*_{\bm{k}_4}U^{22*}_{\bm{k}_1}U^{22*}_{\bm{k}_2}U^{22}_{\bm{k}_3}U^{12}_{\bm{k}_4}-\gamma_{\bm{k}_4}U^{12*}_{\bm{k}_1}U^{12*}_{\bm{k}_2}U^{12}_{\bm{k}_3}U^{22}_{\bm{k}_4})\\
V^{(3)}&=\frac{-J}{4N_L}\left(4\gamma_{\bm{k}_4-\bm{k}_2}U^{12*}_{\bm{k}_1}U^{22*}_{\bm{k}_2}U^{11}_{\bm{k}_3}U^{22}_{\bm{k}_4}+4\gamma_{\bm{k}_3-\bm{k}_2}U^{12*}_{\bm{k}_1}U^{22*}_{\bm{k}_2}U^{21}_{\bm{k}_3}U^{12}_{\bm{k}_4}-2\gamma_{\bm{k}_1}U^{12*}_{\bm{k}_1}U^{22*}_{\bm{k}_2}U^{21}_{\bm{k}_3}U^{22}_{\bm{k}_4}-2\gamma^*_{\bm{k}_1}U^{22*}_{\bm{k}_1}U^{12*}_{\bm{k}_2}U^{11}_{\bm{k}_3}U^{12}_{\bm{k}_4} \right. \\ 
&\left.-\gamma^*_{\bm{k}_4}U^{22*}_{\bm{k}_1}U^{22*}_{\bm{k}_2}U^{21}_{\bm{k}_3}U^{12}_{\bm{k}_4}
 -\gamma^*_{\bm{k}_3}U^{22*}_{\bm{k}_1}U^{22*}_{\bm{k}_2}U^{11}_{\bm{k}_3}U^{22}_{\bm{k}_4}-\gamma_{\bm{k}_4}U^{12*}_{\bm{k}_1}U^{12*}_{\bm{k}_2}U^{11}_{\bm{k}_3}U^{22}_{\bm{k}_4}-\gamma_{\bm{k}_3}U^{12*}_{\bm{k}_1}U^{12*}_{\bm{k}_2}U^{21}_{\bm{k}_3}U^{12}_{\bm{k}_4}\right)\\
 V^{(4)}&=\frac{-J}{4N_L}\left(4\gamma_{\bm{k}_4-\bm{k}_2}U^{11*}_{\bm{k}_1}U^{21*}_{\bm{k}_2}U^{11}_{\bm{k}_3}U^{22}_{\bm{k}_4}+4\gamma_{\bm{k}_3-\bm{k}_2}U^{11*}_{\bm{k}_1}U^{21*}_{\bm{k}_2}U^{21}_{\bm{k}_3}U^{12}_{\bm{k}_4}-2\gamma_{\bm{k}_1}U^{11*}_{\bm{k}_1}U^{21*}_{\bm{k}_2}U^{21}_{\bm{k}_3}U^{22}_{\bm{k}_4}-2\gamma^*_{\bm{k}_1}U^{21*}_{\bm{k}_1}U^{11*}_{\bm{k}_2}U^{11}_{\bm{k}_3}U^{12}_{\bm{k}_4}\right. \\
 &\left. -\gamma^*_{\bm{k}_4}U^{21*}_{\bm{k}_1}U^{21*}_{\bm{k}_2}U^{21}_{\bm{k}_3}U^{12}_{\bm{k}_4} -\gamma^*_{\bm{k}_3}U^{21*}_{\bm{k}_1}U^{21*}_{\bm{k}_2}U^{11}_{\bm{k}_3}U^{22}_{\bm{k}_4}-\gamma_{\bm{k}_4}U^{11*}_{\bm{k}_1}U^{11*}_{\bm{k}_2}U^{11}_{\bm{k}_3}U^{22}_{\bm{k}_4}-\gamma_{\bm{k}_3}U^{11*}_{\bm{k}_1}U^{11*}_{\bm{k}_2}U^{21}_{\bm{k}_3}U^{12}_{\bm{k}_4}\right)\\
 V^{(5)}&=\frac{-J}{4N_L}\left(4\gamma_{\bm{k}_4-\bm{k}_2}U^{12*}_{\bm{k}_1}U^{21*}_{\bm{k}_2}U^{12}_{\bm{k}_3}U^{21}_{\bm{k}_4}+4\gamma_{\bm{k}_3-\bm{k}_2}U^{12*}_{\bm{k}_1}U^{21*}_{\bm{k}_2}U^{22}_{\bm{k}_3}U^{11}_{\bm{k}_4}+4\gamma_{\bm{k}_4-\bm{k}_1}U^{22*}_{\bm{k}_1}U^{11*}_{\bm{k}_2}U^{12}_{\bm{k}_3}U^{21}_{\bm{k}_4}+4\gamma_{\bm{k}_3-\bm{k}_1}U^{22*}_{\bm{k}_1}U^{11*}_{\bm{k}_2}U^{22}_{\bm{k}_3}U^{11}_{\bm{k}_4} \right. \\
 &\left.-2\gamma_{\bm{k}_1}U^{12*}_{\bm{k}_1}U^{21*}_{\bm{k}_2}U^{22}_{\bm{k}_3}U^{21}_{\bm{k}_4}
 -2\gamma_{\bm{k}_2}U^{22*}_{\bm{k}_1}U^{11*}_{\bm{k}_2}U^{22}_{\bm{k}_3}U^{21}_{\bm{k}_4}-2\gamma^*_{\bm{k}_1}U^{22*}_{\bm{k}_1}U^{11*}_{\bm{k}_2}U^{12}_{\bm{k}_3}U^{11}_{\bm{k}_4}
 -2\gamma^*_{\bm{k}_2}U^{12*}_{\bm{k}_1}U^{21*}_{\bm{k}_2}U^{12}_{\bm{k}_3}U^{11}_{\bm{k}_4}
 -2\gamma^*_{\bm{k}_4}U^{22*}_{\bm{k}_1}U^{21*}_{\bm{k}_2}U^{22}_{\bm{k}_3}U^{11}_{\bm{k}_4}
 \right. \\
 &\left.-2\gamma^*_{\bm{k}_3}U^{22*}_{\bm{k}_1}U^{21*}_{\bm{k}_2}U^{12}_{\bm{k}_3}U^{21}_{\bm{k}_4}-2\gamma_{\bm{k}_4}U^{12*}_{\bm{k}_1}U^{11*}_{\bm{k}_2}U^{12}_{\bm{k}_3}U^{21}_{\bm{k}_4}
-2\gamma_{\bm{k}_3}U^{12*}_{\bm{k}_1}U^{11*}_{\bm{k}_2}U^{22}_{\bm{k}_3}U^{11}_{\bm{k}_4}
 \right)\\
 V^{(6)}&=\frac{-J}{4N_L}(4\gamma_{\bm{k}_4-\bm{k}_2}U^{12*}_{\bm{k}_1}U^{22*}_{\bm{k}_2}U^{11}_{\bm{k}_3}U^{21}_{\bm{k}_4}-\gamma_{\bm{k}_1}U^{12*}_{\bm{k}_1}U^{22*}_{\bm{k}_2}U^{21}_{\bm{k}_3}U^{21}_{\bm{k}_4}-\gamma^*_{\bm{k}_1}U^{22*}_{\bm{k}_1}U^{12*}_{\bm{k}_2}U^{11}_{\bm{k}_3}U^{11}_{\bm{k}_4}-\gamma^*_{\bm{k}_4}U^{22*}_{\bm{k}_1}U^{22*}_{\bm{k}_2}U^{21}_{\bm{k}_3}U^{11}_{\bm{k}_4}-\gamma_{\bm{k}_4}U^{12*}_{\bm{k}_1}U^{12*}_{\bm{k}_2}U^{11}_{\bm{k}_3}U^{21}_{\bm{k}_4})\\
V^{(7)}&=\frac{-J}{4N_L}(4\gamma_{\bm{k}_4-\bm{k}_2}U^{11*}_{\bm{k}_1}U^{21*}_{\bm{k}_2}U^{12}_{\bm{k}_3}U^{22}_{\bm{k}_4}-\gamma_{\bm{k}_1}U^{11*}_{\bm{k}_1}U^{21*}_{\bm{k}_2}U^{22}_{\bm{k}_3}U^{22}_{\bm{k}_4}-\gamma^*_{\bm{k}_1}U^{21*}_{\bm{k}_1}U^{11*}_{\bm{k}_2}U^{12}_{\bm{k}_3}U^{12}_{\bm{k}_4}-\gamma^*_{\bm{k}_4}U^{21*}_{\bm{k}_1}U^{21*}_{\bm{k}_2}U^{22}_{\bm{k}_3}U^{12}_{\bm{k}_4}-\gamma_{\bm{k}_4}U^{11*}_{\bm{k}_1}U^{11*}_{\bm{k}_2}U^{12}_{\bm{k}_3}U^{22}_{\bm{k}_4})\\
V^{(8)}&=\frac{-J}{4N_L}\left(4\gamma_{\bm{k}_4-\bm{k}_2}U^{12*}_{\bm{k}_1}U^{21*}_{\bm{k}_2}U^{12}_{\bm{k}_3}U^{22}_{\bm{k}_4}+4\gamma_{\bm{k}_4-\bm{k}_1}U^{22*}_{\bm{k}_1}U^{11*}_{\bm{k}_2}U^{12}_{\bm{k}_3}U^{22}_{\bm{k}_4}-\gamma_{\bm{k}_1}U^{12*}_{\bm{k}_1}U^{21*}_{\bm{k}_2}U^{22}_{\bm{k}_3}U^{22}_{\bm{k}_4}-\gamma_{\bm{k}_2}U^{22*}_{\bm{k}_1}U^{11*}_{\bm{k}_2}U^{22}_{\bm{k}_3}U^{22}_{\bm{k}_4}-\gamma^*_{\bm{k}_1}U^{22*}_{\bm{k}_1}U^{11*}_{\bm{k}_2}U^{12}_{\bm{k}_3}U^{12}_{\bm{k}_4} \right. \\ 
&\left.-\gamma^*_{\bm{k}_2}U^{12*}_{\bm{k}_1}U^{21*}_{\bm{k}_2}U^{12}_{\bm{k}_3}U^{12}_{\bm{k}_4}-2\gamma^*_{\bm{k}_4}U^{22*}_{\bm{k}_1}U^{21*}_{\bm{k}_2}U^{22}_{\bm{k}_3}U^{12}_{\bm{k}_4}
 -2\gamma_{\bm{k}_4}U^{12*}_{\bm{k}_1}U^{11*}_{\bm{k}_2}U^{12}_{\bm{k}_3}U^{22}_{\bm{k}_4}\right)\\
 V^{(9)}&=\frac{-J}{4N_L}\left(4\gamma_{\bm{k}_4-\bm{k}_2}U^{12*}_{\bm{k}_1}U^{21*}_{\bm{k}_2}U^{11}_{\bm{k}_3}U^{21}_{\bm{k}_4}+4\gamma_{\bm{k}_4-\bm{k}_1}U^{22*}_{\bm{k}_1}U^{11*}_{\bm{k}_2}U^{11}_{\bm{k}_3}U^{21}_{\bm{k}_4}-\gamma_{\bm{k}_1}U^{12*}_{\bm{k}_1}U^{21*}_{\bm{k}_2}U^{21}_{\bm{k}_3}U^{21}_{\bm{k}_4}-\gamma_{\bm{k}_2}U^{22*}_{\bm{k}_1}U^{11*}_{\bm{k}_2}U^{21}_{\bm{k}_3}U^{21}_{\bm{k}_4}-\gamma^*_{\bm{k}_1}U^{22*}_{\bm{k}_1}U^{11*}_{\bm{k}_2}U^{11}_{\bm{k}_3}U^{11}_{\bm{k}_4} \right. \\ 
&\left.-\gamma^*_{\bm{k}_2}U^{12*}_{\bm{k}_1}U^{21*}_{\bm{k}_2}U^{11}_{\bm{k}_3}U^{11}_{\bm{k}_4}-2\gamma^*_{\bm{k}_4}U^{22*}_{\bm{k}_1}U^{21*}_{\bm{k}_2}U^{21}_{\bm{k}_3}U^{11}_{\bm{k}_4}
 -2\gamma_{\bm{k}_4}U^{12*}_{\bm{k}_1}U^{11*}_{\bm{k}_2}U^{11}_{\bm{k}_3}U^{21}_{\bm{k}_4}\right)
\end{aligned}
\end{equation}
We have employed the unitary transformation $U$:
\begin{equation}
\begin{aligned}[b]
U=\frac{1}{\sqrt{2}}
\begin{pmatrix} 
\sqrt{1+\frac{B}{A}}e^{i\frac{\phi_{\bm{k}}}{2}}&\sqrt{1-\frac{B}{A}}e^{i\frac{\phi_{\bm{k}}}{2}}\\ 
\sqrt{1-\frac{B}{A}}e^{-i\frac{\phi_{\bm{k}}}{2}}&-\sqrt{1+\frac{B}{A}}e^{-i\frac{\phi_{\bm{k}}}{2}}
\end{pmatrix}.
\end{aligned}
\end{equation}

\twocolumngrid
\section{Microscopic theory of electric polarization in parametric pumping}\label{appen:E}
The microscopic theory of electric polarization theory in pumping magnon can be as early as fifty-five years ago. In Moriya's paper, the spin Hamiltonian under the external electric field $\bm{E}$ was given by
\begin{equation}
\begin{aligned}[b]
H=H_0+\bm{E}(t)\cdot\bm{P},
\end{aligned}
\end{equation}
where $\bm{P}$ is the electric dipole, which can be written as $\bm{P}=\sum_{i}\bm{p}_i+\sum_{j,l}\bm{p}_{jl}+\cdots$. $\bm{p}_i$ is the electric dipole moment associated with the $i$th lattice spin and $\bm{p}_{jl}$ the one associated with the ($j$, $l$) pair of ionic spins.
\begin{equation}
\begin{aligned}[b]
p^{\alpha}_i&=\sum_{\beta\gamma}K^{\alpha,\beta\gamma}_iS_{i\beta}S_{i\gamma}\\
p^{\alpha}_{jl}&=\pi^{\alpha}_{jl}\bm{S}_j\cdot\bm{S}_l+\sum_{\beta\gamma}\Gamma^{\alpha,(\beta\gamma)}_{jl}S_{j\beta}S_{l\gamma}+\sum_{\beta\gamma}D^{\alpha,[\beta\gamma]}_{jl}S_{j\beta}S_{l\gamma}.
\end{aligned}
\end{equation}
where the single spin terms $P_j$ are related to the linear Stark effect and vanish when the site has a center of inversion. $\pi_{jl}$ and $\Gamma_{jl}$ vanish when there is an inversion symmetry of the lattice. But with a DM interaction-induced anisotropy in ferromagnets, $\pi_{jl}$ can also give rise to nonzero contribution. The antisymmetric part $D^{\alpha,[\beta\gamma]}$ general does not vanish. Thus, to have a larger coupling strength with external light, the inversion symmetry should be broken. We note that there is a small staggered onsite lattice potential applied in our honeycomb lattice. 
Since the A and B sublattices are not the centers of inversion, the linear Stark terms can survive. Thus single spin terms contribute to the polarization tensor, which is given in the Hamiltonian:
\begin{equation}
\begin{aligned}[b] 
H=\sum_{i}\bm{E}(t)\cdot\bm{p}_i=\sum_{i,\alpha\beta\gamma}E^{\alpha}(t)K^{\alpha,\beta\gamma}_iS_{i\beta}S_{i\gamma}
\end{aligned}
\end{equation}
Since we concentrate on the case that lacks of inversion symmetry, the honeycomb lattice has a $C_3$ point-group symmetry which guarantees the finite value of the single spin contribution. The microscopic origin of spin-dependent electric polarization can be tracked by the spin–orbit coupling (SOC)\cite{Matsumoto2017}. The estimation of coefficient $K^{\alpha,\beta\gamma}_i$ due to SOC-induced $d-p$ orbital hybridization is given by\cite{Jia2006, Jia2007, Matsumoto2017}:
\begin{equation}
\begin{aligned}[b] 
K^{\alpha,\beta\gamma}=\eta_0 eb.
\end{aligned}
\end{equation}
where $b$ is the effective electric dipole by the consequence of the hybridization\cite{Matsumoto2017}, and $\eta_0$ is a constant that is related to the different point symmetries of the lattices. The product of external EM field $\bm{E}(t)$ and the coefficient tensor $K^{\alpha,\beta\gamma}$ gives rise to the coupling strength. Thus, the coupling strength $\epsilon$ by the single spin terms is defined as: 
\begin{equation}
\begin{aligned}[b]
\epsilon=\bm{E}\cdot \bm{K}^{\alpha,\beta\gamma}_i=\eta_0 eb E_0.
\end{aligned}
\end{equation}

Apart from the leading-order single-spin terms, the paired-spin terms will also contribute to the electric polarization. Following the spirit in Ref.\cite{Zhu2014}, we consider the electric polarization in the smallest triangle formed by three spins ($\bm{S}_i$,$\bm{S}_j$ and $\bm{S}_k$, ($\bm{S}_i$, $\bm{S}_j$) and ($\bm{S}_j$, $\bm{S}_k$) are nearest spin pairs, while ($\bm{S}_i$, $\bm{S}_k$) is the next-nearest pair) in a honeycomb lattice; thus we have the polarization $\bm{P}_{ij}$ the bond-$(i,j)$\cite{Malz2019}
\begin{equation}
\begin{aligned}[b]
\bm{p}_{ij}=\bm{p}_{0,ij}\left[\bm{S}_i\cdot\mathfrak{J}_{ij}\bm{S}_j\cos\theta_{ijk}+\bm{n}_{ijk}\cdot \bm{S}_i\times\mathfrak{J}_{ij}\bm{S}_j\sin\theta_{ijk}\right],
\end{aligned}
\end{equation}
and 
\begin{equation}
\begin{aligned}[b]
&\mathfrak{J}_{ij}\bm{S}_j=\\
&\cos(2\theta_{ij})\bm{S}_j+\sin(2\theta_{ij})(\bm{S}_j\times \bm{n}_{ij})+2\sin^2(\theta_{ij})\bm{n}_{ij}(\bm{n}_{ij}\cdot\bm{S}_j).
\end{aligned}
\end{equation}
where $\bm{p}_{0,ij}=8ea\frac{t_{ij}t_{jk}t_{ki}}{U^3}(\bm{e}_{jk}-\bm{e}_{ki})$, $\bm{e}_{ij}$ is the vector on nearest neighbour bonds pointing from site-$i$ to site-$j$, $\theta_{ij}$ is the spin-orbit angle, and $\bm{n}_{ij}$ is the unit vector that can be treated as the effective Zeeman field by spin-orbit interaction. Since $(k,i)$ is the next-nearest spin pair, the hopping integral $t_{ki}=s_0t_{ij}$ with $s_0$ is the ratio constant dependents on the lattice geometries. The coupling strength by the paired spin terms is thus given by 
\begin{equation}
\begin{aligned}[b]
\epsilon=\bm{E}\cdot \bm{p}_{0,ij}=8ea\frac{t_{ij}t_{jk}t_{ki}}{U^3}E_0\approx\frac{8s_0eat^3}{U^3}E_0.
\end{aligned}
\end{equation}
where $t$ is the nearest-neighbour hopping parameter. Finally, the total coupling strength $\epsilon=\left(\frac{8s_0at^3}{U^3}+\eta_0b\right)eE_0$. Due to the smallness of $\frac{t^3}{u^3}$, the coupling strength by single spin terms is the leading order term.

\bibliographystyle{apsrev4-2}
\bibliography{Int_magnon.bib} 
 
\end{document}